\pdfoutput=1

\documentclass[11pt,twoside,a4paper,cmspaper,final,collab]{cms-tdr}
\def\svnVersion{415962}\def\cmsCernNoTag{CERN-EP/2017-030}\def\cmsCernDate{\today}\def\cmsMessage{Published in the European Physical Journal C as \href{http://dx.doi.org/10.1140/epjc/s10052-017-4984-5}{\doi{10.1140/epjc/s10052-017-4984-5.}}}

\begin{document}\cmsNoteHeader{TOP-14-013}

\hyphenation{had-ron-i-za-tion}
\hyphenation{cal-or-i-me-ter}
\hyphenation{de-vices}
\RCS$Revision: 415904 $
\RCS$HeadURL: svn+ssh://svn.cern.ch/reps/tdr2/papers/TOP-14-013/trunk/TOP-14-013.tex $
\RCS$Id: TOP-14-013.tex 415904 2017-07-12 14:04:33Z ozenaiev $

\newcommand{\PYTHIAsix} {\PYTHIA{}6\xspace}
\newcommand{\HERWIGsix} {\HERWIG{}6\xspace}
\newcommand{\MadSpin} {\textsc{MadSpin}\xspace}
\newcommand{\MadPyt}{ {\MADGRAPH{}+\PYTHIAsix}\xspace}
\newcommand{\PowPyt} {{\POWHEG{}+\PYTHIAsix}\xspace}
\newcommand{\PowHer} {{\POWHEG{}+\HERWIGsix}\xspace}
\newcommand{\MCHer} {{\MCATNLO{}+\HERWIGsix}\xspace}
\newcommand{\mtt} {\ensuremath{M(\ttbar)}\xspace}
\newcommand{\ptt} {\ensuremath{\pt(\PQt)}\xspace}
\newcommand{\yt} {\ensuremath{y(\PQt)}\xspace}
\newcommand{\ytt} {\ensuremath{y(\ttbar)}\xspace}
\newcommand{\detatt} {\ensuremath{\Delta \eta(\PQt,\PAQt)}\xspace}
\newcommand{\dphitt} {\ensuremath{\Delta \phi(\PQt,\PAQt)}\xspace}
\newcommand{\pttt} {\ensuremath{\pt(\ttbar)}\xspace}
\newcommand{\abm} {ABM11\xspace}
\newcommand{\cj} {CJ15\xspace}
\newcommand{\ct} {CT14\xspace}
\newcommand{\herapdf} {HERAPDF2.0\xspace}
\newcommand{\jr} {JR14\xspace}
\newcommand{\mmht} {MMHT2014\xspace}
\newcommand{\nnpdf} {NNPDF3.0\xspace}
\newcommand{\chisq}{\ensuremath{\chi^2}\xspace}
\newcommand{\ndf}{dof\xspace}
\newcommand{\chisqndf}{\ensuremath{\chi^2}/\ndf\xspace}
\newcommand{\xfitter} {\textsc{xFitter}\xspace}
\newcommand{\herafitter} {\textsc{HERAFitter}\xspace}
\newcommand{\applgrid} {\textsc{ApplGrid}\xspace}
\newcommand{\qcdnum} {\textsc{qcdnum}\xspace}
\newcommand{\difftop} {{\textsc{DiffTop}}\xspace}
\newcommand{\fastnlo} {{\textsc{fastNLO}}\xspace}
\newcommand{\lhapdf} {{\textsc{lhapdf}}\xspace}
\newcommand{\Wasymm} {\ensuremath{\PW^\pm} boson charge asymmetry\xspace}

\newlength\cmsFigOne
\setlength\cmsFigOne{0.80\textwidth}
\newlength\cmsFigWidth
\ifthenelse{\boolean{cms@external}}{\setlength\cmsFigWidth{0.98\columnwidth}}{\setlength\cmsFigWidth{0.8\textwidth}}
\ifthenelse{\boolean{cms@external}}{\providecommand{\cmsLeft}{top}}{\providecommand{\cmsLeft}{left}}
\ifthenelse{\boolean{cms@external}}{\providecommand{\cmsRight}{bottom}}{\providecommand{\cmsRight}{right}}
\ifthenelse{\boolean{cms@external}}{\providecommand{\cmsTableSwitch}[1]{#1}}{\providecommand{\cmsTableSwitch}[1]{\resizebox{\textwidth}{!}{#1}}}
\providecommand{\cmsTable}[1]{\resizebox{\textwidth}{!}{#1}}
\cmsNoteHeader{TOP-14-013}
\title{Measurement of double-differential cross sections for top quark pair production in pp collisions at $\sqrt{s} = 8$\TeV and impact on parton distribution functions}
\titlerunning{Double-differential cross sections for top quark pair production}

\date{\today}

\abstract{Normalized double-differential cross sections for top quark pair (\ttbar) production are measured in pp collisions
at a centre-of-mass energy of 8\TeV with the CMS experiment at the LHC. The analyzed data correspond to an integrated luminosity of 19.7\fbinv.
The measurement is performed in the dilepton $\Pe^{\pm}\mu^{\mp}$ final state.
The \ttbar cross section is determined as a function of various pairs of observables characterizing the kinematics of the top quark and \ttbar system.
The data are compared to calculations using perturbative quantum chromodynamics at next-to-leading and approximate next-to-next-to-leading orders.
They are also compared to predictions of Monte Carlo event generators that complement fixed-order computations with parton showers, hadronization, and multiple-parton interactions.
Overall agreement is observed with the predictions, which is improved when the latest global sets of proton parton distribution functions are used.
The inclusion of the measured \ttbar cross sections in a fit of parametrized parton distribution functions is shown to have significant impact on the gluon distribution.}

\hypersetup{
pdfauthor={CMS Collaboration},
pdftitle={Measurement of double-differential cross sections for top quark pair production in pp collisions at sqrt(s) = 8 TeV and impact on parton distribution functions},
pdfsubject={CMS},
pdfkeywords={CMS, top, cross sections, kinematic, double-differential}}

\maketitle

\section{Introduction}
\label{sec:intro}

Understanding the production and properties of the top quark,
discovered in 1995 at the Fermilab Tevatron~\cite{Abachi:1995iq,Abe:1995hr}, is fundamental in testing the standard model
and searching for new phenomena.
A large sample of proton-proton (pp) collision events containing a top quark pair (\ttbar)
has been recorded at the CERN LHC, facilitating precise top quark measurements.
In particular, precise measurements of the \ttbar production cross section as a function of \ttbar kinematic observables
have become possible, which allow for the validation of the most-recent predictions of perturbative quantum chromodynamics (QCD).
At the LHC, top quarks are predominantly produced via gluon-gluon fusion.
Thus, using measurements of the production cross section in a global fit of the parton distribution functions (PDFs)
can help to better determine the gluon distribution at large values of $x$, where $x$ is the fraction of the proton momentum carried by a parton~\cite{Czakon:2013tha,Guzzi:2014wia,Czakon:2016olj}.
In this context, \ttbar measurements are complementary to studies~\cite{Chatrchyan:2012bja,Khachatryan:2014waa,CMSjets8tev}
that exploit inclusive jet production cross sections at the LHC.

Normalized differential cross sections for \ttbar production
have been measured previously
in proton-antiproton collisions
at the Tevatron at a centre-of-mass energy of 1.96\TeV \cite{Aaltonen:2009iz,Abazov:2014vga}
and in pp collisions at the LHC
at $\sqrt{s} = 7$\TeV~\cite{bib:ATLAS,bib:TOP-11-013_paper,bib:ATLASnew,Aaboud:2016iot},
8\TeV~\cite{Khachatryan:2015oqa,Aad:2015mbv,Aaboud:2016iot}, and 13\TeV~\cite{Khachatryan:2016mnb}.
This paper presents the measurement of the normalized double-differential $\ttbar + \mathrm{X}$ production cross section,
where X is inclusive in the number of extra jets in the event but excludes $\ttbar +\Z/\PW/\gamma$ production.
The cross section is measured as a function of observables describing the kinematics of the top quark and \ttbar:
the transverse momentum of the top quark, \ptt, the rapidity of the top quark, \yt,
the transverse momentum, \pttt,
the rapidity, \ytt,
and the invariant mass, \mtt, of \ttbar,
the pseudorapidity between the top quark and antiquark, \detatt, and
the angle between the top quark and antiquark in the transverse plane, \dphitt.
In total, the double-differential \ttbar cross section is measured as a function of six different pairs of kinematic variables.

These measurements provide a sensitive test of the standard model by probing the details of the \ttbar production dynamics.
The double-differential measurement is expected to impose stronger constraints on the gluon distribution than single-differential measurements
owing to the improved resolution of the momentum fractions carried by the two incoming partons.

The analysis uses the data recorded at $\sqrt{s}=8$\TeV by the CMS experiment in 2012, corresponding to an integrated luminosity of $19.7 \pm 0.5\fbinv$.
The measurement is performed using the $\Pe^{\pm}\mu^{\mp}$ decay mode ($\Pe\mu$) of \ttbar,
requiring two oppositely charged leptons and at least two jets.
The analysis largely follows the procedures of the single-differential \ttbar cross section measurement~\cite{Khachatryan:2015oqa}.
The restriction to the $\Pe\mu$ channel provides a pure \ttbar event sample because of the negligible contamination
from $\Z/\gamma^{*}$ processes with same-flavour leptons in the final state.

The measurements are defined at parton level and thus are corrected for the effects of hadronization and detector resolutions and inefficiencies.
A regularized unfolding process is performed simultaneously in bins of the two variables in which the cross sections are measured.
The normalized differential \ttbar cross section is determined by dividing by the measured total inclusive \ttbar production cross section, where
the latter is evaluated by integrating over all bins in the two observables.
The parton level results are compared to different theoretical predictions from
leading-order (LO) and next-to-leading-order (NLO) Monte Carlo (MC) event generators, as well as with fixed-order NLO~\cite{Mangano:1991jk}
and approximate next-to-next-to-leading-order (NNLO)~\cite{Kidonakis:2001nj} calculations using several different PDF sets.
Parametrized PDFs are fitted to the data in a procedure that is referred to as the PDF fit.

{\tolerance=1200
The structure of the paper is as follows:
in Section~\ref{sec:cms} a brief description of the CMS detector is given.
Details of the event simulation are provided in Section~\ref{sec:samples}.
The event selection, kinematic reconstruction, and comparisons between data and simulation are provided in Section~\ref{sec:sel}.
The two-dimensional unfolding procedure is detailed in Section \ref{sec:unfold};
the method to determine the double-differential cross sections is presented in Section~\ref{sec:cs},
and the assessment of the systematic uncertainties is described in Section~\ref{sec:syst}.
The results of the measurement are discussed and compared to theoretical predictions in Section~\ref{sec:results}.
Section~\ref{sec:qcdanalysis} presents the PDF fit.
Finally, Section~\ref{sec:concl} provides a summary.
}

\section{The CMS detector}
\label{sec:cms}

The central feature of the CMS apparatus is a superconducting solenoid of 13\unit{m} length and 6\unit{m} inner diameter, which provides an axial magnetic field of 3.8\unit{T}. Within the field volume are a silicon pixel and strip tracker, a lead tungstate crystal electromagnetic calorimeter (ECAL), and a brass and scintillator hadron calorimeter (HCAL), each composed of a barrel and two endcap sections.
Extensive forward calorimetry complements the coverage provided by the barrel and endcap sections up to $\abs{\eta}<5.2$.
Charged particle trajectories are measured by the inner tracking system, covering a range of $\abs{\eta}<2.5$. The ECAL and HCAL surround the tracking volume, providing high-resolution energy and direction measurements of electrons, photons, and hadronic jets up to $\abs{\eta}<3$. Muons are measured in gas-ionization detectors embedded in the steel flux-return yoke outside the solenoid covering the region $\abs{\eta}<2.4$. The detector is nearly hermetic, allowing momentum balance measurements in the plane transverse to the beam directions. A more detailed description of the CMS detector, together with a definition of the coordinate system and the relevant kinematic variables, can be found in Ref.~\cite{bib:JINST}.

\section{Signal and background modelling}\label{sec:samples}

The $\ttbar$ signal process is simulated using the matrix element event generator \MADGRAPH (version 5.1.5.11)~\cite{MadGraph}, together with the \MadSpin~\cite{bib:madspin} package
for the modelling of spin correlations. The \PYTHIAsix program (version 6.426)~\cite{Sjostrand:2006za} is used to model parton showering and hadronization. In the signal simulation,
the mass of the top quark, $m_{\PQt}$, is fixed to 172.5\GeV. The proton structure is described by the CTEQ6L1 PDF set~\cite{Pumplin:2002vw}.
The same programs are used to model dependencies on the renormalization and factorization scales, $\mu_\mathrm{r}$ and $\mu_\mathrm{f}$, respectively,
the matching threshold between jets produced at the matrix-element level and via parton showering,
and $m_{\PQt}$.

The cross sections obtained in this paper are also compared to theoretical calculations obtained  with  the  NLO
event generators \POWHEG (version 1.0 r1380)~\cite{powhegnew1,powhegnew2,powhegnew3}, interfaced with \PYTHIAsix
or \textsc{Herwig6} (version 6.520)~\cite{bib:HERWIG} for the subsequent parton showering and hadronization, and \MCATNLO (version 3.41)~\cite{mcatnlo}, interfaced with \HERWIG.
Both \PYTHIAsix and \HERWIGsix include a modelling of multiple-parton interactions and the underlying event.
The \PYTHIAsix Z2* tune~\cite{bib:Z2startune} is used to characterize the underlying event in both the \ttbar and the background simulations.
The \HERWIGsix AUET2 tune~\cite{bib:auet2tune} is used to model the underlying event in the \PowHer simulation,
while the default tune is used in the \MCHer simulation.
The PDF sets CT10~\cite{Lai:2010vv} and CTEQ6M~\cite{Pumplin:2002vw} are used for \POWHEG and \MCATNLO, respectively.
Additional simulated event samples generated with \POWHEG and interfaced with \PYTHIAsix or \HERWIGsix are
used to assess the systematic uncertainties related to the modelling
of the hard-scattering process and hadronization, respectively, as described in Section~\ref{sec:syst}.

The production of W and Z/$\gamma^{*}$ bosons with additional jets, respectively referred to as W+jets and Z/$\gamma^{*}$+jets in the following,
and $\ttbar +\Z/\PW/\gamma$ backgrounds are simulated using \MADGRAPH,
while $\PW$ boson plus associated single top quark production ($\PQt\PW$) is simulated using \POWHEG.
The showering and hadronization is modelled with \PYTHIAsix for these processes.
Diboson (WW, WZ, and ZZ) samples, as well as QCD multijet backgrounds, are produced with \PYTHIAsix.
All of the background simulations are normalized to the fixed-order theoretical predictions as described in Ref.~\cite{Khachatryan:2015oqa}.
The CMS detector response is simulated using \GEANTfour (version~9.4)~\cite{Agostinelli:2002hh}.
 \section{Event selection}\label{sec:sel}

The event selection follows closely the one reported in Ref.~\cite{Khachatryan:2015oqa}.
The top quark decays almost exclusively into a W boson and a bottom quark,
and only events in which the two W bosons decay into exactly one electron and one muon and corresponding neutrinos are considered.
Events are triggered by requiring one electron and one muon of opposite charge, one of which is required to have $\pt > 17$\GeV and the other $\pt > 8$\GeV.

Events are reconstructed using a particle-flow (PF) technique~\cite{CMS-PAS-PFT-09-001,CMS-PAS-PFT-10-001},
which combines signals from all subdetectors to enhance the reconstruction and identification of the individual particles observed in pp~collisions.
An interaction vertex~\cite{trkpas}
is required within 24\unit{cm} of the detector centre along the beam line direction,
and within 2\unit{cm} of the beam line in the transverse plane.
Among all such vertices, the primary vertex of an event
is identified as the one with the largest value
of the sum of the $\pt^2$ of the associated tracks.
Charged hadrons from pileup events, \ie those originating from additional pp interactions within the same or nearby bunch crossing, are subtracted on an event-by-event basis.
Subsequently, the remaining neutral-hadron component from pileup is accounted for through jet energy corrections~\cite{bib:PUSubtraction}.

Electron candidates are reconstructed from a combination of the track momentum at the primary vertex,
the corresponding energy deposition in the ECAL, and the energy sum of all bremsstrahlung photons associated with the track~\cite{bib:ele2013}.
Muon candidates are reconstructed using the track information from the silicon tracker and the muon system.
An event is required to contain at least two oppositely charged leptons, one electron and one muon,
each with $\pt > 20\GeV$ and $\abs{\eta} < 2.4$.
Only the electron and the muon with the highest \pt are considered for the analysis. 
The invariant mass of the selected electron and muon must be larger than 20\GeV to suppress events from decays of heavy-flavour resonances.
The leptons are required to be isolated with $I_\text{rel}\leq 0.15$ inside a cone
in $\eta$-$\phi$ space of $\Delta R = \sqrt{(\Delta\eta)^{2} + (\Delta\phi)^{2}} = 0.3$ around the lepton track,
where $\Delta\eta$ and $\Delta\phi$ are the differences in pseudorapidity and azimuthal angle (in radians), respectively, between the directions of the lepton and any other particle.
The parameter $I_\text{rel}$ is the relative isolation parameter defined as the sum of transverse energy deposits inside the cone from charged and neutral hadrons, and photons,
relative to the lepton $\pt$, corrected for pileup effects.
The efficiencies of the lepton isolation were determined in $Z$ boson data samples using the ``tag-and-probe'' method
of Ref.~\cite{Chatrchyan:2011cm},
and are found to be well described by the simulation for both electrons and muons.
The overall difference between data and simulation
is estimated to be ${<} 2\%$ for electrons,
and ${<}1\%$ for muons.
The simulation is adjusted for this by using correction factors parametrized
as a function of the lepton \pt and $\eta$ and applied to simulated events,
separately for electrons and muons.

Jets are reconstructed by clustering the PF candidates using the anti-\kt clustering algorithm~\cite{Cacciari:2008gp,Cacciari:2011ma}
with a distance parameter $R = 0.5$.
Electrons and muons passing less-stringent selections on lepton kinematic quantities and isolation,
relative to those specified above, are identified but excluded from clustering.
A jet is selected if it has $\pt > 30\GeV$ and $\abs{\eta} < 2.4$.
Jets originating from the hadronization of b quarks (b jets) are identified using an algorithm~\cite{bib:btag004}
that provides a b tagging discriminant by combining secondary vertices and track-based lifetime information.
This provides a b tagging efficiency of ${\approx}80$--85\% for b jets
and a mistagging efficiency of  ${\approx}10\%$ for jets originating from gluons, as well as u, d, or s quarks,
and ${\approx}30$--40\% for jets originating from c quarks~\cite{bib:btag004}.
Events are selected if they contain at least two jets, and at least one of these jets is b-tagged.
These requirements are chosen to reduce the background contribution while keeping a large fraction of the \ttbar signal.
The \PQb tagging efficiency is adjusted in the simulation with the correction factors parametrized
as a function of the jet \pt and $\eta$.

The missing transverse momentum vector is defined as the projection on the plane perpendicular
to the beams of the negative vector sum of the momenta of all PF particles in an event~\cite{bib:MET}.
Its magnitude is referred to as \ptmiss.
To mitigate the pileup effects on the \ptmiss resolution,
a multivariate correction is used
where the measured momentum is separated into components that originate from the primary and from other interaction vertices~\cite{bib:mvamet}.
No selection requirement on \ptmiss is applied.

The \ttbar kinematic properties are determined from the four-momenta of the decay products using
the same kinematic reconstruction method~\cite{Abbott:1997fv,LSpaper} as that of the single-differential \ttbar measurement~\cite{Khachatryan:2015oqa}.
The six unknown quantities are the three-momenta of the two neutrinos, which are reconstructed by imposing the following six kinematic constraints: \pt conservation in the event and
the masses of the W bosons, top quark, and top antiquark.
The top quark and antiquark are required to have a mass of 172.5\GeV.
It is assumed that the \ptmiss in the event results from the two neutrinos in the top quark and antiquark decay chains. 
To resolve the ambiguity due to multiple algebraic solutions of the equations for the neutrino momenta, 
the solution with the smallest invariant mass of the \ttbar system is taken.
The reconstruction is performed 100 times, each time randomly smearing the measured energies and directions
of the reconstructed leptons and jets within their resolution.
This smearing recovers events that yielded no solution because of measurement fluctuations.
The three-momenta of the two neutrinos are determined as a weighted average over all the smeared solutions.
For each solution, the weight is calculated based on the expected invariant mass spectrum of a lepton and a bottom jet 
as the product of two weights for the top quark and antiquark decay chains. 
All possible lepton-jet combinations in the event are considered. 
Combinations are ranked based on the presence of b-tagged jets in the assignments, \ie a combination with 
both leptons assigned to b-tagged jets is preferred over those with one or no b-tagged jet. 
Among assignments with equal number of b-tagged jets, the one with the highest average weight is chosen.
Events with no solution after smearing are discarded.
The method yields an average reconstruction efficiency of ${\approx}95\%$,
which is determined in simulation as the fraction of selected signal events (which include only direct \ttbar decays via the $\Pe^{\pm}\mu^{\mp}$ channel, i.e.\ excluding cascade decays via $\tau$ leptons) passing the kinematic reconstruction.
The overall difference in this efficiency between data and simulation is estimated to be ${\approx}1\%$,
and a corresponding correction factor is applied to the simulation~\cite{Korolthesis}.
A more detailed description of the kinematic reconstruction procedure can be found in Ref.~\cite{Korolthesis}.

In total, 38\,569 events are selected in the data.
The signal contribution to the event sample is 79.2\%, as estimated from the simulation.
The remaining fraction of events is dominated by \ttbar decays other than via the $\Pe^{\pm}\mu^{\mp}$ channel (14.2\%).
Other sources of background are single top quark production (3.6\%), Z/$\gamma^{*}$+jets events (1.4\%), associated $\ttbar +\Z/\PW/\gamma$ production (1.1\%),
and a negligible (${<} 0.5\%$) fraction of diboson, W+jets, and QCD multijet events.

\begin{figure*}[htbp]
\centering
  \includegraphics[width=0.40\textwidth]{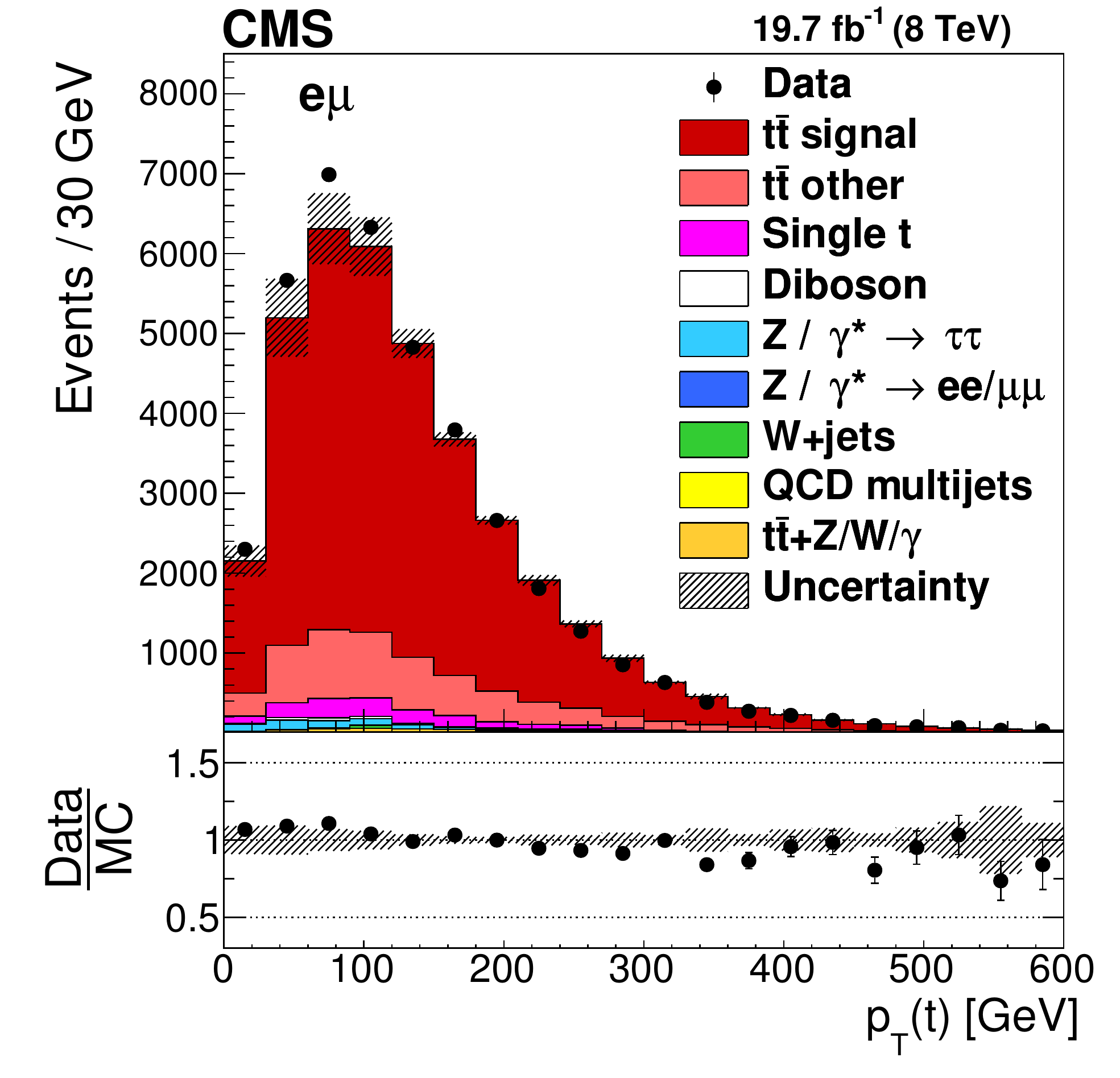}
  \includegraphics[width=0.40\textwidth]{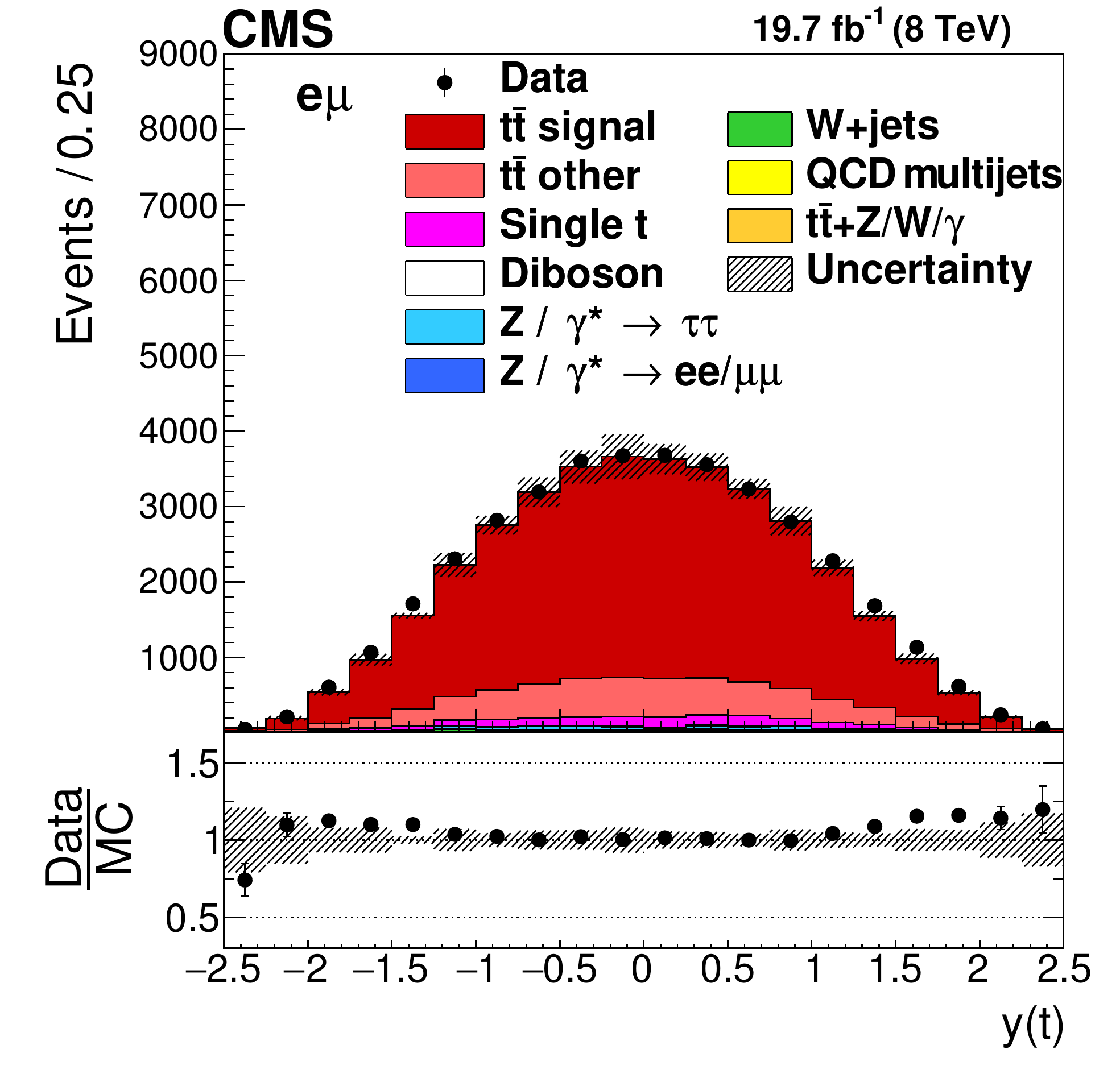}
  \includegraphics[width=0.40\textwidth]{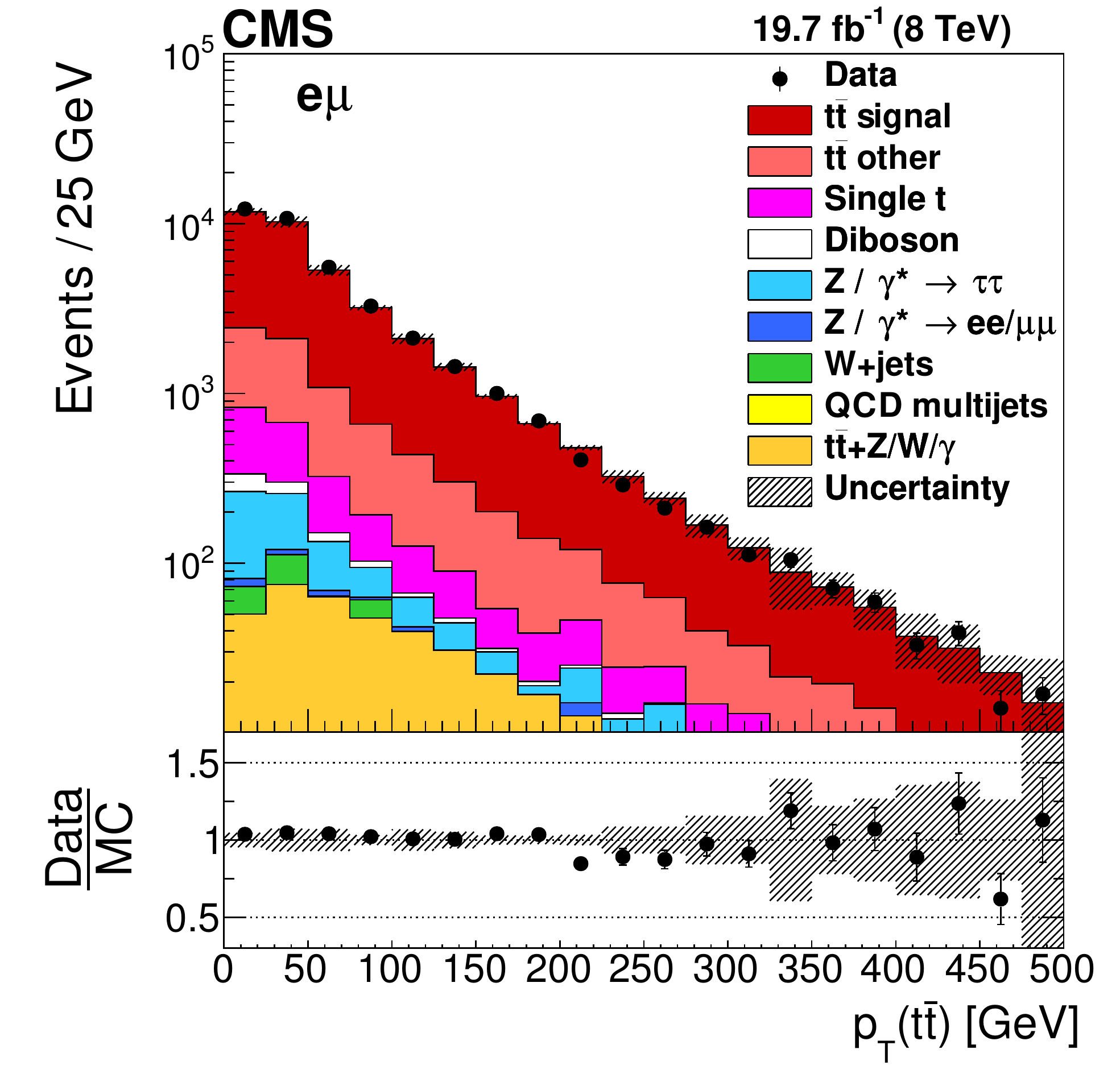}
  \includegraphics[width=0.40\textwidth]{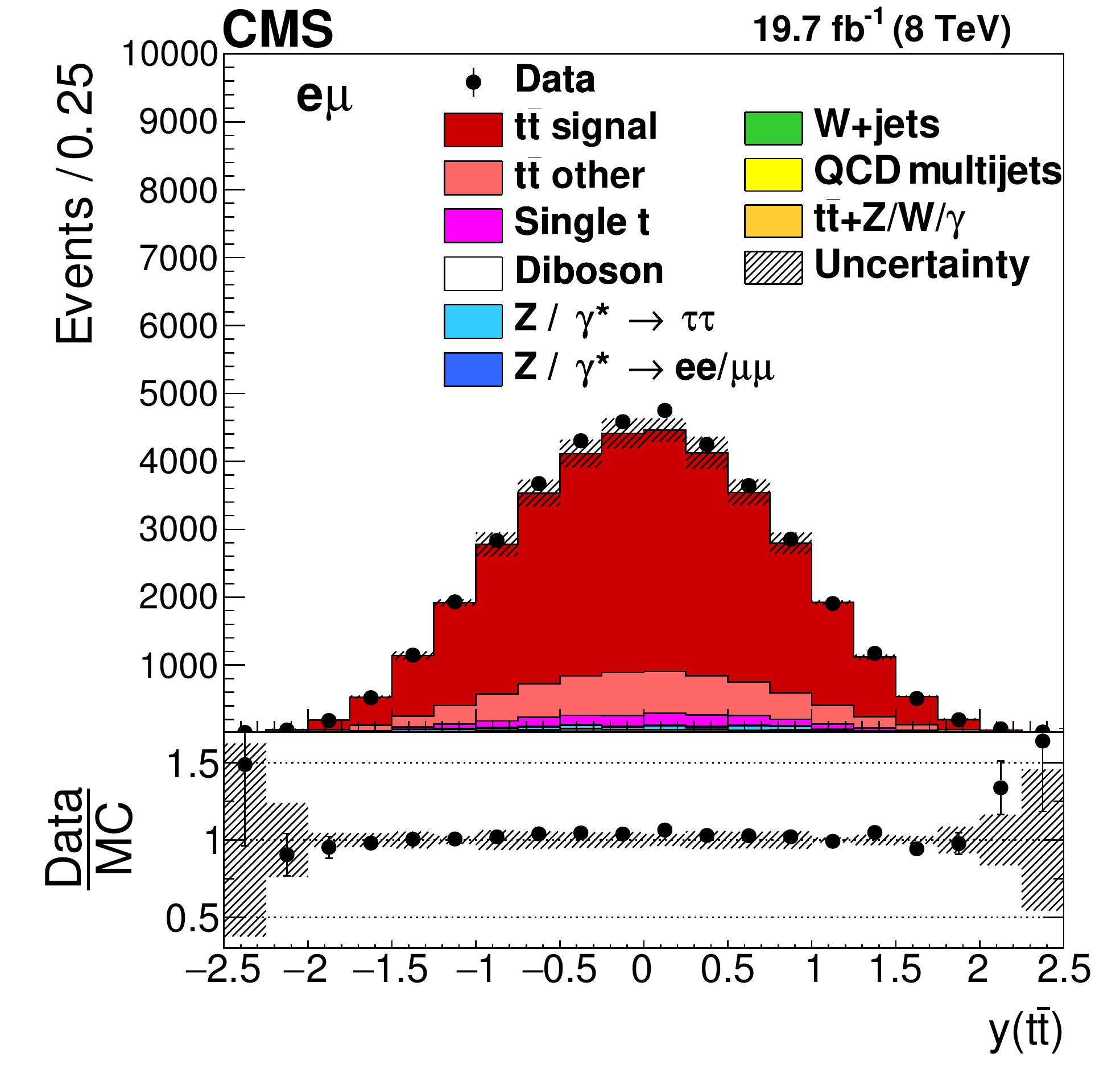}
  \includegraphics[width=0.40\textwidth]{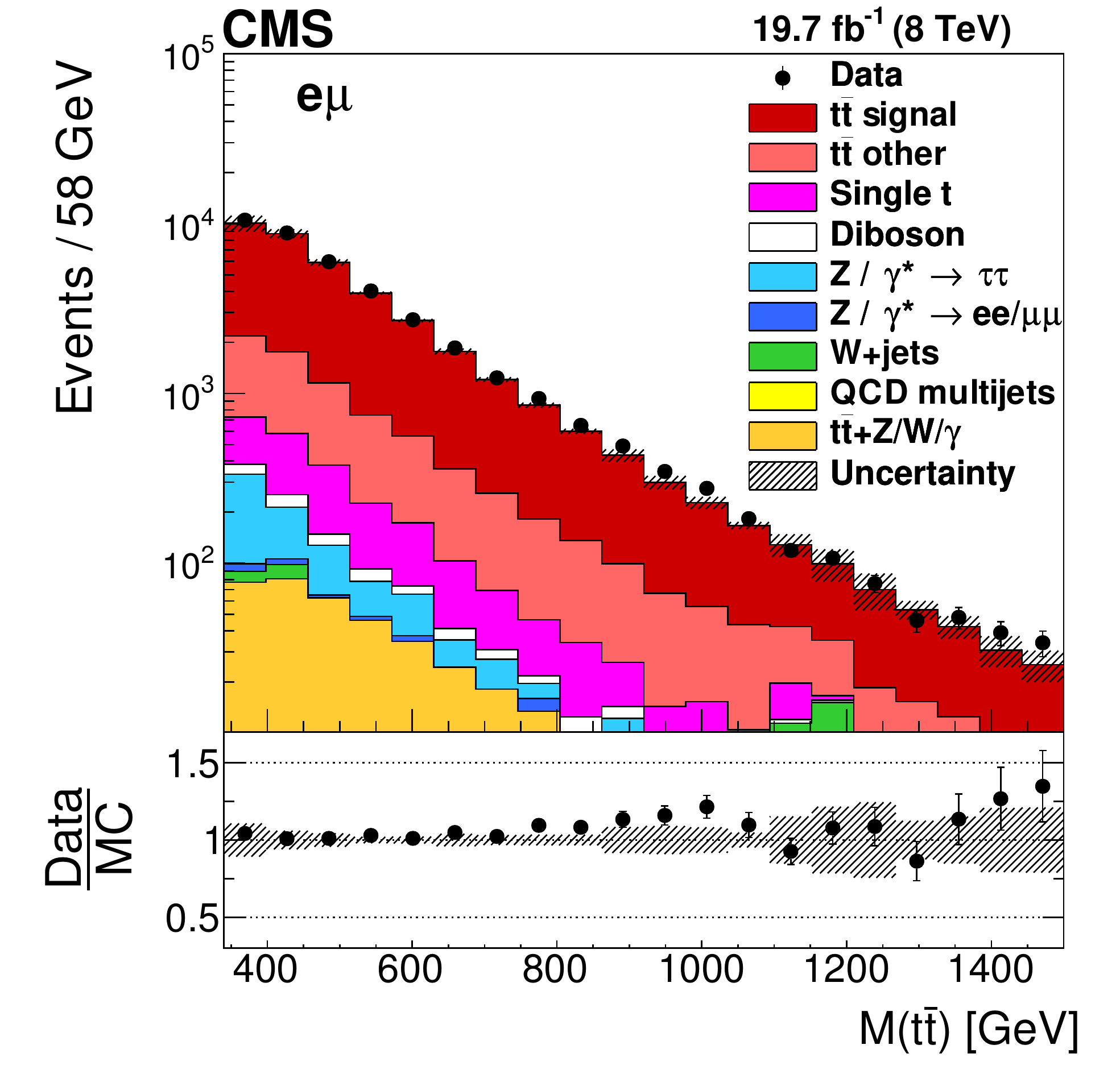}
\caption{Distributions of $\pt(\PQt)$ (upper left), $y(\PQt)$ (upper right), $\pt(\ttbar)$ (middle left), $y(\ttbar)$ (middle right), and \mtt (lower)
 in selected events after the kinematic reconstruction. The experimental data with the vertical bars corresponding to their statistical uncertainties
 are plotted together with distributions of simulated signal and different background processes.
 The hatched regions correspond to the shape uncertainties in the signal and backgrounds (cf.\ Section~\ref{sec:syst}).
 The lower panel in each plot shows the ratio of the observed data event yields to those expected in the simulation.
 }
\label{fig:CPkinTop}
\end{figure*}

Figure~\ref{fig:CPkinTop} shows the distributions of the reconstructed top quark and \ttbar kinematic variables.
In general, the data are reasonably well described by the simulation,
however some trends are visible.
In particular, the simulation shows a harder \ptt spectrum than the data,
as observed in previous measurements~\cite{bib:TOP-11-013_paper,bib:ATLASnew,Aaboud:2016iot,Khachatryan:2015oqa,Aad:2015mbv,Aaboud:2016iot,Khachatryan:2016mnb}.
The \ytt distribution is found to be less central in the simulation than in the data,
while an opposite behavior is observed in the \yt distribution.
The \mtt and \pttt distributions are overall well described by the simulation.

 \section{Signal extraction and unfolding}\label{sec:unfold}

The number of signal events, $N^\text{sig}_i$, is extracted from the data in the $i$th bin of the reconstructed observables using
\begin{equation}\label{eq:bgsub}
 N^\text{sig}_i = N^\text{sel}_i - N^\text{bkg}_i, \quad 1 \leq i \leq n,
\end{equation}
where $n$ denotes the total number of bins,
$N^\text{sel}_i$ is the number of selected events in the $i$th bin, and
$N^\text{bkg}_i$ corresponds to the expected number of background events in this bin, except for $\ttbar$ final states other than the signal.
The latter are dominated by events
in which one or both of the intermediate W bosons decay into $\tau$ leptons with subsequent decay into an electron or muon.
Since these events arise from the same \ttbar production process as the signal, the normalisation of this background is fixed to that of the signal.
The expected signal fraction is defined as the ratio of the number of selected \ttbar signal events to the total number of selected \ttbar events
(\ie the signal and all other \ttbar events) in simulation.
This procedure avoids the dependence on the total inclusive \ttbar cross section used in the normalization of the simulated signal sample.

The signal yields $N^\text{sig}_i$, determined in each $i$th bin of the reconstructed kinematic variables,
may contain entries that were originally produced in other bins and have migrated because of the imperfect resolutions.
This effect can be described as
\begin{equation}\label{eq:UnfoldProb}
 {M}^\text{sig}_i = \sum_{j = 1}^{m} A_{ij}{M}_{j}^\text{unf}, \quad 1 \leq i \leq n,
\end{equation}
where $m$ denotes the total number of bins in the true distribution, and ${M}_{j}^\text{unf}$ is the number of events in the $j$th  bin of the true distribution from data.
The quantity ${M}^\text{sig}_i$ is the expected number of events at detector level in the $i$th  bin,
and $A_{ij}$ is a matrix of probabilities describing
the migrations from the $j$th  bin of the true distribution to the $i$th  bin of the detector-level distribution, including acceptance and detector efficiencies.
In this analysis, the migration matrix $A_{ij}$ is defined such that the true level corresponds
to the full phase space (with no kinematic restrictions) for \ttbar production at parton level. 
At the detector level a binning is chosen in the same kinematic ranges as at the true level, 
but with the total number of bins typically a few times larger. 
The kinematic ranges of all variables are chosen such that the fraction of events that migrate into the regions outside the measured range is very small. 
It was checked that the inclusion of overflow bins outside the kinematic ranges does not significantly alter the unfolded results.
The migration matrix $A_{ij}$ is taken from the signal simulation.
The observed event counts $N^\text{sig}_i$ may be different from ${M}_i^\text{sig}$ owing to statistical fluctuations.

The estimated value of ${M}_{j}^\text{unf}$, designated as $\hat{M_{j}}^\text{unf}$,
is found using the TUnfold algorithm~\cite{Schmitt:2012kp}.
The unfolding of multidimensional distributions is performed by mapping the multidimensional arrays to one-dimensional arrays internally~\cite{Schmitt:2012kp}.
The unfolding is realized by a \chisq minimization and includes an additional \chisq term representing the Tikhonov regularization~\cite{Tikhonov:1963}.
The regularization reduces the effect of the statistical fluctuations present in $N^\text{sig}_i$
on the high-frequency content of $\hat{M_{j}}^\text{unf}$.
The regularization strength is chosen such that the global correlation coefficient is minimal~\cite{Schmitt:2016orm}.
For the measurements presented here, this choice results in a small contribution from the regularization term to the total $\chi^2$, on the order of 1\%.
A more detailed description of the unfolding procedure can be found in Ref.~\cite{Korolthesis}.
 \section{Cross section determination}\label{sec:cs}

The normalized double-differential cross sections of \ttbar production are measured in the full \ttbar kinematic phase space at parton level.
The number of unfolded signal events $\hat{M}^\text{unf}_{ij}$ in bin $i$ of variable $x$ and bin $j$ of variable $y$
is used to define the normalized double-differential cross sections of the \ttbar production process,
\begin{equation}\label{eq:ddxsecdef}
 \left(\frac{1}{\sigma} \frac{\rd^{2}\sigma}{\rd x\,\rd y}\right)_{ij} = \frac{1}{\sigma} \, \frac{1}{\Delta x_{i}} \,\frac{1}{\Delta y_{j}} \, \frac{\hat{M}^\text{unf}_{ij}} {\mathcal{B} \, \mathcal{L}},
\end{equation}
where $\sigma$ is the total cross section, which is evaluated by integrating $(\rd^{2}\sigma/\rd x\,\rd y)_{ij}$ over all bins.
The branching fraction of \ttbar into $\Pe\mu$ final state is taken to be $\mathcal{B} = 2.3\%$~\cite{Olive:2016xmw},
and $\mathcal{L}$ is the integrated luminosity of the data sample.
The bin widths of the $x$ and $y$ variables are denoted by $\Delta x_{i}$ and $\Delta y_{j}$, respectively. 
The bin widths are chosen based on the resolution, such that the purity and the stability of each bin is generally above 30\%.
For a given bin, the purity is defined as the fraction of events in the \ttbar signal simulation that are generated and reconstructed in the same bin
with respect to the total number of events reconstructed in that bin.
To evaluate the stability, the number of events in the \ttbar signal simulation that are generated and reconstructed in a given bin are divided
by the total number of reconstructed events generated in the bin.

\section{Systematic uncertainties}\label{sec:syst}

The measurement is affected by systematic uncertainties that originate
from detector effects and from the modelling of the processes. Each source of systematic uncertainty
is assessed individually by changing in the simulation the corresponding efficiency, resolution, or scale by its uncertainty, using
a prescription similar to the one followed in Ref.~\cite{Khachatryan:2015oqa}.
For each change made, the cross section determination is repeated,
and the difference with respect to the nominal result in each bin is taken as the systematic uncertainty.

To account for the pileup uncertainty, the value of the total $\Pp\Pp$ inelastic cross section,
which is used to estimate the mean number of additional pp interactions, is varied by ${\pm}5\%$~\cite{bib:ppInelXSec}.
The data-to-simulation correction factors for $\PQb$ tagging and mistagging efficiencies are varied within their uncertainties~\cite{bib:btag004}
as a function of the $\pt$ and $\abs{\eta}$ of the jet, following the procedure described in Ref.~\cite{Khachatryan:2015oqa}.
The data-to-simulation correction factors for the trigger efficiency, determined relatively to the triggers based on \ptmiss,
are varied within their uncertainty of 1\%.
The systematic uncertainty related to the kinematic reconstruction of top quarks is
assessed by varying the MC correction factor by its estimated uncertainty of ${\pm}1$\%~\cite{Korolthesis}.
For the uncertainties related to the jet energy scale, the jet energy is varied in the simulation within its uncertainty~\cite{Khachatryan:2016kdb}.
The uncertainty owing to the limited knowledge of the jet energy resolution is determined
by changing the latter in the simulation by ${\pm}1$ standard deviation in different $\eta$ regions~\cite{bib:JME-10-011:JES}.
The normalizations of the background processes are varied by 30\% to account for the corresponding uncertainty.
The uncertainty in the integrated luminosity of 2.6\%~\cite{CMS-PAS-LUM-13-001} is propagated to the measured cross sections.

The impact of theoretical assumptions on the measurement is determined
by repeating the analysis replacing the standard \MADGRAPH \ttbar simulation with
simulated samples in which specific parameters or assumptions are altered.
The PDF systematic uncertainty is estimated by
reweighting the \MADGRAPH \ttbar signal sample according to the uncertainties in the CT10 PDF set,
evaluated at 90\% confidence level (CL)~\cite{Lai:2010vv}, and then rescaled to 68\% CL.
To estimate the uncertainty related to the choice of the tree-level multijet scattering model used in \MADGRAPH,
the results are recalculated using an alternative prescription for interfacing NLO calculations with parton showering as implemented
in \POWHEG.
For $\mu_\mathrm{r}$ and $\mu_\mathrm{f}$, two samples are used with the scales being simultaneously increased or decreased by a factor of two
relative to their common nominal value $\mu_\mathrm{r} = \mu_\mathrm{f} = \sqrt{\smash[b]{m^2_{\PQt} + \Sigma \pt^2}}$,
where the sum is over all additional final-state partons in the matrix element.
The effect of additional jet
production is studied by varying in \MADGRAPH the matching threshold between jets produced at the matrix-element level and via parton showering.
The uncertainty in the effect of the initial- and final-state radiation on the signal efficiency is covered by the uncertainty
in $\mu_\mathrm{r}$ and $\mu_\mathrm{f}$, as well as in the matching threshold.
The samples generated with \PowHer and \PowPyt are used
to estimate the uncertainty related to the choice of the showering and hadronization model.
The effect due to
uncertainties in $m_{\PQt}$ is estimated using simulations with altered top quark masses.
The cross section differences observed for an $m_{\PQt}$ variation
of 1\GeV around the central value of 172.5\GeV used in the simulation is quoted as the uncertainty.

The total systematic uncertainty is estimated by adding all the contributions described above in quadrature, separately for positive and negative cross section variations.
If a systematic uncertainty results in two cross section variations of the same sign, the largest one is taken, while the opposite variation is set to zero.
 \section{Results}\label{sec:results}
{\tolerance=1200
Normalized differential \ttbar cross sections are measured as a function of pairs of variables representing the kinematics of the top quark
(only the top quark is taken and not the top antiquark, thus avoiding any double counting of events),
and \ttbar system, defined in Section~\ref{sec:intro}:
$[\ptt, \yt]$,
$[\yt, \mtt]$,
$[\ytt, \mtt]$,
$[\detatt, \mtt]$,
$[\pttt, \mtt]$,
and $[\dphitt, \mtt]$.
These pairs are chosen in order to obtain representative combinations that are sensitive to different aspects of the \ttbar production dynamics,
as will be discussed in the following.
\par}

In general, the systematic uncertainties are of similar size to the statistical uncertainties.
The dominant systematic uncertainties are those in the signal modelling,
which also are affected by the statistical uncertainties in the simulated samples that are used for the evaluation of these uncertainties.
The largest experimental systematic uncertainty is the jet energy scale.
The measured double-differential normalized \ttbar cross sections are compared in Figs.~\ref{fig:mc_yt_ptt}--\ref{fig:nlo_mtt_dphitt} to theoretical predictions obtained using
different MC generators and fixed-order QCD calculations.
The numerical values of the measured cross sections and their uncertainties are provided in Appendix~\ref{sec:app}.
 \subsection{Comparison to MC models}
\label{sec:thmc}

\begin{figure*}
  \centering
  \includegraphics[width=\cmsFigOne]{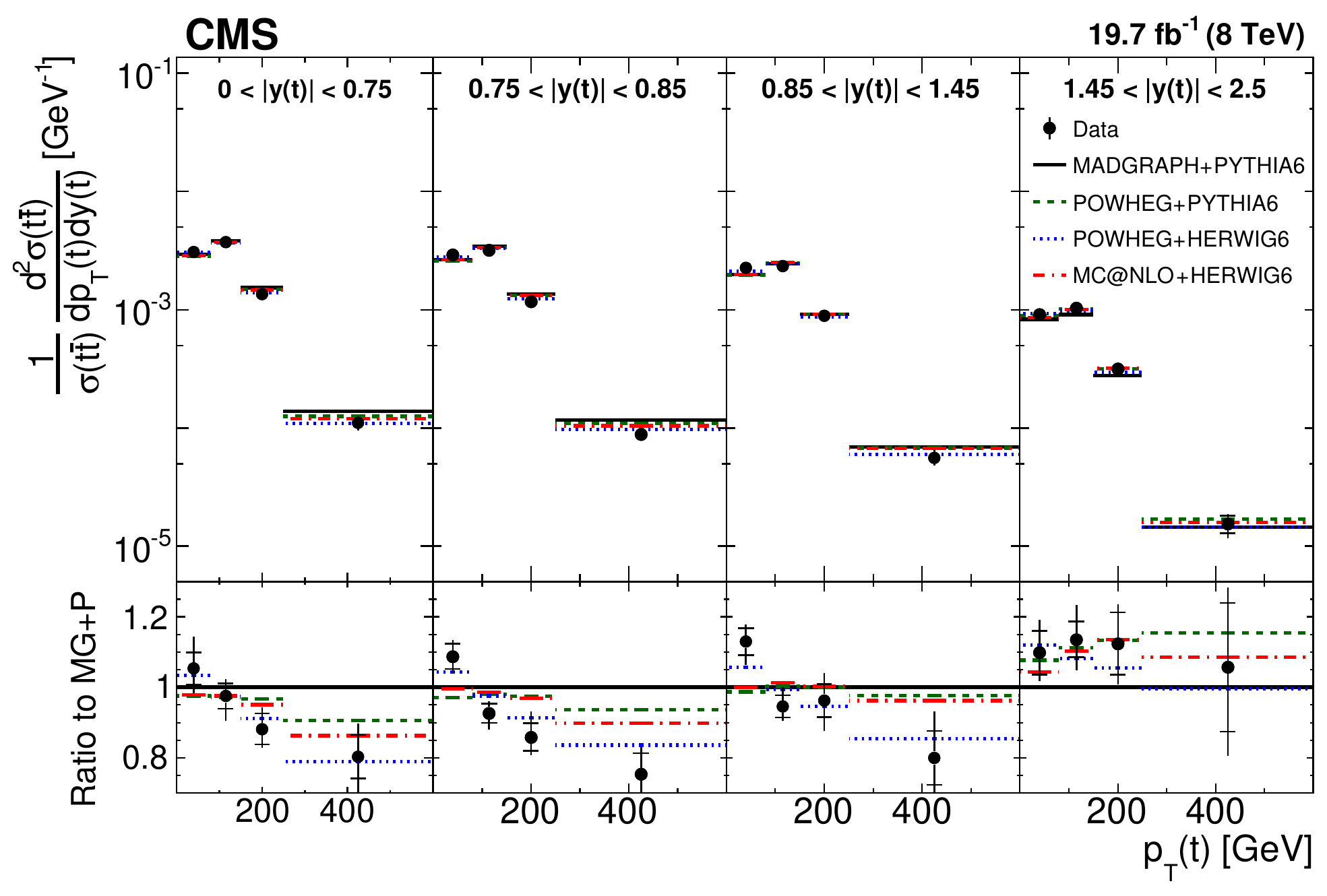}
  \caption{Comparison of the measured normalized \ttbar double-differential cross section as a function of \ptt in different $\abs{\yt}$ ranges
  to MC predictions calculated using \MadPyt, \PowPyt, \PowHer, and \MCHer.
  The inner vertical bars on the data points represent the statistical uncertainties and the full bars include also the systematic uncertainties added in quadrature.
  In the bottom panel, the ratios of the data and other simulations to the \MADGRAPH+\PYTHIAsix (MG+P) predictions are shown.}
  \label{fig:mc_yt_ptt}
\end{figure*}

{\tolerance=2800
In Fig.~\ref{fig:mc_yt_ptt}, the \ptt distribution is compared in different ranges of $\abs{\yt}$ to predictions from \MadPyt, \PowPyt, \PowHer, and \MCHer.
The data distribution is softer than that of the MC expectation over almost the entire $\yt$ range, except at high $\abs{\yt}$ values.
The disagreement level is the strongest for \MadPyt, while \PowHer describes the data best.
\par}

\begin{figure*}
  \centering
  \includegraphics[width=\cmsFigOne]{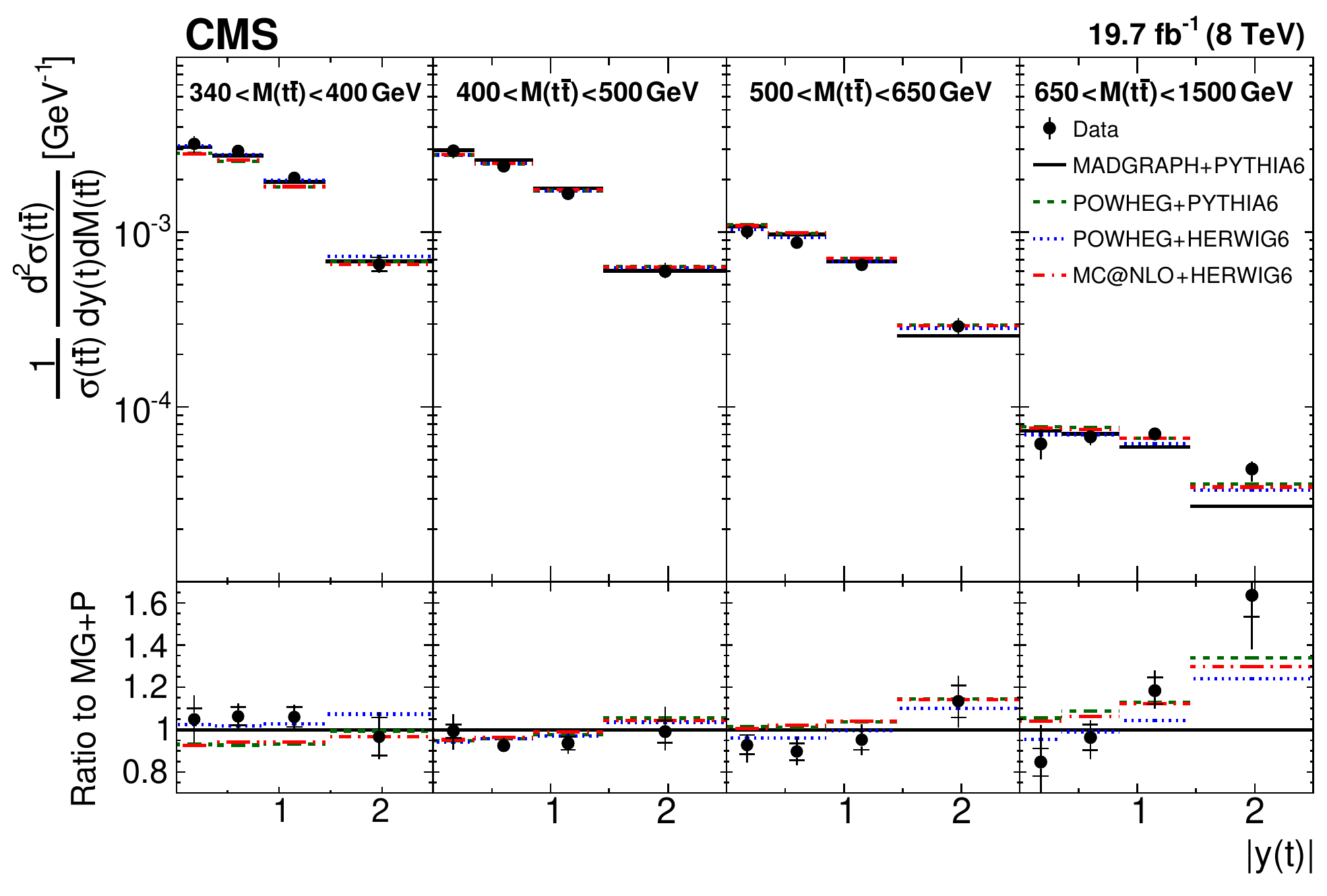}
  \caption{Comparison of the measured normalized \ttbar double-differential cross section as a function of $\abs{\yt}$ in different \mtt ranges to MC predictions. Details can be found in the caption of Fig.~\ref{fig:mc_yt_ptt}.}
  \label{fig:mc_mtt_yt}
\end{figure*}

\begin{figure*}
  \centering
  \includegraphics[width=\cmsFigOne]{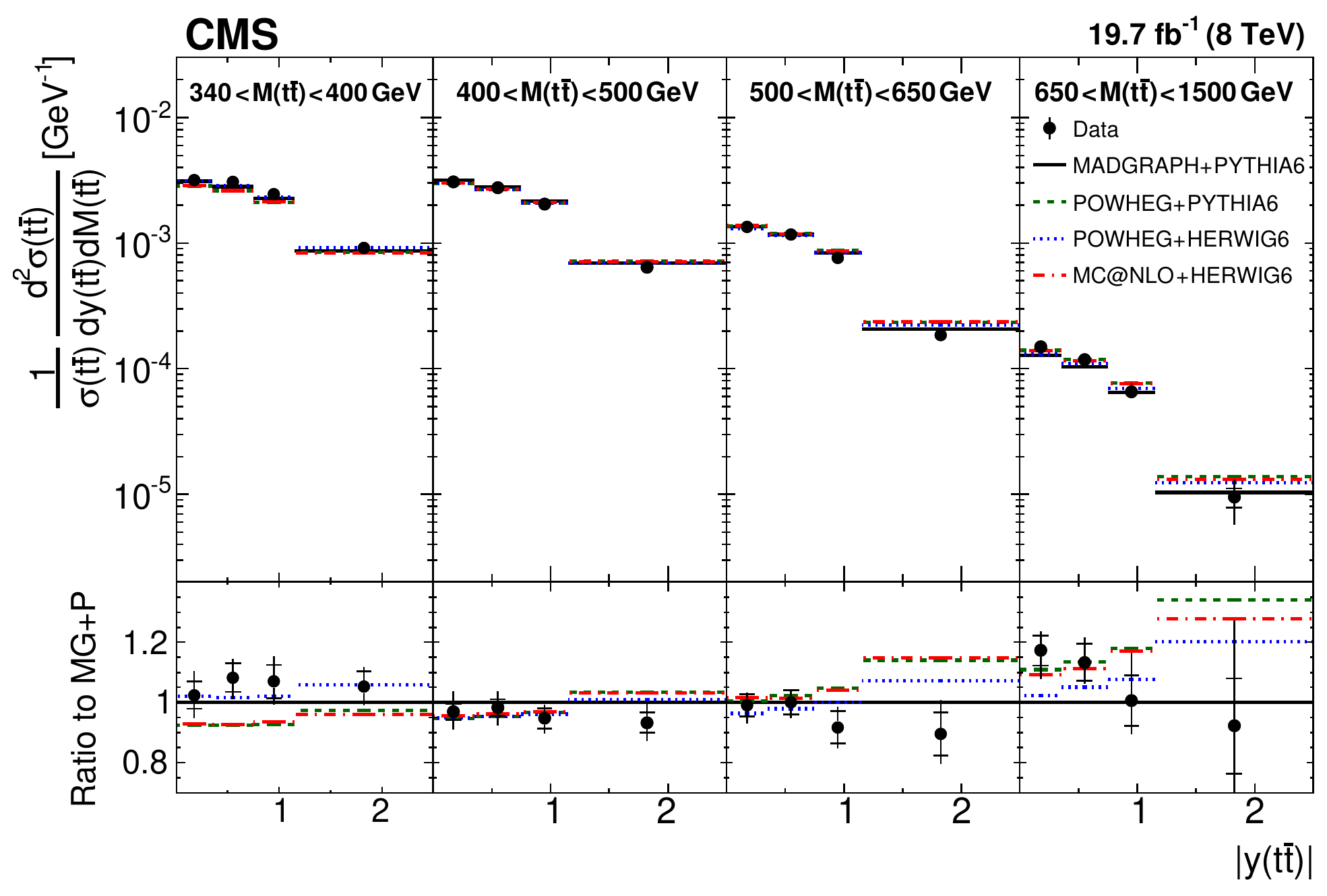}
  \caption{Comparison of the measured normalized \ttbar double-differential cross section as a function of $\abs{\ytt}$ in different of \mtt ranges to MC predictions. Details can be found in the caption of Fig.~\ref{fig:mc_yt_ptt}.}
  \label{fig:mc_mtt_ytt}
\end{figure*}

Figures~\ref{fig:mc_mtt_yt} and \ref{fig:mc_mtt_ytt} illustrate the distributions of $\abs{\yt}$ and $\abs{\ytt}$ in different \mtt ranges compared to the same set of MC models.
While the agreement between the data and MC predictions is good in the lower ranges of \mtt, the simulation starts to deviate from the data at higher \mtt,
where the predictions are more central than the data for \yt and less central for \ytt.

\begin{figure*}
  \centering
  \includegraphics[width=\cmsFigOne]{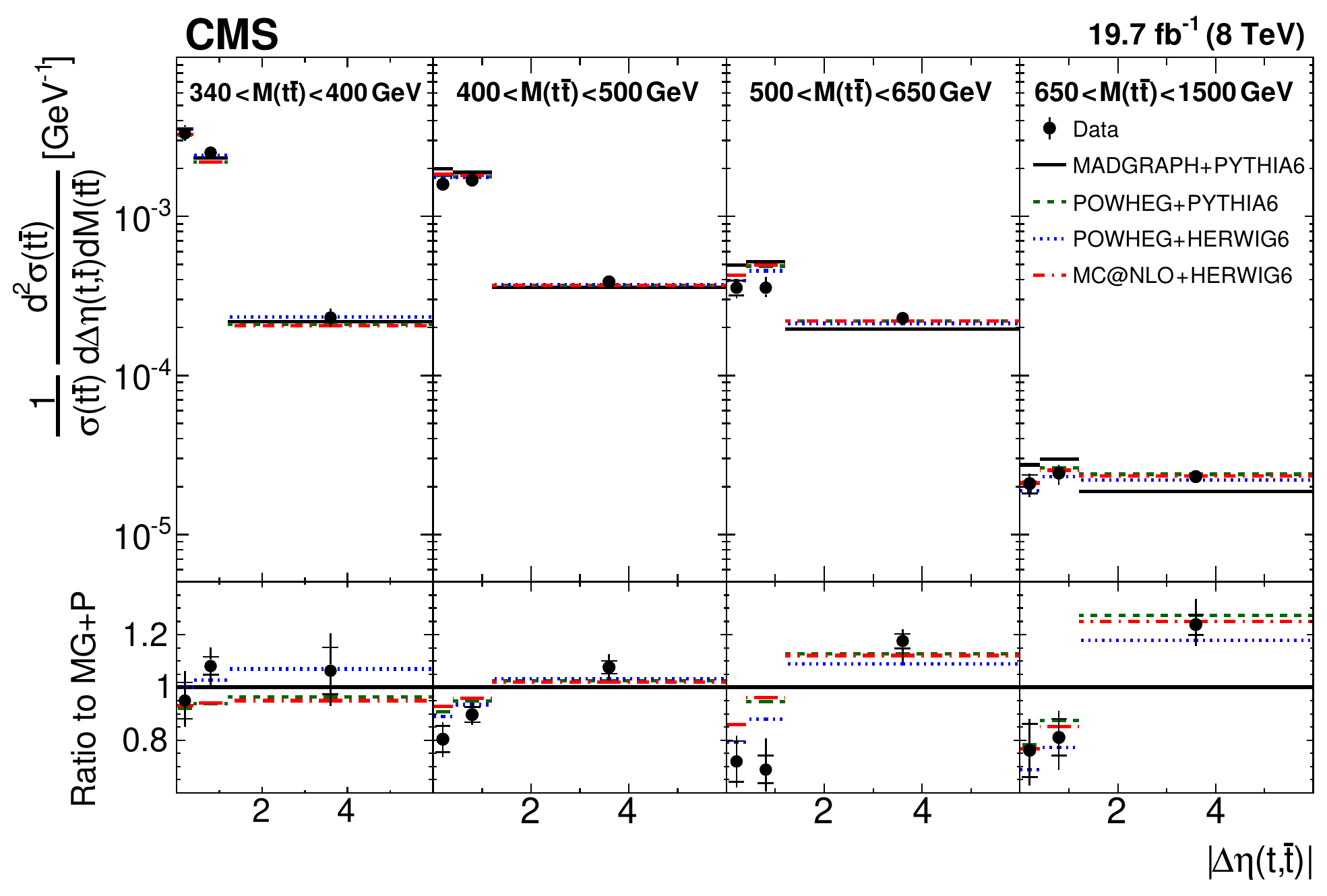}
  \caption{Comparison of the measured normalized \ttbar double-differential cross section as a function of \detatt in different \mtt ranges to MC predictions. Details can be found in the caption of Fig.~\ref{fig:mc_yt_ptt}.}
  \label{fig:mc_mtt_detatt}
\end{figure*}

In Fig.~\ref{fig:mc_mtt_detatt}, the \detatt distribution is compared in the same \mtt ranges to the MC predictions.
For all generators there is a discrepancy between the data and simulation for the medium \mtt bins, where the predicted \detatt values are too low.
The disagreement is the strongest for \MadPyt.

\begin{figure*}
  \centering
  \includegraphics[width=\cmsFigOne]{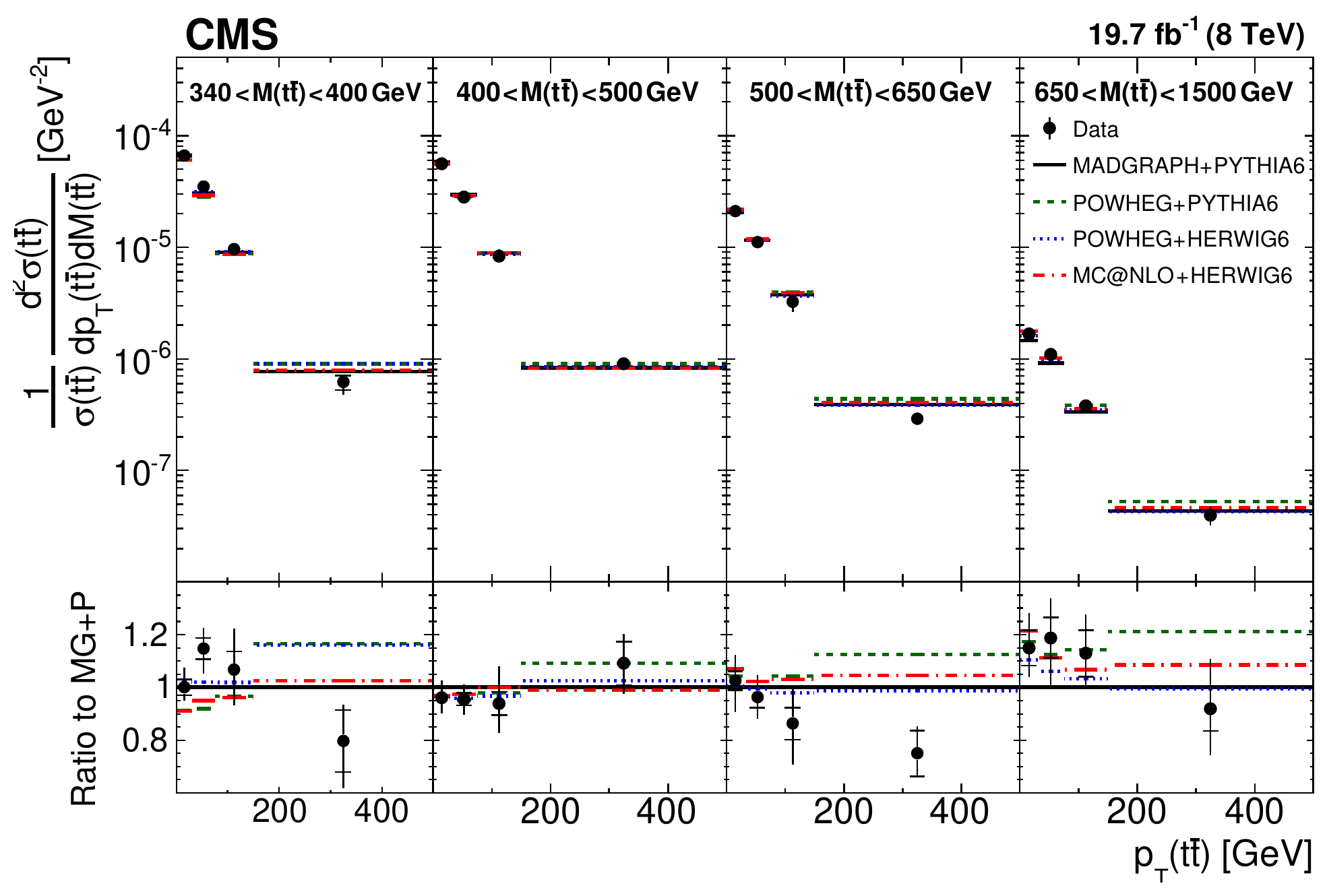}
  \caption{Comparison of the measured normalized \ttbar double-differential cross section as a function of \pttt in different \mtt ranges to MC predictions. Details can be found in the caption of Fig.~\ref{fig:mc_yt_ptt}.}
  \label{fig:mc_mtt_pttt}
\end{figure*}

\begin{figure*}
  \centering
  \includegraphics[width=\cmsFigOne]{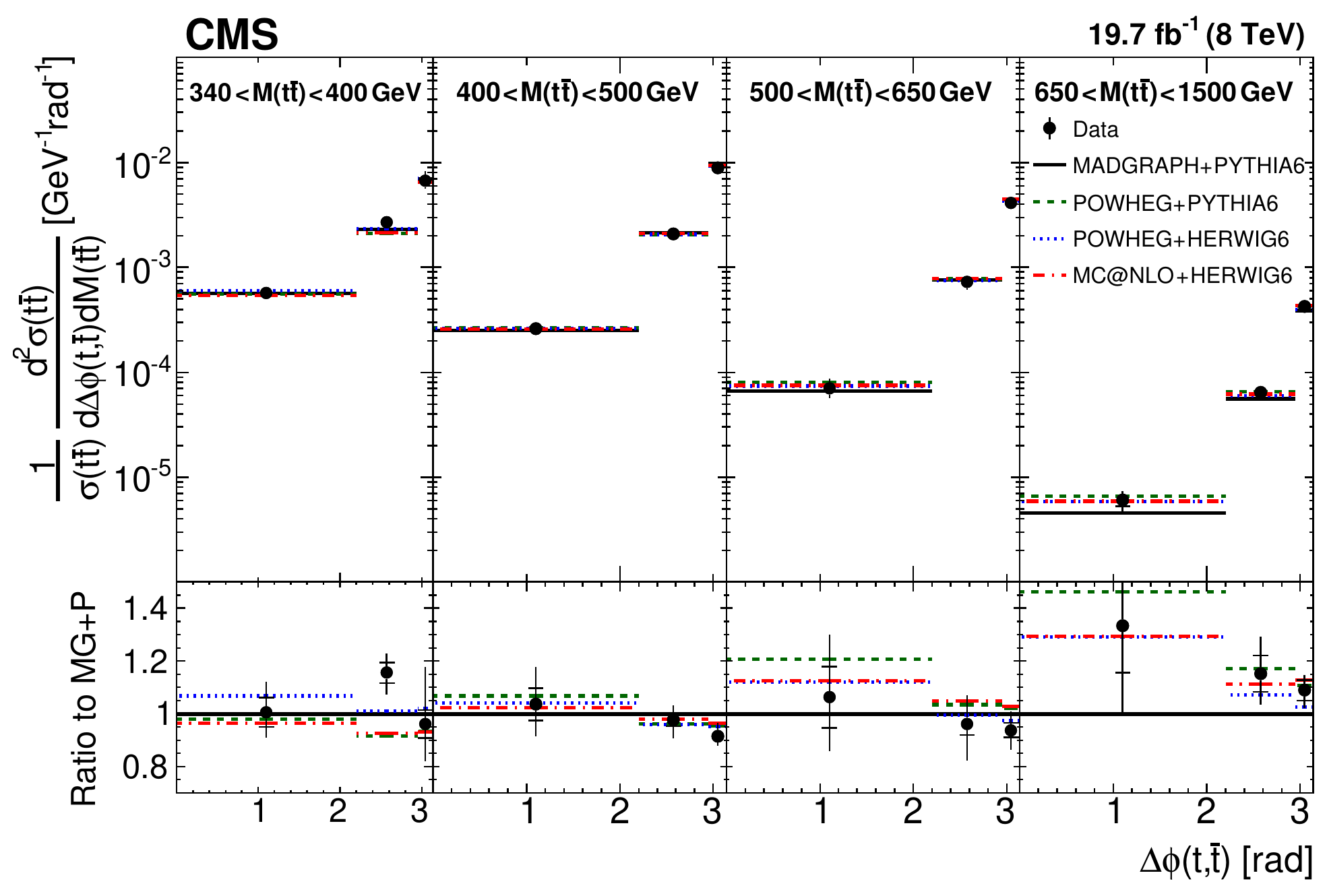}
  \caption{Comparison of the measured normalized \ttbar double-differential cross section as a function of \dphitt in different \mtt ranges to MC predictions. Details can be found in the caption of Fig.~\ref{fig:mc_yt_ptt}.}
  \label{fig:mc_mtt_dphitt}
\end{figure*}

Figures~\ref{fig:mc_mtt_pttt} and \ref{fig:mc_mtt_dphitt} illustrate the comparison of the distributions of \pttt and \dphitt in the same \mtt ranges to the MC models.
For the \pttt distribution (Fig.~\ref{fig:mc_mtt_pttt}), which is sensitive to radiation,
none of the MC generators provide a good description.
The largest differences are observed between the data and \PowPyt for the highest values of \pttt, where the predictions lie above the data.
For the \dphitt distribution (Fig.~\ref{fig:mc_mtt_dphitt}),
all MC models describe the data reasonably well.

In order to perform a quantitative comparison of the measured cross sections to all considered MC generators,
\chisq values are calculated as follows:
\begin{equation}
 \chisq = \mathbf{R}^{T}_{N-1} \mathbf{Cov}^{-1}_{N-1} \mathbf{R}_{N-1},
\end{equation}
where $\mathbf{R}_{N-1}$ is the column vector of the residuals calculated as the difference of the measured cross sections and the corresponding predictions
obtained by discarding one of the $N$ bins, and $\mathbf{Cov}_{N-1}$ is the $(N-1)\times(N-1)$ submatrix obtained from the full covariance matrix by
discarding the corresponding row and column.
The matrix $\mathbf{Cov}_{N-1}$ obtained in this way is invertible, while the original covariance matrix $\mathbf{Cov}$ is singular.
This is because for normalized cross sections one loses one degree of freedom, as can be deduced from Eq.~(\ref{eq:ddxsecdef}).
The covariance matrix $\mathbf{Cov}$ is calculated as:
\begin{equation}
\label{eq:covmat}
\mathbf{Cov} = \mathbf{Cov}^\text{unf} + \mathbf{Cov}^\text{syst},
\end{equation}
where $\mathbf{Cov}^\text{unf}$ and $\mathbf{Cov}^\text{syst}$ are the covariance matrices accounting for the statistical uncertainties from the unfolding,
and the systematic uncertainties, respectively.
The systematic covariance matrix $\mathbf{Cov}^\text{syst}$ is calculated as:
\ifthenelse{\boolean{cms@external}}{
\begin{multline}
\mathbf{Cov}^\text{syst}_{ij} = \sum_{k}C_{j,k}C_{i,k}\\
 + \frac{1}{2} \left( \sum_{k'}C^{+}_{j,k'}C^{+}_{i,k'} + \sum_{k'}C^{-}_{j,k'}C^{-}_{i,k'} \right),\\ 1 \le i \le N, \quad 1 \le j \le N,
\end{multline}}{
\begin{equation}
\mathbf{Cov}^\text{syst}_{ij} = \sum_{k}C_{j,k}C_{i,k} + \frac{1}{2} \left( \sum_{k'}C^{+}_{j,k'}C^{+}_{i,k'} + \sum_{k'}C^{-}_{j,k'}C^{-}_{i,k'} \right), \quad 1 \le i \le N, \quad 1 \le j \le N,
\end{equation}}
where $C_{i,k}$ stands for the systematic uncertainty from source $k$ in the $i$th  bin, which consists of one variation only,
and $C^{+}_{i,k'}$ and $C^{-}_{i,k'}$ stand for the positive and negative variations, respectively, of the systematic uncertainty due to source $k'$ in the $i$th  bin.
The sums run over all sources of the corresponding systematic uncertainties.
All systematic uncertainties are treated as additive,
i.e.\ the relative uncertainties are used to scale the corresponding measured value in the construction of $\mathbf{Cov}^\text{syst}$.
This treatment is consistent with the cross section normalization.
The cross section measurements for different pairs of observables are statistically and systematically correlated.
No attempt is made to quantify the correlations between bins from different double-differential
distributions. Thus, quantitative comparisons between theoretical predictions and the data can only be made for individual distributions.

The obtained \chisq values, together with the corresponding numbers of degrees of freedom (\ndf), are listed in Table~\ref{tab:chi2mc}.
From these values one can conclude that
none of the considered MC generators is able to correctly describe all distributions.
In particular, for $[\detatt, \mtt]$ and $[\pttt, \mtt]$, the \chisq values are relatively large for all MC generators.
The best agreement with the data is provided by \PowHer.

\begin{table*}
  \topcaption{The \chisq values and dof of the measured normalized double-differential
      \ttbar cross sections with respect to the various MC predictions.}
  \label{tab:chi2mc}
  \centering
  \begin{tabular}{lccccc}
   \multirow{2}{*}{Cross section} & \multirow{3}{*}{\ndf} & \multicolumn{4}{c}{\chisq} \\
   \cline{3-6}
 &  & \MADGRAPH    & \POWHEG   & \POWHEG   & \MCATNLO \\[-0.4ex]
     {variables}&&+\PYTHIAsix&+\PYTHIAsix&+\HERWIGsix&+\HERWIGsix\\
  \hline
    $[\ptt, \yt]$      & $   15$ & 96  & 58   & 14  & 46  \\
    $[\yt, \mtt]$      & $   15$ & 53  & 20   & 13  & 21  \\
    $[\ytt, \mtt]$  	 & $   15$ & 19  & 21   & 15  & 22  \\
    $[\detatt, \mtt]$  & $   11$ & 163 & 33   & 20  & 39  \\
    $[\pttt, \mtt]$ 	 & $   15$ & 31  & 83   & 30  & 33  \\
    $[\dphitt, \mtt]$  & $   11$ & 21  & 21   & 10  & 17  \\
  \end{tabular}
\end{table*}

\subsection{Comparison to fixed-order calculations}
\label{sec:comparison}
Fixed-order theoretical calculations for fully differential cross sections in inclusive \ttbar production
are publicly available at NLO $O(\alpha_s^3)$ in the fixed-flavour number scheme~\cite{Mangano:1991jk}, where $\alpha_s$ is the strong coupling strength.
The exact fully differential NNLO $O(\alpha_s^4)$ calculations for \ttbar production have recently appeared in the literature~\cite{Czakon:2015owf,Czakon:2016dgf},
but are not yet publicly available.
For higher orders, the cross sections as functions of single-particle kinematic variables have been calculated at approximate NNLO $O(\alpha_s^4)$~\cite{Kidonakis:2001nj}
and next-to-next-to-next-to-leading-order $O(\alpha_s^5)$~\cite{Kidonakis:2014pja}, using methods of threshold resummation beyond the leading-logarithmic accuracy.

The measured cross sections are compared with NLO QCD predictions based on several PDF sets.
The predictions are calculated using the \MCFM program (version 6.8)~\cite{Campbell:2010ff} and a number of the latest PDF sets, namely:
ABM11~\cite{Alekhin:2012ig}, CJ15~\cite{Accardi:2016qay}, CT14~\cite{Dulat:2015mca}, HERAPDF2.0~\cite{Abramowicz:2015mha}, JR14~\cite{Jimenez-Delgado:2014twa},
MMHT2014~\cite{Harland-Lang:2014zoa}, and NNPDF3.0~\cite{Ball:2014uwa},
available via the \lhapdf interface (version 6.1.5)~\cite{Buckley:2014ana}.
The number of active flavours is set to $n_f = 5$ and
the top quark pole mass $m_{\PQt} = 172.5$\GeV is used.
The effect of using $n_f = 6$ in the PDF evolution, i.e.\ treating the top quark as a massless parton above threshold (as was done, e.g.\ in HERAPDF2.0~\cite{Abramowicz:2015mha}),
has been checked and the differences were found to be ${<}0.1\%$ (also see the corresponding discussion in Ref.~\cite{Ball:2014uwa}).
The renormalization and factorization scales
are chosen to be $\mu_\mathrm{r} = \mu_\mathrm{f} = \sqrt{\smash[b]{m_{\PQt}^2+[\pt^2({\PQt})+\pt^2(\PAQt)]/2}}$, whereas
$\alpha_s$ is set to the value used for the corresponding PDF extraction.
The theoretical uncertainty is estimated by varying $\mu_\mathrm{r}$ and $\mu_\mathrm{f}$ independently up and down by a factor of 2, subject to the additional restriction
that the ratio $\mu_\mathrm{r} / \mu_\mathrm{f}$ be between 0.5 and 2~\cite{Cacciari:2008zb} (referred to hereafter as scale uncertainties).
These uncertainties are supposed to estimate the missing higher-order corrections.
The PDF uncertainties are taken into account in the theoretical predictions for each PDF set.
The PDF uncertainties of \cj~\cite{Accardi:2016qay} and \ct~\cite{Dulat:2015mca}, evaluated at 90\% CL, are rescaled to the 68\% CL.
The uncertainties in the normalized \ttbar cross sections originating from $\alpha_s$ and $m_{\PQt}$ are found to be negligible (${<}1\%$) compared to the current data precision and thus are not considered.

For the double-differential cross section as a function of \ptt and \yt, approximate NNLO predictions~\cite{Kidonakis:2001nj} are obtained using the \difftop program~\cite{Guzzi:2014wia,Guzzi:2014tna,Guzzi:2013noa}.
In this calculation, the scales are set to $\mu_\mathrm{r} = \mu_\mathrm{f} = \sqrt{\smash[b]{m_{\PQt}^2+\pt^2({\PQt})}}$ and NNLO variants of the PDF sets are used.
For the ABM PDFs, the recent version ABM12~\cite{Alekhin:2013nda} is used, which is available only at NNLO.
Predictions using \difftop are not available for the rest of the measured cross sections that involve \ttbar kinematic variables.

A quantitative comparison of the measured double-differential cross sections to the theoretical predictions is performed by evaluating the \chisq values, as described in Section~\ref{sec:thmc}.
The results are listed in Tables~\ref{tab:chi2nlo} and \ref{tab:chi2nnlo} for the NLO and approximate NNLO calculations, respectively.
For the NLO predictions, additional \chisq values are reported including the corresponding PDF uncertainties,
i.e.\ Eq.~(\ref{eq:covmat}) becomes $\mathbf{Cov} = \mathbf{Cov}^\text{unf} + \mathbf{Cov}^\text{syst} + \mathbf{Cov}^\mathrm{PDF}$,
where $\mathbf{Cov}^\mathrm{PDF}$ is a covariance matrix that accounts for the PDF uncertainties.
Theoretical uncertainties from scale variations
are not included in this \chisq calculation.
The NLO predictions with recent global PDFs using LHC data, namely \mmht, \ct, and \nnpdf, are found to describe
the \ptt, \yt, and \ytt cross sections reasonably, as illustrated by the \chisq values.
The \cj PDF set also provides a good description of these cross sections,
although it does not include LHC data~\cite{Accardi:2016qay}.
The \abm, \jr, and \herapdf sets yield a poorer description of the data.
Large differences between the data and the nominal NLO calculations are observed for the \detatt, \pttt, and \dphitt cross sections.
It is noteworthy that the scale uncertainties in the predictions,
which are of comparable size or exceed the experimental uncertainties, are not taken into account in the \chisq calculations.
The \pttt and \dphitt normalized cross sections are represented at LO $O(\alpha_s^2)$ by delta functions, and nontrivial shapes appear at $O(\alpha_s^3)$,
thus resulting in large NLO scale uncertainties~\cite{Mangano:1991jk}.
Compared to the NLO predictions, the approximate NNLO predictions using NNLO PDF sets (where available) provide an improved description
of the \ptt cross sections in different $\abs{\yt}$ ranges.

\begin{table*}
  \topcaption{The \chisq values and dof of the double-differential normalized \ttbar cross sections with respect to NLO $O(\alpha_s^3)$ theoretical calculations~\cite{Mangano:1991jk} using different PDF sets.
  The \chisq values that include PDF uncertainties are shown in parentheses.
  }
  \label{tab:chi2nlo}
   \centering
  \renewcommand*{\arraystretch}{1.35}
     \cmsTable{\begin{tabular}{lcccccccc}
  {Cross section} &  \multirow{2}{*}{\ndf} & \multicolumn{7}{c}{\chisq NLO $O(\alpha_s^3)$ (including PDF uncertainties)} \\
  \cline{3-9}
	variables &						 &      { \herapdf }     &        { \mmht }     &    {  \ct} &    {  \nnpdf} &  { ABM11}&    { JR14} &     { CJ15 }\\
  \hline
    $[\ptt, \yt]$      & $   15$ & 46 (40) & 26 (24) & 24 (21) & 28 (25) & $ 62$ (51) & 47 ($   47$) & $    27$ ($   24$)  \\
    $[\yt, \mtt]$      & $   15$ & $    52$ ($   44$) & $    22$ ($   20$) & $    19$ ($   18$) & $    14$ ($   14$) & $    71$ ($   55$) & $    44$ ($   44$) & $    26$ ($   24$)  \\
    $[\ytt, \mtt]$  	 & $   15$ & $    29$ ($   25$) & $    15$ ($   15$) & $    16$ ($   15$) & $    10$ ($   10$) & $    42$ ($   31$) & $    25$ ($   25$) & $    16$ ($   16$)  \\
    $[\detatt, \mtt]$  & $   11$ & $    46$ ($   43$) & $    31$ ($   31$) & $    32$ ($   31$) & $    45$ ($   42$) & $    48$ ($   44$) & $    39$ ($   39$) & $    33$ ($   33$)  \\
    $[\pttt, \mtt]$ 	 & $   15$ & $   485$ ($  429$) & $   377$ ($  310$) & $   379$ ($  264$) & $   251$ ($  212$) & $   553$ ($  426$) & $   428$ ($  415$) & $   413$ ($  398$)  \\
    $[\dphitt, \mtt]$  & $   11$ & $   354$ ($  336$) & $   293$ ($  272$) & $   296$ ($  259$) & $   148$ ($  143$) & $   386$ ($  335$) & $   329$ ($  324$) & $   312$ ($  308$)  \\
  \end{tabular}
  }
\end{table*}

\begin{table*}
  \topcaption{The \chisq values and dof of the double-differential normalized \ttbar cross sections with respect to approximate NNLO $O(\alpha_s^4)$ theoretical calculations~\cite{Kidonakis:2001nj,Guzzi:2014wia,Guzzi:2014tna,Guzzi:2013noa} using different PDF sets.
  }
  \label{tab:chi2nnlo}
  \centering
   \begin{tabular}{l*{7}{c}}
  {Cross section} &  \multirow{2}{*}{\ndf} & \multicolumn{6}{c}{\chisq approximate NNLO $O(\alpha_s^4)$} \\
  \cline{3-8}
  variables &  &     \herapdf      &        \mmht      &               \ct &            \nnpdf &           ABM12&               JR14             \\
  \hline
    $[\ptt, \yt]$      & $   15$ & 22 & 11 & 13 & 15 & 54 & 44 \\
  \end{tabular}

\end{table*}

\begin{figure*}
  \centering
  \includegraphics[width=\cmsFigOne]{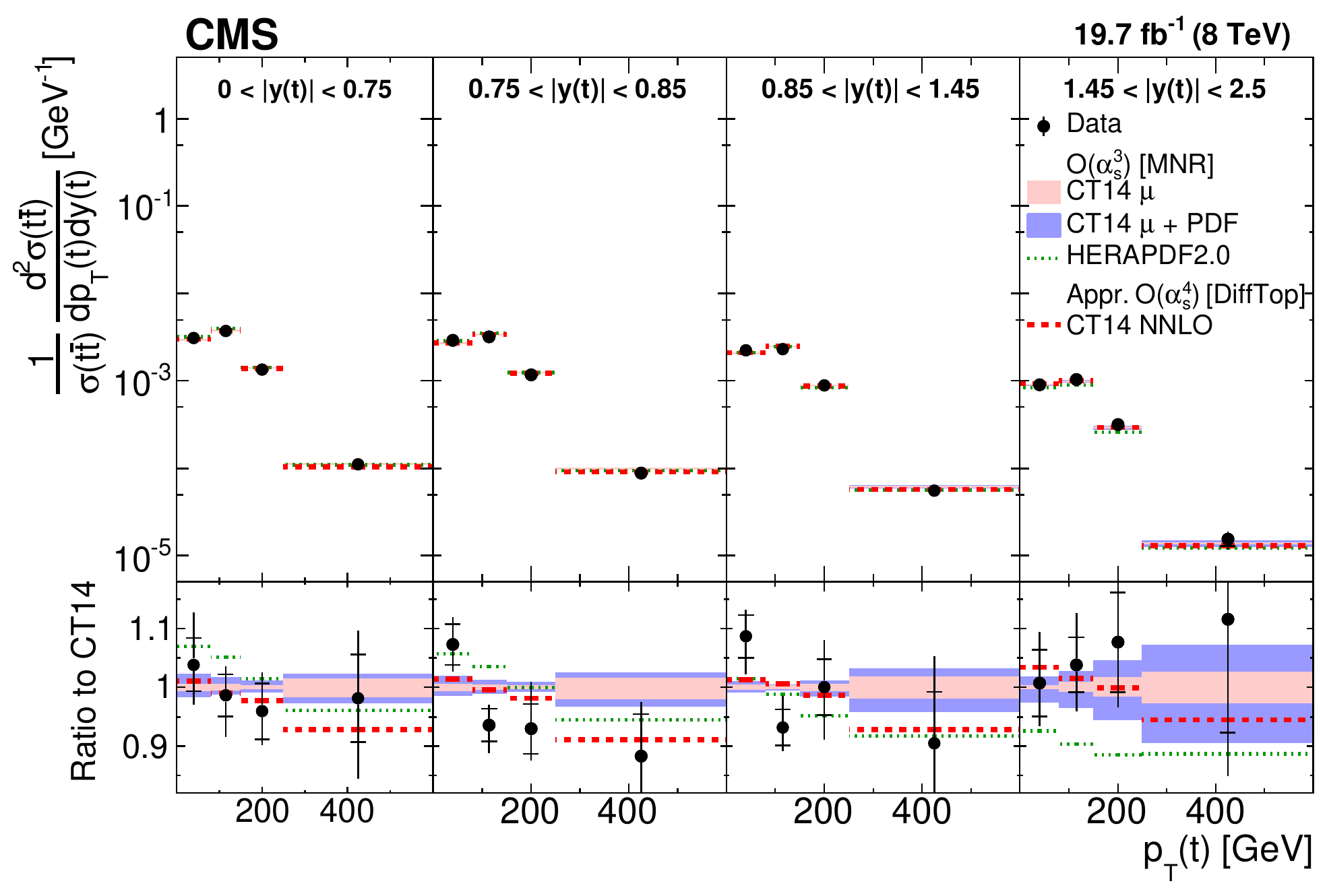}
  \caption{Comparison of the measured normalized \ttbar double-differential cross section as a function of \ptt in different $\abs{\yt}$ ranges to NLO $O(\alpha_s^3)$ (MNR)
  predictions calculated with \ct and \herapdf, and approximate NNLO $O(\alpha_s^4)$ (\difftop) prediction calculated with \ct.
  The inner vertical bars on the data points represent the statistical uncertainties and the full bars include also the systematic uncertainties added in quadrature.
  The light band shows the scale uncertainties ($\mu$) for the NLO predictions using \ct,
  while the dark band includes also the PDF uncertainties added in quadrature ($\mu + \mathrm{PDF}$).
  The dotted line shows the NLO predictions calculated with \herapdf.
  The dashed line shows the approximate NNLO predictions calculated with \ct.
  In the bottom panel, the ratios of the data and other calculations to the NLO prediction using \ct are shown.
  }
  \label{fig:nlo_yt_ptt}
\end{figure*}

\begin{figure*}
  \centering
  \includegraphics[width=\cmsFigOne]{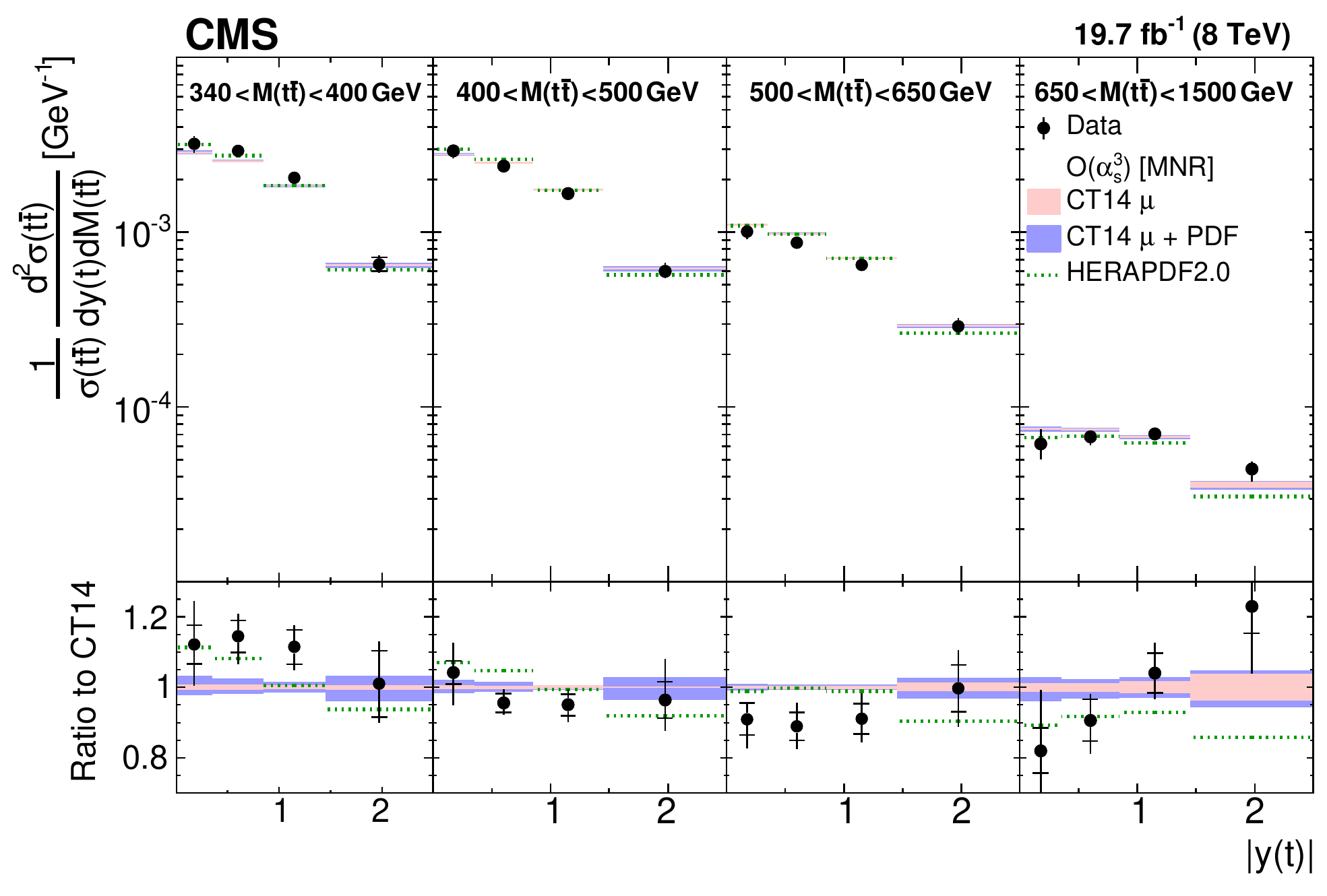}
  \caption{Comparison of the measured normalized \ttbar double-differential cross section as a function of $\abs{\yt}$ in different \mtt ranges to NLO $O(\alpha_s^3)$ predictions. Details can be found in the caption of Fig.~\ref{fig:nlo_yt_ptt}. Approximate NNLO $O(\alpha_s^4)$ predictions are not available for this cross section.}
  \label{fig:nlo_mtt_yt}
\end{figure*}

\begin{figure*}
  \centering
  \includegraphics[width=\cmsFigOne]{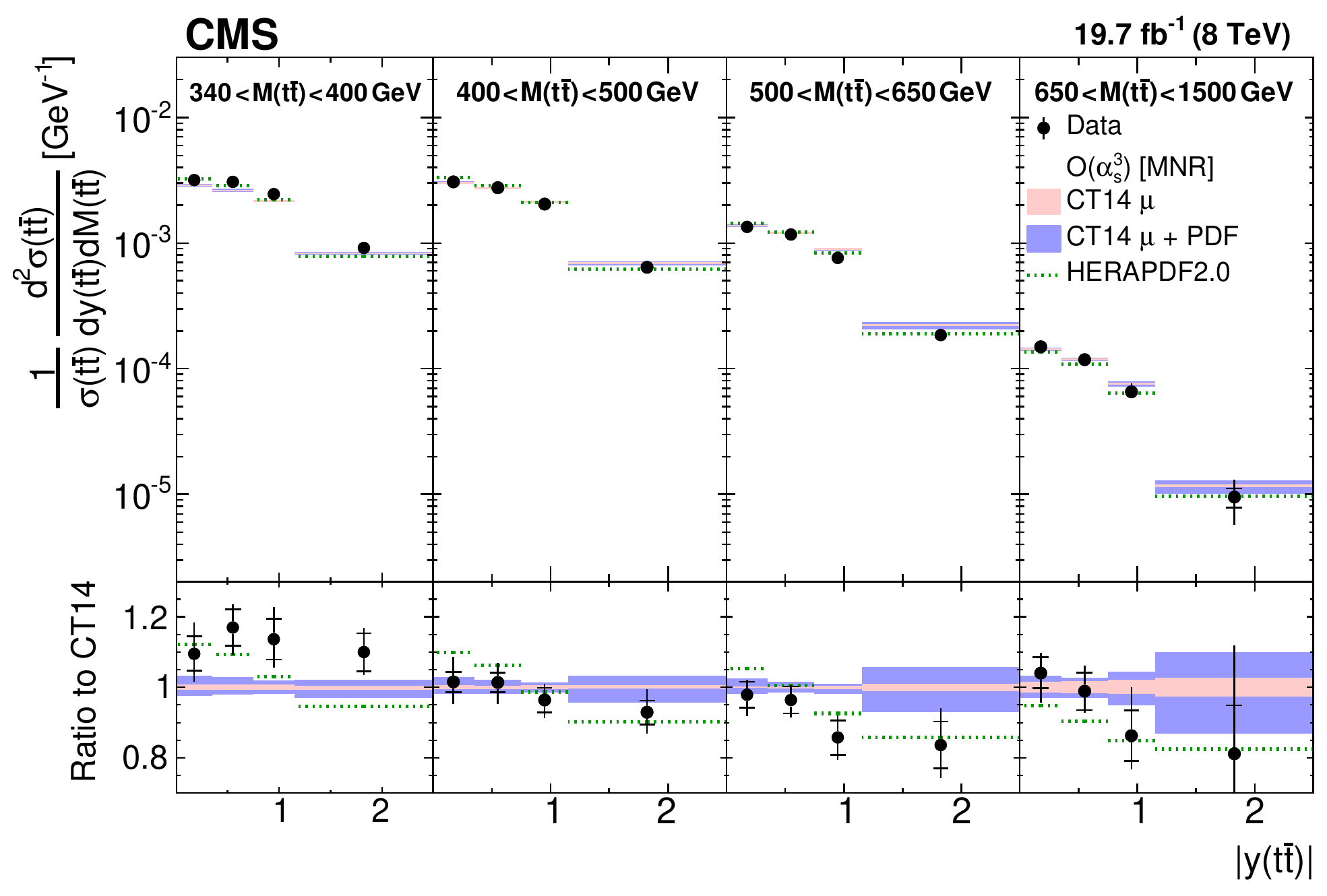}
  \caption{Comparison of the measured normalized \ttbar double-differential cross section as a function of $\abs{\ytt}$ in different \mtt ranges to NLO $O(\alpha_s^3)$ predictions. Details can be found in the caption of Fig.~\ref{fig:nlo_yt_ptt}. Approximate NNLO $O(\alpha_s^4)$ predictions are not available for this cross section.}
  \label{fig:nlo_mtt_ytt}
\end{figure*}

\begin{figure*}
  \centering
  \includegraphics[width=\cmsFigOne]{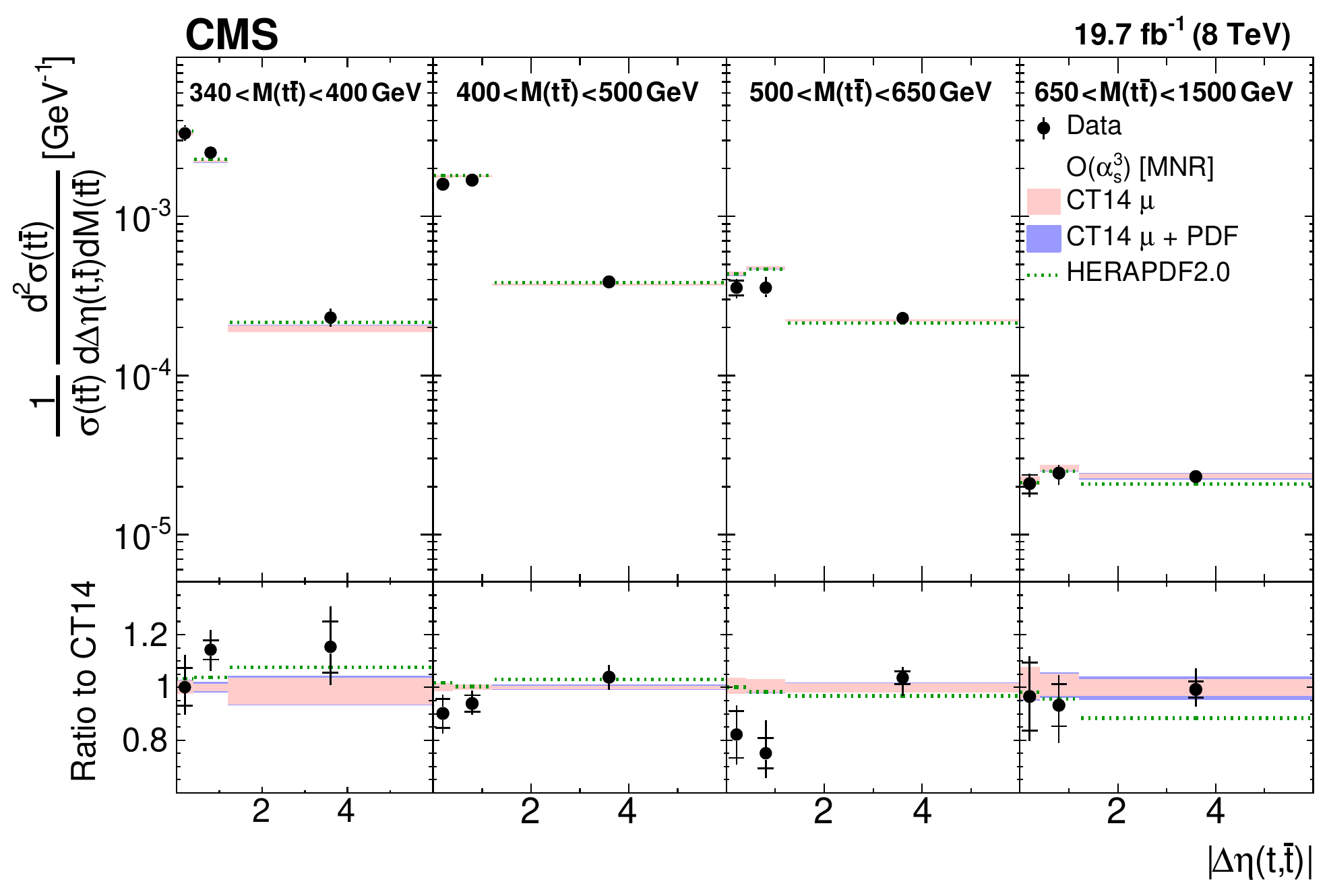}
  \caption{Comparison of the measured normalized \ttbar double-differential cross section as a function of \detatt in different \mtt ranges to NLO $O(\alpha_s^3)$ predictions. Details can be found in the caption of Fig.~\ref{fig:nlo_yt_ptt}. Approximate NNLO $O(\alpha_s^4)$ predictions are not available for this cross section.}
  \label{fig:nlo_mtt_detatt}
\end{figure*}

\begin{figure*}
  \centering
  \includegraphics[width=\cmsFigOne]{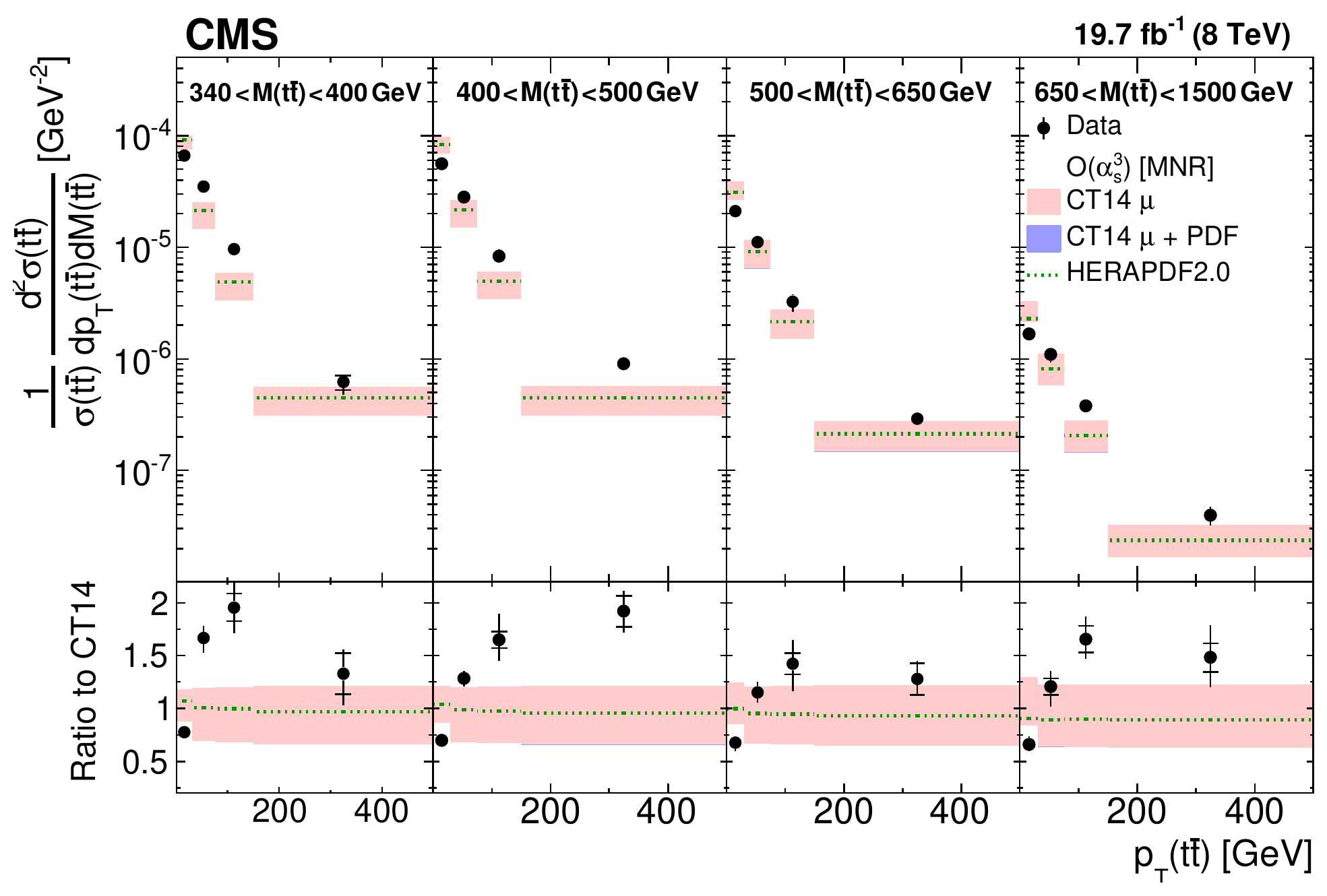}
  \caption{Comparison of the measured normalized \ttbar double-differential cross section as a function of \pttt in different \mtt ranges to NLO $O(\alpha_s^3)$ predictions. Details can be found in the caption of Fig.~\ref{fig:nlo_yt_ptt}. Approximate NNLO $O(\alpha_s^4)$ predictions are not available for this cross section.}
  \label{fig:nlo_mtt_pttt}
\end{figure*}

\begin{figure*}
  \centering
  \includegraphics[width=\cmsFigOne]{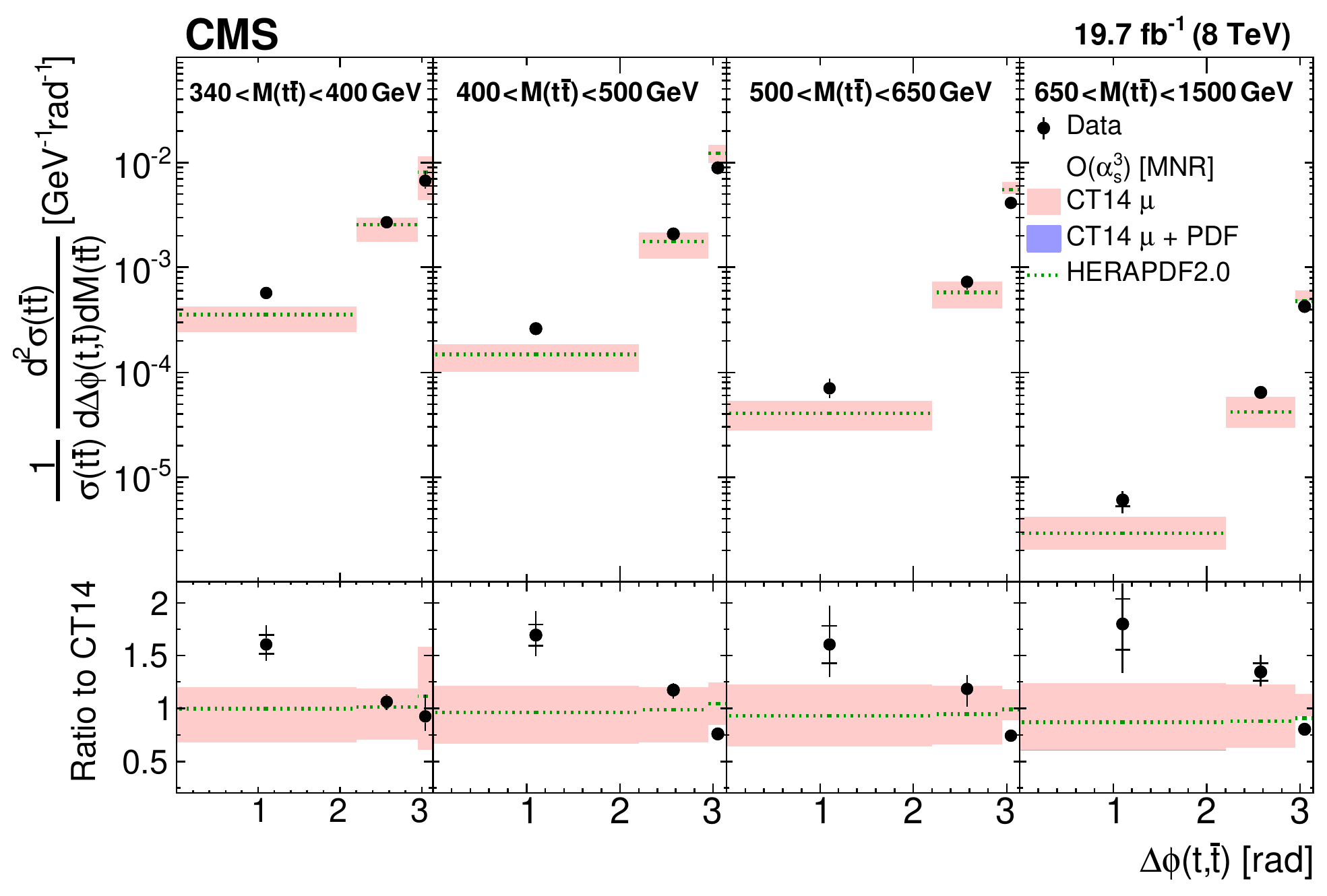}
  \caption{Comparison of the measured normalized \ttbar double-differential cross section as a function of \dphitt in different \mtt ranges to NLO $O(\alpha_s^3)$ predictions. Details can be found in the caption of Fig.~\ref{fig:nlo_yt_ptt}. Approximate NNLO $O(\alpha_s^4)$ predictions are not available for this cross section.}
  \label{fig:nlo_mtt_dphitt}
\end{figure*}

To visualize the comparison of the measurements to the theoretical predictions, the results obtained
using the NLO and approximate NNLO calculations with the \ct PDF set are compared
to the measured \ptt cross sections in different $\abs{\yt}$ ranges in Fig.~\ref{fig:nlo_yt_ptt}.
To further illustrate the sensitivity to PDFs, the nominal values of the NLO predictions
using \herapdf are shown as well.
Similar comparisons, in regions of \mtt, for the $\abs{\yt}$, $\abs{\ytt}$, \detatt, \pttt, and \dphitt cross sections
are presented in Figs.~\ref{fig:nlo_mtt_yt}--\ref{fig:nlo_mtt_dphitt}.
Considering the scale uncertainties in the predictions, the agreement between the measurement and predictions is reasonable for all distributions.
For the \ptt, \yt, and \ytt cross sections, the scale uncertainties in the predictions reach 4\% at maximum.
They increase to 8\% for the \detatt cross section, and vary within $20\text{--}50\%$ for the \pttt and \dphitt cross sections,
where larger differences between data and predictions are observed.
For the \ptt, \yt, and \ytt cross sections, the PDF uncertainties as estimated from the \ct PDF set are of the same size or larger than the scale uncertainties.
The \herapdf predictions are mostly outside the total \ct uncertainty band, showing also some visible shape differences with respect to \ct.
The approximate NNLO predictions provide an improved description of the \ptt shape.

The data-to-theory comparisons illustrate the power of the measured normalized cross sections
as a function of $[\ptt, \yt]$, $[\yt, \mtt]$, and $[\ytt, \mtt]$ to eventually distinguish between modern PDF sets.
Such a study is performed on these data and described in the next section.
The remaining measured normalized cross sections as a function of $[\detatt, \mtt]$, $[\pttt, \mtt]$,
and $[\dphitt, \mtt]$ could be used for this purpose as well, once higher-order QCD calculations become publicly available to match the data precision.
Moreover, since the latter distributions are more sensitive to QCD radiation, they will provide additional input
in testing improvements to the perturbative calculations.

\section{The PDF fit}
\label{sec:qcdanalysis}

The double-differential normalized \ttbar cross sections are used in a PDF fit at NLO, together with the combined HERA
inclusive deep inelastic scattering (DIS) data~\cite{Abramowicz:2015mha} and the CMS measurement of the \Wasymm at $\sqrt{s} = 8$\TeV~\cite{Khachatryan:2016pev}.
The fitted PDFs are also compared to the ones obtained in the recently published CMS measurement of inclusive jet production at 8\TeV~\cite{CMSjets8tev}.
The \xfitter program (formerly known as \herafitter)~\cite{Alekhin:2014irh} (version 1.2.0), an open-source QCD fit framework for PDF determination, is used.
The precise HERA DIS data, obtained from the combination of individual H1 and ZEUS results,
are directly sensitive to the valence and sea quark distributions and probe the gluon
distribution through scaling violations. Therefore, these data form the core of all PDF fits.
The CMS \Wasymm data provide further constraints on the valence quark distributions, as discussed in Ref.~\cite{Khachatryan:2016pev}.
The measured double-differential normalized \ttbar cross sections are included in the fit to constrain the gluon distribution at high $x$ values.
The typical probed $x$ values can be estimated using the LO kinematic relation
$x = (\mtt/\sqrt{s})\exp{[\pm y(\ttbar)]}$.
Therefore, the present measurement is expected to be sensitive to $x$ values in the region $0.01 \lesssim x \lesssim 0.25$,
as estimated using the highest or lowest $\abs{\ytt}$ or \mtt bins and taking the low or high bin edge where the cross section is largest (see Table~\ref{tab:mtt_ytt_xsec}).

\subsection{Details of the PDF fit}
\label{sec:qcdanalysis:details}
{\tolerance=800
The scale evolution of partons is calculated through DGLAP
equations~\cite{Dokshitzer:1977sg,Gribov:1972ri,Altarelli:1977zs,Curci:1980uw,Furmanski:1980cm,Moch:2004pa,Vogt:2004mw} at NLO,
as implemented in the \qcdnum program~\cite{Botje:2010ay} (version 17.01.11).
The Thorne--Roberts~\cite{Thorne:1997ga,Thorne:2006qt,Thorne:2012az} variable-flavour number scheme at NLO is used for the treatment of the heavy-quark contributions.
The number of flavours is set to 5, with c and b quark mass parameters $M_{\PQc}= 1.47$\GeV and $M_{\PQb} = 4.5$\GeV~\cite{Abramowicz:2015mha}.
The theoretical predictions for the \Wasymm data are calculated at NLO~\cite{Campbell:1999ah}
using the \MCFM program, which is interfaced with
\applgrid (version 1.4.70)~\cite{Carli:2010rw}, as described in Ref.~\cite{Khachatryan:2016pev}.
For the DIS and \Wasymm data $\mu_\mathrm{r}$ and $\mu_\mathrm{f}$ are set to $Q$,
which denotes the four-momentum transfer in the case of the DIS data,
and the mass of the $\PW^{\pm}$ boson in the case of the \Wasymm.
The theoretical predictions for the \ttbar cross sections are calculated as described in Section~\ref{sec:comparison}
and included in the fit using the \MCFM and \applgrid programs.
The strong coupling strength is set to $\alpha_s(m_{\Z}) = 0.118$.
The $Q^2$ range of the HERA data is restricted to $Q^2 > Q^2_\text{min} = 3.5\GeV^2$~\cite{Abramowicz:2015mha}.
\par}

{The procedure for the determination of the PDFs follows the approach of \herapdf~\cite{Abramowicz:2015mha}.
The parametrized PDFs are the gluon distribution $x\Pg(x)$, the valence quark distributions $x\cPqu_v(x)$ and $x\cPqd_v(x)$, and
the $\cPqu$- and $\cPqd$-type antiquark distributions $x\overline{U}(x)$ and $x\overline{D}(x)$. At the initial QCD evolution scale $\mu_\mathrm{f0}^2 = 1.9\GeV^2$, the PDFs are parametrized as:
\ifthenelse{\boolean{cms@external}}{
\begin{equation}\begin{aligned}
x\Pg(x) =& A_{\Pg} x^{B_{\Pg}}\,(1-x)^{C_{\Pg}}\, (1 + E_{\Pg} x^2 + F_{\Pg} x^3) \\
&- A'_{\Pg} x^{B'_{\Pg}}\,(1-x)^{C'_{\Pg}},\\
x\cPqu_v(x) =& A_{\cPqu_v}x^{B_{\cPqu_v}}\,(1-x)^{C_{\cPqu_v}}\,(1+D_{\cPqu_v}x+E_{\cPqu_v}x^2) ,\\
x\cPqd_v(x) =& A_{\cPqd_v}x^{B_{\cPqd_v}}\,(1-x)^{C_{\cPqd_v}},\\
x\overline{U}(x)=& A_{\overline{U}}x^{B_{\overline{U}}}\, (1-x)^{C_{\overline{U}}}\, (1+D_{\overline{U}}x+F_{\overline{U}}x^3), \\
x\overline{D}(x)=& A_{\overline{D}}x^{B_{\overline{D}}}\, (1-x)^{C_{\overline{D}}},
\end{aligned}
\label{eq:dv}
\end{equation}
}{
\begin{equation}\begin{aligned}
x\Pg(x) &= A_{\Pg} x^{B_{\Pg}}\,(1-x)^{C_{\Pg}}\, (1 + E_{\Pg} x^2 + F_{\Pg} x^3) - A'_{\Pg} x^{B'_{\Pg}}\,(1-x)^{C'_{\Pg}},\\
x\cPqu_v(x) &= A_{\cPqu_v}x^{B_{\cPqu_v}}\,(1-x)^{C_{\cPqu_v}}\,(1+D_{\cPqu_v}x+E_{\cPqu_v}x^2) ,\\
x\cPqd_v(x) &= A_{\cPqd_v}x^{B_{\cPqd_v}}\,(1-x)^{C_{\cPqd_v}},\\
x\overline{U}(x)&= A_{\overline{U}}x^{B_{\overline{U}}}\, (1-x)^{C_{\overline{U}}}\, (1+D_{\overline{U}}x+F_{\overline{U}}x^3), \\
x\overline{D}(x)&= A_{\overline{D}}x^{B_{\overline{D}}}\, (1-x)^{C_{\overline{D}}},
\end{aligned}
\label{eq:dv}
\end{equation}
}
\tolerance=1200
assuming the relations $x\overline{U}(x) = x\cPaqu(x)$ and $x\overline{D}(x) = x\cPaqd(x) + x\cPaqs(x)$.
Here, $x\cPaqu(x)$, $x\cPaqd(x)$, and $x\cPaqs(x)$ are the up, down, and strange antiquark distributions, respectively.
The sea quark distribution is defined as x$\Sigma(x)=x\cPaqu(x)+x\cPaqd(x)+x\cPaqs(x)$.
The normalization parameters $A_{\cPqu_{{v}}}$, $A_{\cPqd_{v}}$, and $A_{\cPg}$ are determined by the QCD sum rules.
The $B$ and $B'$ parameters determine the PDFs at small $x$,
and the $C$ parameters describe the shape of the distributions as $x\,{\to}\,1$.
The parameter $C'_{\Pg}$ is fixed to 25~\cite{Martin:2009iq}.
Additional constraints $B_{\overline{{U}}} = B_{\overline{{D}}}$ and
$A_{\overline{{U}}} = A_{\overline{{D}}}(1 - f_{\cPqs})$ are imposed to ensure the same normalization
for the $x\cPaqu$ and $x\cPaqd$ distributions as $x \to 0$.
The strangeness fraction $f_{\cPqs} = x\cPaqs/( x\cPaqd + x\cPaqs)$ is fixed to
$f_{\cPqs}=0.4$ as in the \herapdf analysis~\cite{Abramowicz:2015mha}.
This value is consistent with the determination of the
strangeness fraction when using the CMS measurements of $\PW+{\PQc}$ production~\cite{Chatrchyan:2013mza}.
\par}

The parameters in Eq.~(\ref{eq:dv}) are selected by first fitting with all $D$, $E$, and $F$ parameters set to zero,
and then including them independently one at a time in the fit.
The improvement in the $\chi^2$ of the fit is monitored
and the procedure is stopped when no further improvement is observed. This leads to an 18-parameter fit.
The $\chi^2$ definition used for the HERA DIS data follows that of Eq.~(32) in Ref.~\cite{Abramowicz:2015mha}.
It includes an additional logarithmic term that is relevant when the estimated statistical and uncorrelated systematic uncertainties in the data
are rescaled during the fit~\cite{Aaron:2012qi}.
For the CMS \Wasymm and \ttbar data presented here a $\chi^2$ definition without such a logarithmic term is employed.
The full covariance matrix representing the statistical and uncorrelated systematic uncertainties of the data is used in the fit.
The correlated systematic uncertainties are treated through nuisance parameters.
For each nuisance parameter a penalty term is added to the \chisq, representing the prior knowledge of the parameter.
The treatment of the experimental uncertainties for the HERA DIS and CMS \Wasymm data follows the prescription given
in Refs.~\cite{Abramowicz:2015mha} and~\cite{Khachatryan:2016pev}, respectively.
The treatment of the experimental uncertainties in the \ttbar double-differential cross section measurements follows the prescription given in Section~\ref{sec:thmc}.
The experimental systematic uncertainties owing to the PDFs are omitted in the PDF fit.

The PDF uncertainties are estimated according to the general approach of \herapdf~\cite{Abramowicz:2015mha} in which
the fit, model, and parametrization uncertainties are taken into account.
Fit uncertainties are determined using the tolerance criterion of $\Delta\chi^2 =1$.
Model uncertainties arise from the variations in the values assumed for the b and c quark mass parameters
of $4.25\leq M_{\PQb}\leq 4.75\GeV$ and $1.41\leq M_{\PQc}\leq 1.53\GeV$,
the strangeness fraction $0.3 \leq f_{\cPqs} \leq 0.4$, and the value of $Q^2_{\text{min}}$ imposed on the HERA data.
The latter is varied within $2.5 \leq Q^2_{\text{min}}\leq 5.0\GeV^2$, following Ref.~\cite{Abramowicz:2015mha}.
The parametrization uncertainty is estimated by extending the functional form in Eq.~(\ref{eq:dv})
of all parton distributions with additional parameters $D$, $E$, and $F$ added one at a time.
Furthermore, $\mu_\mathrm{f0}^2$ is changed to $1.6\GeV^2$ and $2.2\GeV^2$.
The parametrization uncertainty is constructed as an envelope at each $x$ value,
built from the maximal differences between the PDFs resulting from the central fit and all parametrization variations.
This uncertainty is valid in the $x$ range covered by the PDF fit to the data.
The total PDF uncertainty is obtained by adding the fit, model, and parametrization uncertainties in quadrature.
In the following, the quoted uncertainties correspond to 68\% CL.

\subsection{Impact of the double-differential \texorpdfstring{\ttbar}{ttbar} cross section measurements}
{\tolerance=1200
The PDF fit is first performed using only the HERA DIS and CMS \Wasymm data.
To demonstrate the added value of the double-differential normalized \ttbar cross sections,
$[\ptt, \yt]$, $[\yt, \mtt]$, and $[\ytt, \mtt]$ measurements are added to the fit one at a time.
The global and partial \chisq values for all variants of the fit are listed in Table~\ref{tab:chi2fit},
illustrating the consistency among the input data.
The DIS data show \chisqndf values slightly larger than unity.
This is similar to what is observed and investigated in Ref.~\cite{Abramowicz:2015mha}.
Fit results consistent with those from Ref.~\cite{Khachatryan:2016pev} are obtained using the \Wasymm measurements.
\par}
    \begin{table*}
    \topcaption{The global and partial \chisqndf values for all variants of the PDF fit.
      The variant of the fit that uses the DIS and \Wasymm data only is denoted as `Nominal fit'.
      Each double-differential \ttbar cross section is added ($+$) to the nominal data, one at a time.
      For the HERA measurements, the energy of the proton beam, $E_{\Pp}$, is listed for each data set, with the electron energy being $E_{\Pe}=27.5\GeV$,
      CC and NC stand for charged and neutral current, respectively.
      The correlated \chisq and the log-penalty $\chi^2$ entries refer to the \chisq contributions
      from the nuisance parameters and from the logarithmic term, respectively, as described in the text.
      }
    \label{tab:chi2fit}
	  \centering
	\begin{tabular}{lcccc}
    \multirow{2}{*}{Data sets} & \multicolumn{4}{c}{\chisqndf} \\
    \cline{2-5}
	    & Nominal fit & $+ [\ptt, \yt]$   & $+ [\yt, \mtt]$   & $ + [\ytt, \mtt]$  \\
    \hline
	  CMS double-differential \ttbar     &  & $10 / 15$& $7.4 / 15$& $7.6 / 15$  \\
	  HERA CC $\Pem\Pp$, $E_{\Pp}=920$\GeV & $57 / 42$& $56 / 42$& $56 / 42$& $57 / 42$  \\
	  HERA CC $\Pep\Pp$, $E_{\Pp}=920$\GeV & $44 / 39$& $44 / 39$& $44 / 39$& $43 / 39$  \\
	  HERA NC $\Pem\Pp$, $E_{\Pp}=920$\GeV & $219 / 159$& $219 / 159$& $219 / 159$& $218 / 159$  \\
	  HERA NC $\Pep\Pp$, $E_{\Pp}=920$\GeV & $440 / 377$& $437 / 377$& $439 / 377$& $441 / 377$  \\
	  HERA NC $\Pep\Pp$, $E_{\Pp}=820$\GeV & $69 / 70$& $68 / 70$& $68 / 70$& $69 / 70$  \\
	  HERA NC $\Pep\Pp$, $E_{\Pp}=575$\GeV & $221 / 254$& $220 / 254$& $221 / 254$& $221 / 254$  \\
	  HERA NC $\Pep\Pp$, $E_{\Pp}=460$\GeV & $219 / 204$& $219 / 204$& $219 / 204$& $219 / 204$  \\
	  CMS $\PW^{\pm}$ asymmetry & $4.7 / 11$& $4.6 / 11$& $4.8 / 11$& $4.9 / 11$  \\
	  \hline
	  Correlated $\chi^2$  & 82& 87& 91& 89  \\
	  Log-penalty $\chi^2$  & $-2.5$& $+2.6$& $-2.2$& $-3.3$  \\
	  \hline
	  Total \chisqndf & $1352 / 1138$& $1368 / 1153$& $1368 / 1153$& $1366 / 1153$  \\
		\end{tabular}	
	\end{table*}

\begin{figure*}
  \centering
  \includegraphics[width=0.495\textwidth]{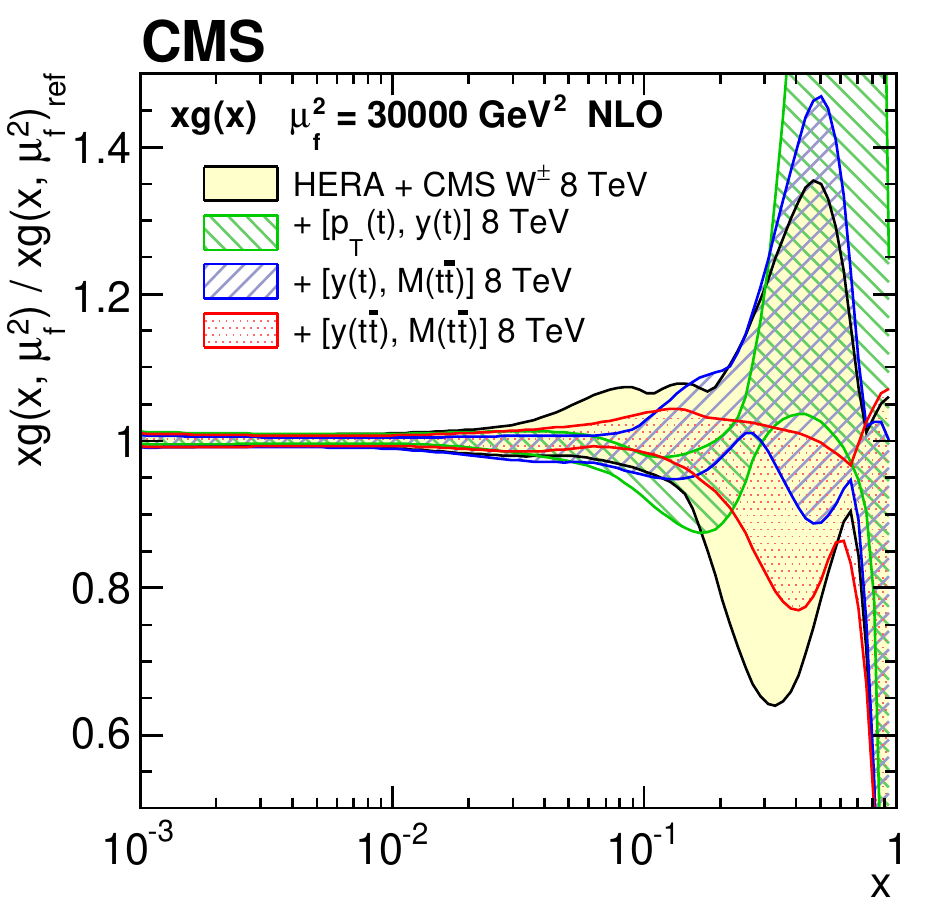}
  \includegraphics[width=0.495\textwidth]{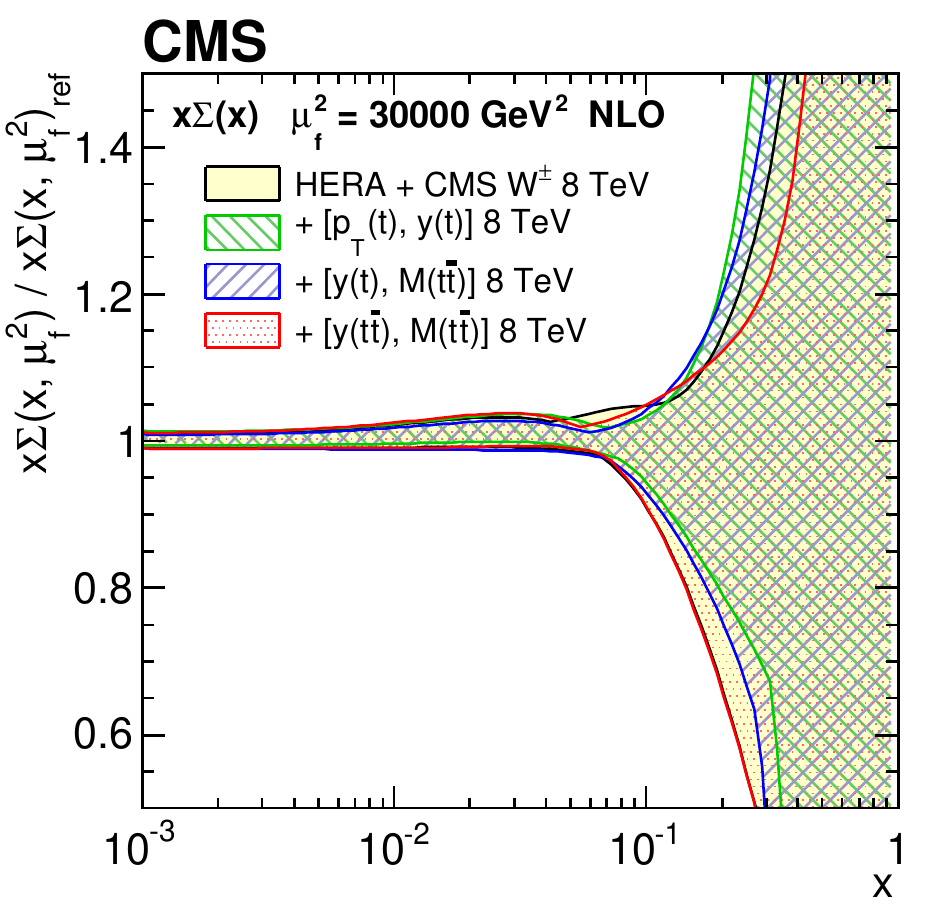}\\
  \includegraphics[width=0.495\textwidth]{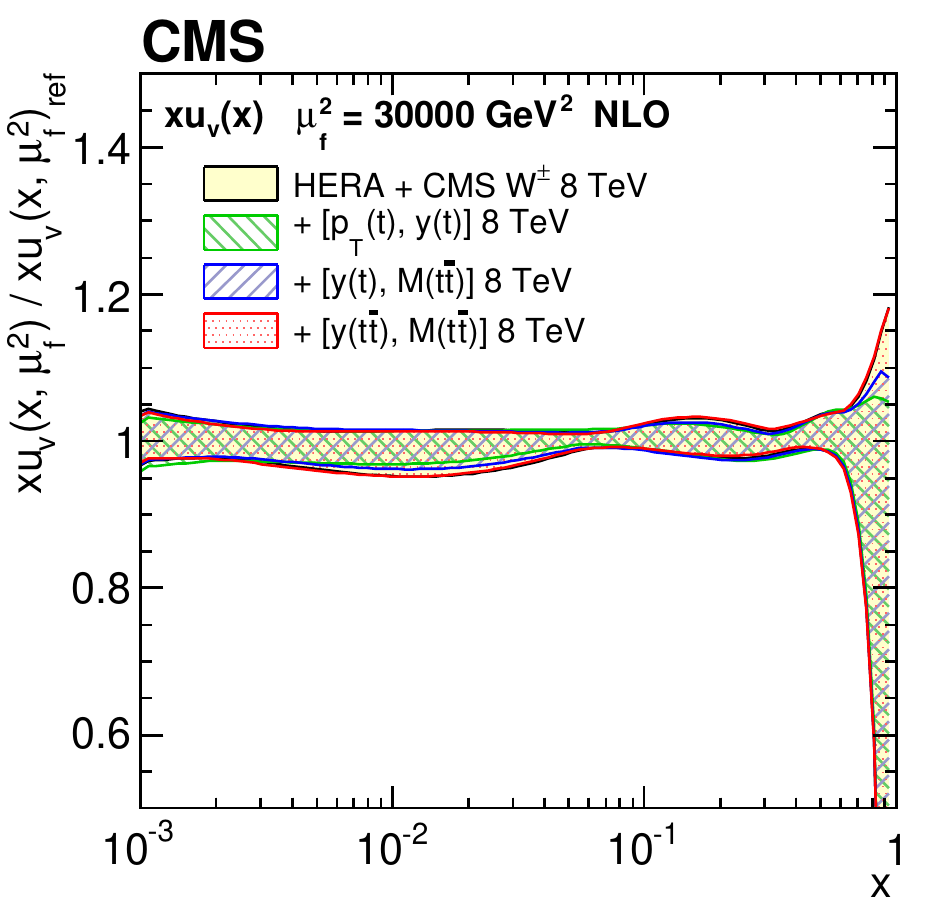}
  \includegraphics[width=0.495\textwidth]{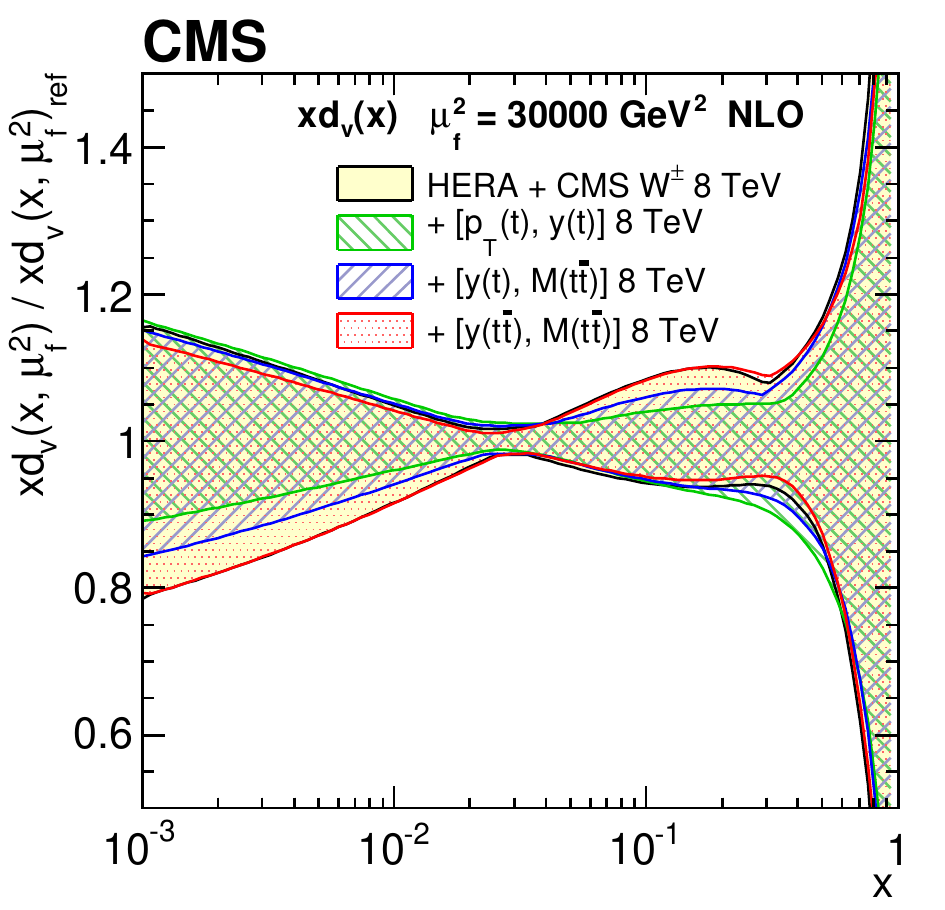}
  \caption{The gluon (upper left), sea quark (upper right), u valence quark (lower left), and d valence quark (lower right) PDFs
    at $\mu_\mathrm{f}^2=30\,000\GeV^2$, as obtained in all variants of the PDF fit,
    normalized to the results from the fit using the HERA DIS and CMS \Wasymm measurements only.
    The shaded, hatched, and dotted areas represent the total uncertainty in each of the fits.}
  \label{fig:pdf_bands}
\end{figure*}

\begin{figure*}
  \centering
  \includegraphics[width=0.495\textwidth]{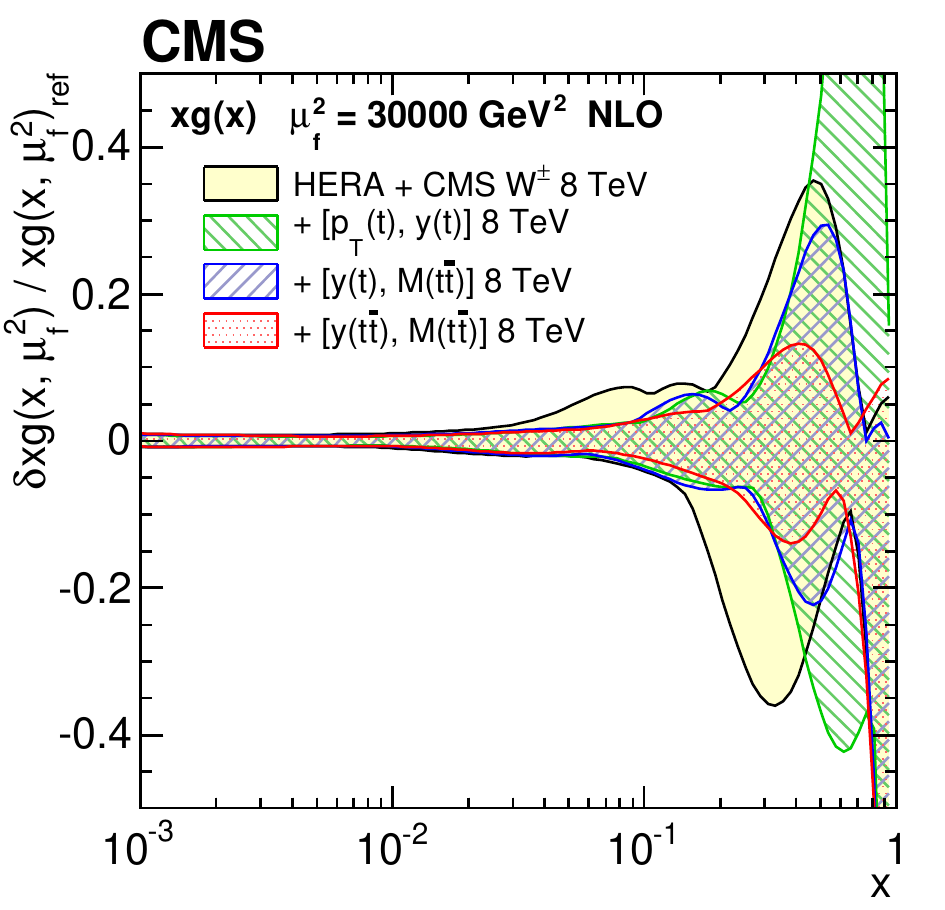}
  \includegraphics[width=0.495\textwidth]{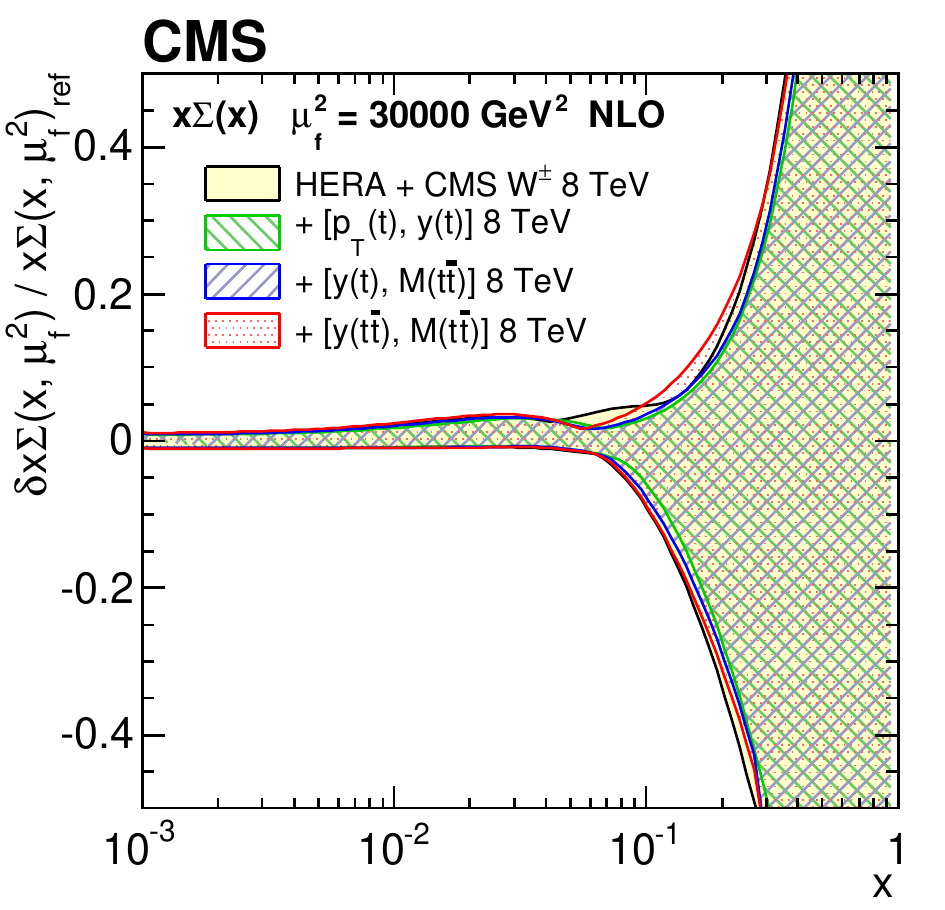}\\
  \includegraphics[width=0.495\textwidth]{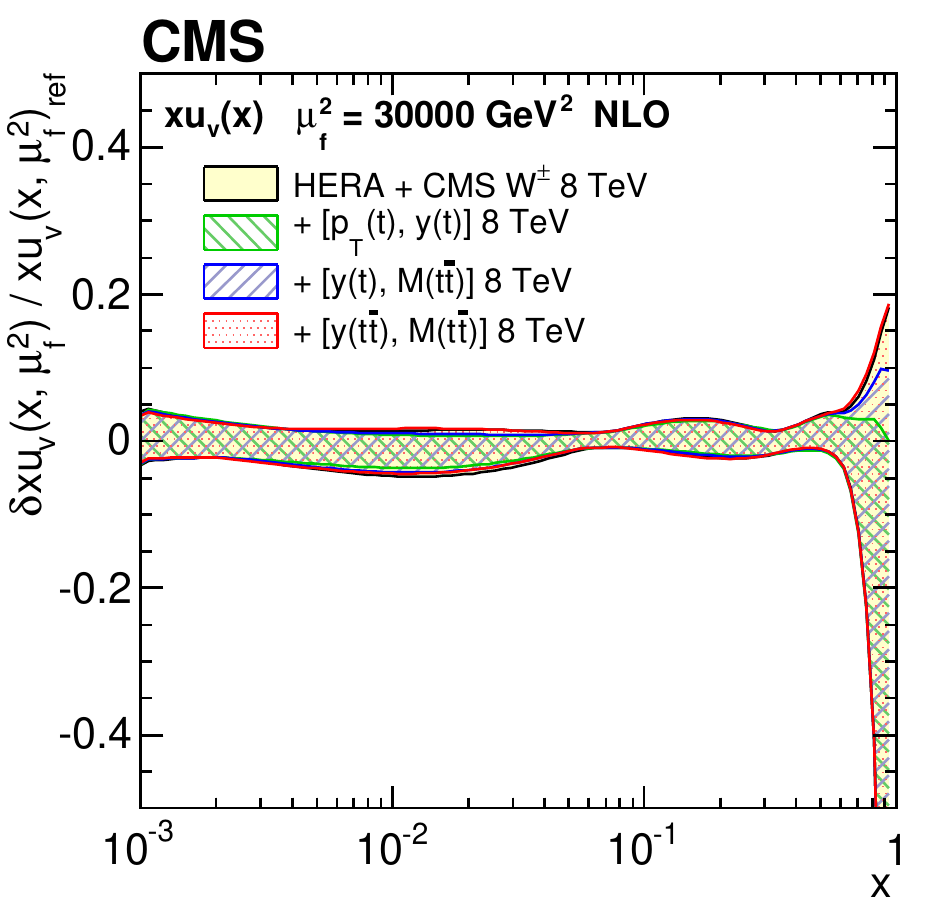}
  \includegraphics[width=0.495\textwidth]{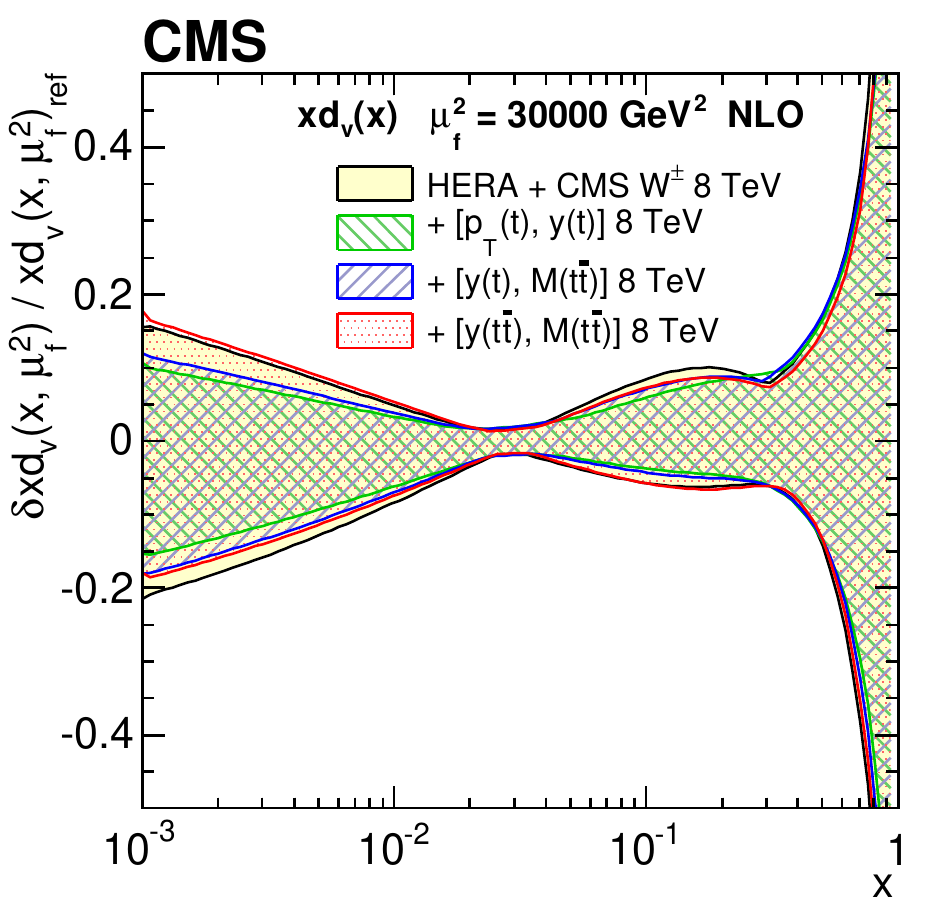}
  \caption{Relative total uncertainties of the gluon (upper left), sea quark (upper right),
    u valence quark (lower left), and d valence quark (lower right) distributions
    at $\mu_\mathrm{f}^2=30\,000\GeV^2$, shown by shaded, hatched, and dotted areas, as obtained in all variants of the PDF fit.}
  \label{fig:pdf_relunc}
\end{figure*}

{\tolerance=1200
The resulting gluon, valence quark, and sea quark distributions
are shown in Fig.~\ref{fig:pdf_bands} at the scale $\mu_\mathrm{f}^2=30\,000\GeV^2 \simeq m_{\PQt}^2$ relevant for \ttbar production.
For a direct comparison, the distributions for all variants of the fit are normalized to the results from the fit using only the DIS and \Wasymm data.
The reduction of the uncertainties is further illustrated in Fig.~\ref{fig:pdf_relunc}.
The uncertainties in the gluon distribution at $x>0.01$ are significantly reduced once the \ttbar data are included in the fit.
The largest improvement comes from the $[\ytt, \mtt]$ cross section by which the total gluon PDF uncertainty is reduced by more than a factor of two at $x \simeq 0.3$.
This value of $x$ is at the edge of kinematic reach of the current \ttbar measurement.
At higher values $x \gtrsim 0.3$, the gluon distribution is not directly constrained by the data and should be considered as an extrapolation
that relies on the PDF parametrization assumptions.
No substantial effects on the valence quark and sea quark distributions are observed.
The variation of $\mu_\mathrm{r}$ and $\mu_\mathrm{f}$ in the prediction of the normalized \ttbar cross sections has been performed
and the effect on the fitted PDFs is found to be well within the total uncertainty.
\par}

\begin{figure}
  \centering
  \includegraphics[width=0.49\textwidth]{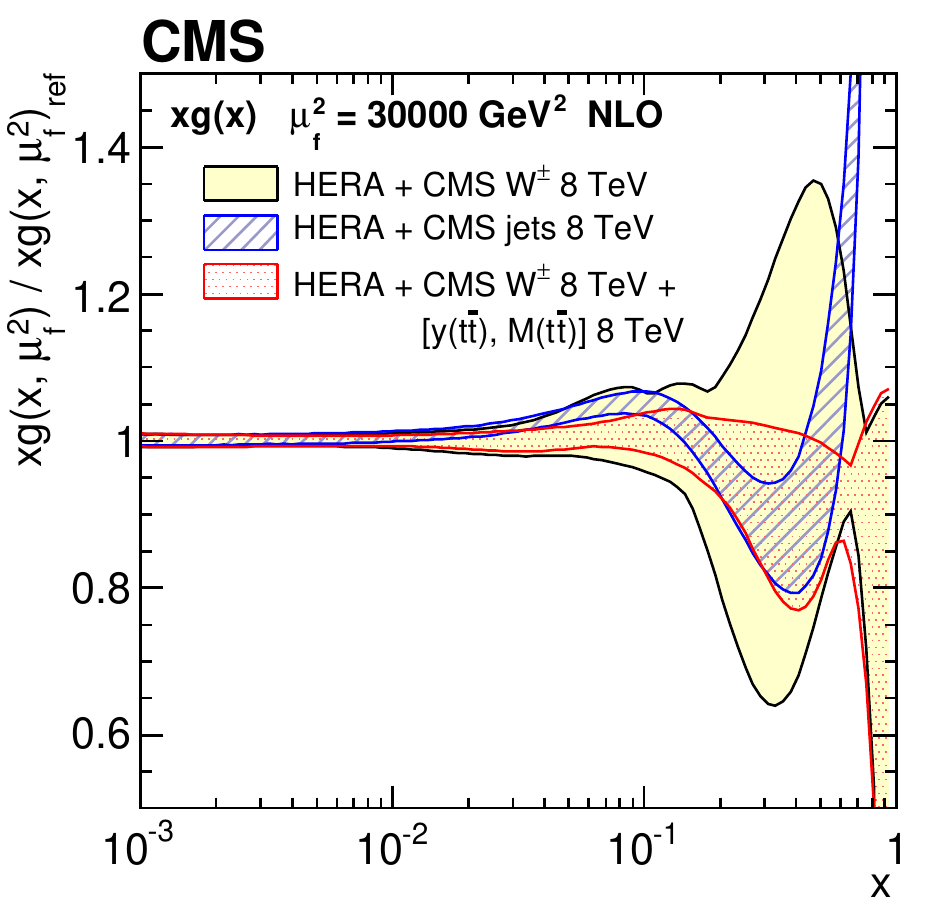}
  \caption{
    The gluon distribution at $\mu_\mathrm{f}^2=30\,000\GeV^2$, as obtained from the PDF fit to the HERA DIS data
    and CMS \Wasymm measurements (shaded area), the CMS inclusive jet production cross sections (hatched area),
    and the \Wasymm plus the double-differential \ttbar cross section (dotted area).
    All presented PDFs are normalized to the results from the fit using the DIS and \Wasymm measurements.
    The shaded, hatched, and dotted areas represent the total uncertainty in each of the fits.
  }
  \label{fig:pdf_vsjets}
\end{figure}

The gluon distribution obtained from fitting the measured $[\ytt, \mtt]$ cross section is compared in Fig.~\ref{fig:pdf_vsjets}
to the one obtained in a similar study using the CMS measurement of inclusive jet production at 8\TeV~\cite{CMSjets8tev}.
The two results are in agreement in the probed $x$ range.
The constraints provided by the double-differential \ttbar measurement are competitive with those from the inclusive jet data.

\subsection{Comparison to the impact of single-differential \texorpdfstring{\ttbar}{ttbar} cross section measurements}
\begin{figure*}
  \centering
  \includegraphics[width=0.495\textwidth]{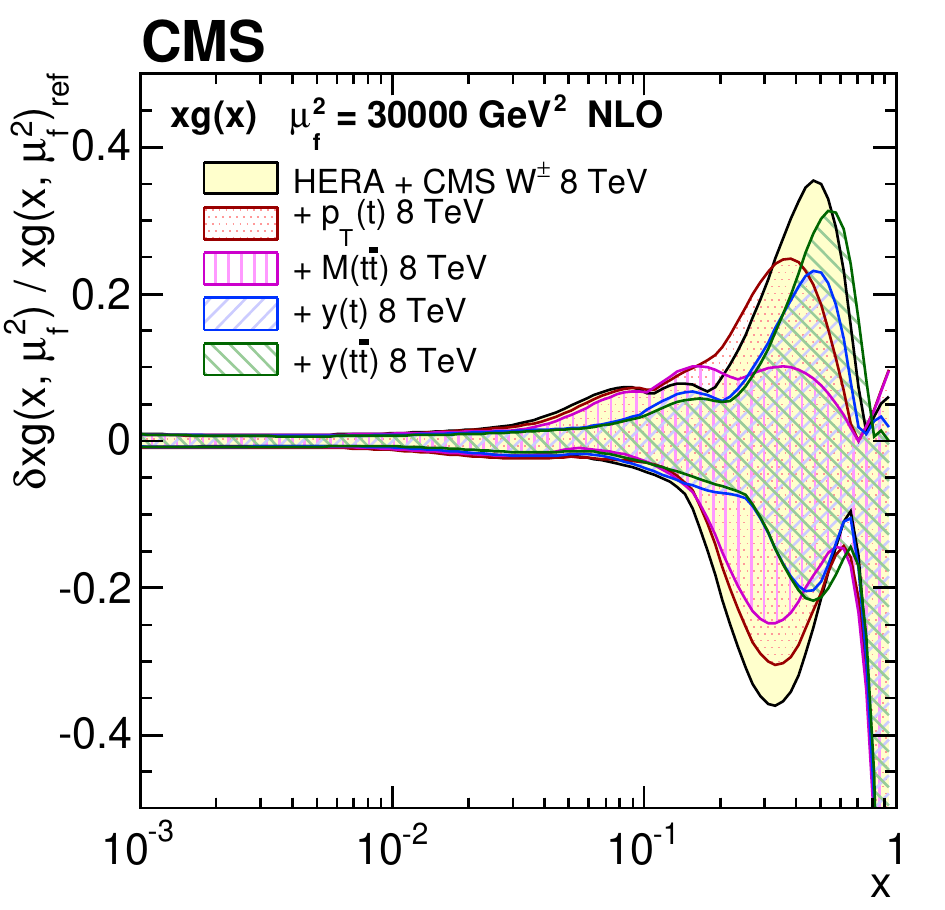}
  \includegraphics[width=0.495\textwidth]{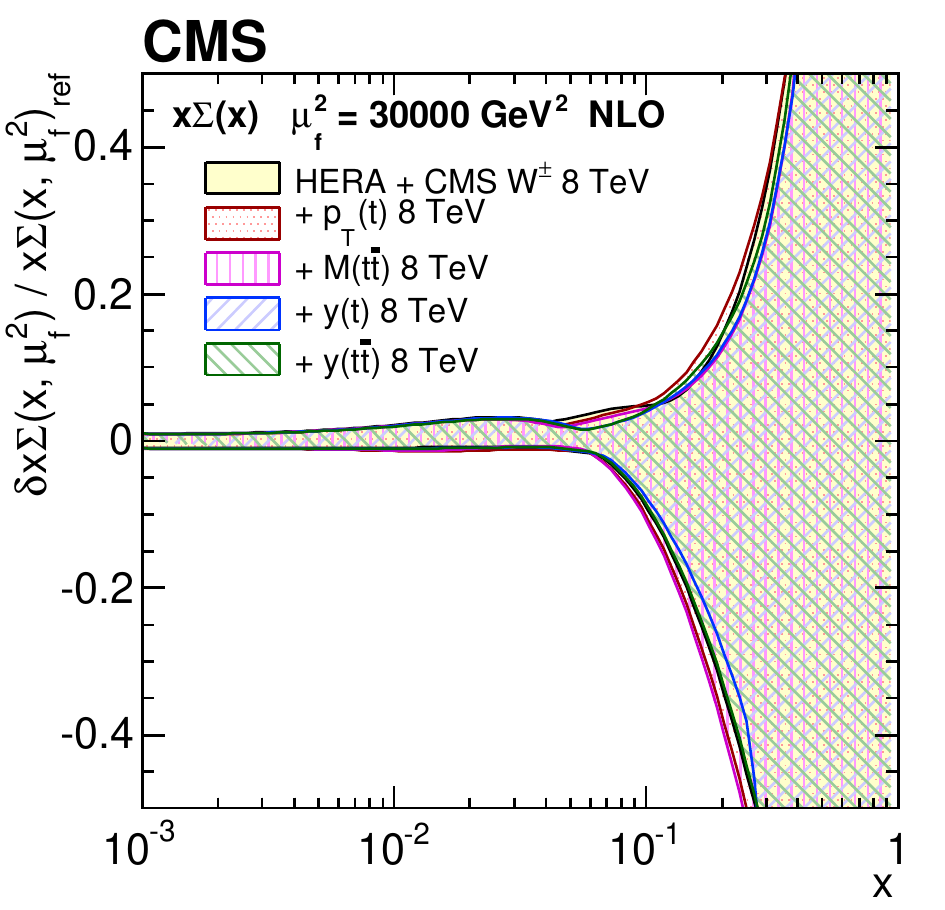}\\
  \includegraphics[width=0.495\textwidth]{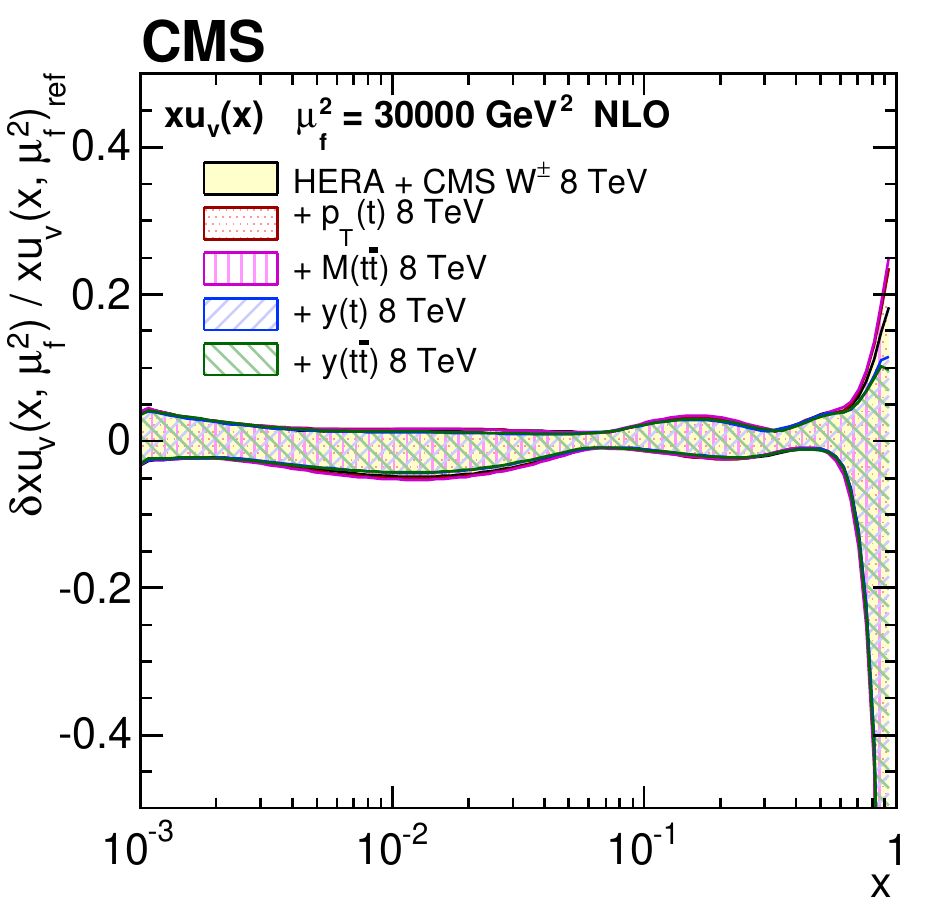}
  \includegraphics[width=0.495\textwidth]{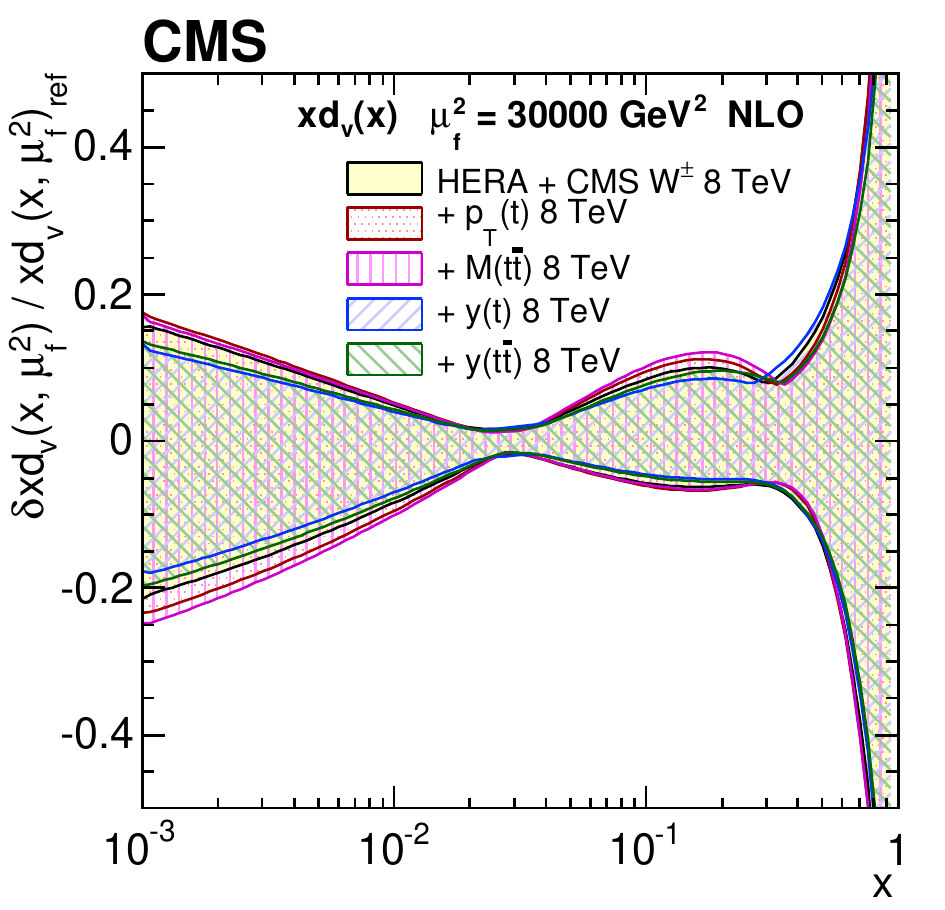}
  \caption{The same as in Fig.~\ref{fig:pdf_relunc} for the variants of the PDF fit using the single-differential \ttbar cross sections.}
  \label{fig:pdf_relunc_1d}
\end{figure*}

The power of the double-differential \ttbar measurement in fitting PDFs is compared with that of the single-differential analysis,
where the \ttbar cross section is measured as a function of \ptt, \yt, \ytt, and \mtt, employing in one dimension
the same procedure described in this paper.
The measurements are added, one at a time, to the HERA DIS and CMS \Wasymm data in the PDF fit.
The reduction of the uncertainties for the resulting PDFs is illustrated in Fig.~\ref{fig:pdf_relunc_1d}.
Similar effects are observed from all measurements, with the largest impact coming from \yt and \ytt.
For the single-differential \ttbar data one can extend the studies using the approximate NNLO
calculations~\cite{Kidonakis:2001nj,Guzzi:2014wia,Guzzi:2014tna,Guzzi:2013noa}.
An example, using the \yt distribution, is presented in Appendix B.

\begin{figure}
  \centering
  \includegraphics[width=0.49\textwidth]{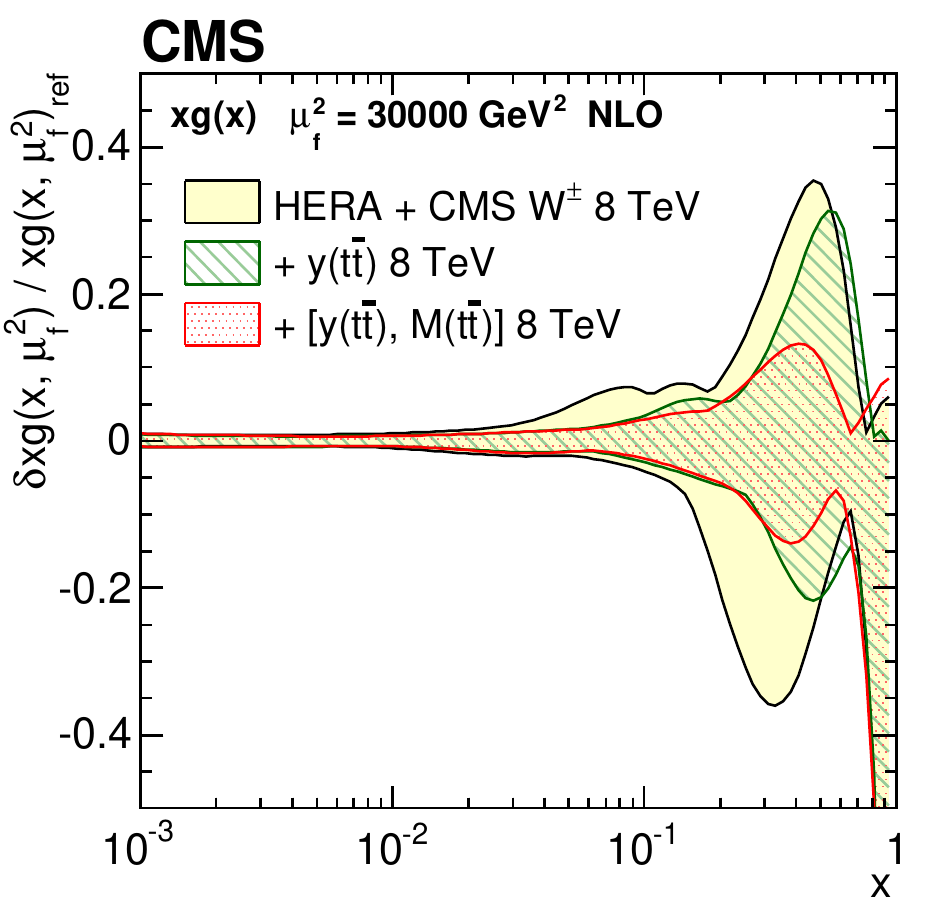}
  \caption{Relative total uncertainties of the gluon distribution
    at $\mu_\mathrm{f}^2=30\,000\GeV^2$, shown by shaded (or hatched) bands, as obtained in the PDF fit using the DIS and \Wasymm data only,
    as well as single- and double-differential \ttbar cross sections.}
  \label{fig:pdf_relunc_1d_2d}
\end{figure}

A comparison of the PDF uncertainties from the double-differential cross section as a function of $[\ytt, \mtt]$,
and single-differential cross section as a function of \ytt is presented in Fig.~\ref{fig:pdf_relunc_1d_2d}.
Only the gluon distribution is shown, since no substantial impact on the other distributions
is observed (see Figs.~\ref{fig:pdf_bands}, \ref{fig:pdf_relunc}, and \ref{fig:pdf_relunc_1d}).
The total gluon PDF uncertainty becomes noticeably smaller once the double-differential cross sections are included.
The observed improvement makes future PDF fits at NNLO using
the fully differential calculations~\cite{Czakon:2015owf,Czakon:2016dgf},
once they become available, very interesting.

\section{Summary}
\label{sec:concl}

A measurement of normalized double-differential \ttbar production cross sections in pp~collisions at $\sqrt{s}=8\TeV$ has been presented.
The measurement is performed in the $\Pe^{\pm}\mu^{\mp}$ final state, using data collected with the CMS detector at the LHC, corresponding to an integrated luminosity of 19.7\fbinv.
The normalized \ttbar cross section is measured in the full phase space as a function of different pairs of kinematic variables
describing the top quark or \ttbar system.
None of the tested MC models is able to correctly describe all the double-differential distributions.
The data exhibit a softer transverse momentum \ptt distribution, compared to the Monte Carlo predictions, as was reported in previous single-differential \ttbar cross section measurements.
The double-differential studies reveal a broader distribution of rapidity \yt at high \ttbar invariant mass \mtt
and a larger pseudorapidity separation \detatt at moderate \mtt in data compared to simulation.
The data are in reasonable agreement with next-to-leading-order predictions of quantum chromodynamics using recent sets of parton distribution functions (PDFs).

The measured double-differential cross sections have been incorporated into a PDF fit, together with other data from HERA and the LHC.
Including the \ttbar data, one observes
a significant reduction in the uncertainties in the gluon distribution at large values of parton momentum fraction $x$,
in particular when using the double-differential \ttbar cross section as a function of \ytt and \mtt.
The constraints provided by these data are competitive with those from inclusive jet data.
This improvement exceeds that from using single-differential \ttbar cross section data,
thus strongly suggesting the use of the double-differential \ttbar measurements in PDF fits.

\begin{acknowledgments}
\hyphenation{Bundes-ministerium Forschungs-gemeinschaft Forschungs-zentren Rachada-pisek} We congratulate our colleagues in the CERN accelerator departments for the excellent performance of the LHC and thank the technical and administrative staffs at CERN and at other CMS institutes for their contributions to the success of the CMS effort. In addition, we gratefully acknowledge the computing centres and personnel of the Worldwide LHC Computing Grid for delivering so effectively the computing infrastructure essential to our analyses. Finally, we acknowledge the enduring support for the construction and operation of the LHC and the CMS detector provided by the following funding agencies: the Austrian Federal Ministry of Science, Research and Economy and the Austrian Science Fund; the Belgian Fonds de la Recherche Scientifique, and Fonds voor Wetenschappelijk Onderzoek; the Brazilian Funding Agencies (CNPq, CAPES, FAPERJ, and FAPESP); the Bulgarian Ministry of Education and Science; CERN; the Chinese Academy of Sciences, Ministry of Science and Technology, and National Natural Science Foundation of China; the Colombian Funding Agency (COLCIENCIAS); the Croatian Ministry of Science, Education and Sport, and the Croatian Science Foundation; the Research Promotion Foundation, Cyprus; the Secretariat for Higher Education, Science, Technology and Innovation, Ecuador; the Ministry of Education and Research, Estonian Research Council via IUT23-4 and IUT23-6 and European Regional Development Fund, Estonia; the Academy of Finland, Finnish Ministry of Education and Culture, and Helsinki Institute of Physics; the Institut National de Physique Nucl\'eaire et de Physique des Particules~/~CNRS, and Commissariat \`a l'\'Energie Atomique et aux \'Energies Alternatives~/~CEA, France; the Bundesministerium f\"ur Bildung und Forschung, Deutsche Forschungsgemeinschaft, and Helmholtz-Gemeinschaft Deutscher Forschungszentren, Germany; the General Secretariat for Research and Technology, Greece; the National Scientific Research Foundation, and National Innovation Office, Hungary; the Department of Atomic Energy and the Department of Science and Technology, India; the Institute for Studies in Theoretical Physics and Mathematics, Iran; the Science Foundation, Ireland; the Istituto Nazionale di Fisica Nucleare, Italy; the Ministry of Science, ICT and Future Planning, and National Research Foundation (NRF), Republic of Korea; the Lithuanian Academy of Sciences; the Ministry of Education, and University of Malaya (Malaysia); the Mexican Funding Agencies (BUAP, CINVESTAV, CONACYT, LNS, SEP, and UASLP-FAI); the Ministry of Business, Innovation and Employment, New Zealand; the Pakistan Atomic Energy Commission; the Ministry of Science and Higher Education and the National Science Centre, Poland; the Funda\c{c}\~ao para a Ci\^encia e a Tecnologia, Portugal; JINR, Dubna; the Ministry of Education and Science of the Russian Federation, the Federal Agency of Atomic Energy of the Russian Federation, Russian Academy of Sciences, the Russian Foundation for Basic Research and the Russian Competitiveness Program of NRNU MEPhI; the Ministry of Education, Science and Technological Development of Serbia; the Secretar\'{\i}a de Estado de Investigaci\'on, Desarrollo e Innovaci\'on, Programa Consolider-Ingenio 2010, Plan de Ciencia, Tecnolog\'ia e Innovaci\'on 2013-2017 del Principado de Asturias and Fondo Europeo de Desarrollo Regional, Spain; the Swiss Funding Agencies (ETH Board, ETH Zurich, PSI, SNF, UniZH, Canton Zurich, and SER); the Ministry of Science and Technology, Taipei; the Thailand Center of Excellence in Physics, the Institute for the Promotion of Teaching Science and Technology of Thailand, Special Task Force for Activating Research and the National Science and Technology Development Agency of Thailand; the Scientific and Technical Research Council of Turkey, and Turkish Atomic Energy Authority; the National Academy of Sciences of Ukraine, and State Fund for Fundamental Researches, Ukraine; the Science and Technology Facilities Council, UK; the US Department of Energy, and the US National Science Foundation.

Individuals have received support from the Marie-Curie programme and the European Research Council and EPLANET (European Union); the Leventis Foundation; the A. P. Sloan Foundation; the Alexander von Humboldt Foundation; the Belgian Federal Science Policy Office; the Fonds pour la Formation \`a la Recherche dans l'Industrie et dans l'Agriculture (FRIA-Belgium); the Agentschap voor Innovatie door Wetenschap en Technologie (IWT-Belgium); the Ministry of Education, Youth and Sports (MEYS) of the Czech Republic; the Council of Scientific and Industrial Research, India; the HOMING PLUS programme of the Foundation for Polish Science, cofinanced from European Union, Regional Development Fund, the Mobility Plus programme of the Ministry of Science and Higher Education, the National Science Center (Poland), contracts Harmonia 2014/14/M/ST2/00428, Opus 2014/13/B/ST2/02543, 2014/15/B/ST2/03998, and 2015/19/B/ST2/02861, Sonata-bis 2012/07/E/ST2/01406; the National Priorities Research Program by Qatar National Research Fund; the Programa Clar\'in-COFUND del Principado de Asturias; the Thalis and Aristeia programmes cofinanced by EU-ESF and the Greek NSRF; the Rachadapisek Sompot Fund for Postdoctoral Fellowship, Chulalongkorn University and the Chulalongkorn Academic into Its 2nd Century Project Advancement Project (Thailand); and the Welch Foundation, contract C-1845.
\end{acknowledgments}
\clearpage
\bibliography{auto_generated}

\appendix
\section{Values of the normalized double-differential cross sections}
\label{sec:app}
Tables~\ref{tab:yt_ptt_xsec} to~\ref{tab:mtt_dphitt_syst} provide the measured \ttbar double-differential cross sections for all pairs of variables,
including their correlation matrices of statistical uncertainties and detailed breakdown of systematic uncertainties.
The b tagging systematic uncertainty is obtained by combining in quadrature variations of the data-to-simulation correction factors
as a function of \pt and $\abs{\eta}$, performed separately for jets originating from \PQb quarks and other partons, as presented in Sections~\ref{sec:sel} and \ref{sec:syst}.
The PDF systematic uncertainty is obtained by combining in quadrature variations corresponding to the 52 eigenvectors of the CT10 PDF set~\cite{Lai:2010vv}.

\begin{table*}
\centering
\topcaption{The measured normalized \ttbar double-differential cross sections in different bins of \yt and \ptt, along with their relative statistical and systematic uncertainties.}
\label{tab:yt_ptt_xsec}
\renewcommand*{\arraystretch}{1.40}
\begin{tabular}{*{6}{c}}
$\abs{\yt}$  & $\ptt$ [\GeVns{}] & $\frac{1}{\sigma(\ttbar)} \frac{\rd ^{2}\sigma(\ttbar)}{\rd \yt\rd \ptt}$ [\GeVns{}$^{-1}$] & stat. [\%] & syst. [\%] & bin \\
\hline
\multirow{4}{*}{$      0$--$   0.35$} & $      0$--$     80$ & $3.08 \times 10^{-3}$ & $ 4.4$ & ${}_{ -4.9}^{ +7.4}$ & $ 1$\\
                                      & $     80$--$    150$ & $3.71 \times 10^{-3}$ & $ 3.6$ & ${}_{ -6.2}^{ +3.5}$ & $ 2$\\
                                      & $    150$--$    250$ & $1.36 \times 10^{-3}$ & $ 5.0$ & ${}_{ -3.4}^{ +4.8}$ & $ 3$\\
                                      & $    250$--$    600$ & $1.11 \times 10^{-4}$ & $ 7.7$ & ${}_{-11.7}^{ +9.0}$ & $ 4$\\
\hline
\multirow{4}{*}{$   0.35$--$   0.85$} & $      0$--$     80$ & $2.90 \times 10^{-3}$ & $ 3.2$ & ${}_{ -2.9}^{ +3.0}$ & $ 5$\\
                                      & $     80$--$    150$ & $3.17 \times 10^{-3}$ & $ 3.0$ & ${}_{ -4.2}^{ +2.3}$ & $ 6$\\
                                      & $    150$--$    250$ & $1.17 \times 10^{-3}$ & $ 4.5$ & ${}_{ -3.8}^{ +7.3}$ & $ 7$\\
\end{tabular}
\end{table*}
 \begin{table*}
\centering
\topcaption{The correlation matrix of statistical uncertainties for the normalized \ttbar double-differential cross sections as a function of \yt and \ptt. The values are expressed as percentages. For bin indices see Table~\ref{tab:yt_ptt_xsec}.}
\label{tab:yt_ptt_corr}
\renewcommand*{\arraystretch}{1.40}
\tabcolsep0.40mm
\cmsTable{
\begin{tabular}{c|rrrrrrrrrrrrrrrr}
\quad bin\quad &$     1$&$     2$&$     3$&$     4$&$     5$&$     6$&$     7$&$     8$&$     9$&$    10$&$    11$&$    12$&$    13$&$    14$&$    15$&$    16$ \\
\hline
$ 1$&$+100.0$&$ -18.5$&$ -16.4$&$  +5.0$&$ -31.8$&$ -23.4$&$  +8.6$&$  -2.2$&$ -46.5$&$  +1.4$&$  +7.5$&$  -3.8$&$ +11.3$&$  +4.7$&$  -7.9$&$  +1.8$ \\
$ 2$&        &$+100.0$&$ -16.1$&$  -4.1$&$ -24.3$&$ -10.6$&$ -16.3$&$  +5.2$&$  -0.8$&$ -22.5$&$  +5.0$&$  -0.3$&$  +3.6$&$  -1.4$&$  -0.5$&$  -1.1$ \\
$ 3$&        &        &$+100.0$&$ -35.2$&$ +11.7$&$ -17.4$&$ -23.5$&$  +1.8$&$  +7.9$&$  +4.6$&$  -6.6$&$  +4.2$&$  -9.8$&$  -0.9$&$  +2.2$&$  -1.4$ \\
$ 4$&        &        &        &$+100.0$&$  -1.7$&$  +7.1$&$  +2.0$&$ -23.0$&$  -3.7$&$  +0.3$&$  +3.5$&$  +1.3$&$  -1.2$&$  -3.0$&$  -1.2$&$  +0.1$ \\
$ 5$&        &        &        &        &$+100.0$&$  -4.0$&$ -19.0$&$  +5.3$&$ +30.9$&$ -17.8$&$  -3.4$&$  +2.4$&$ -46.9$&$  -4.0$&$ +10.3$&$  -4.3$ \\
$ 6$&        &        &        &        &        &$+100.0$&$  +1.5$&$  -9.3$&$ -19.3$&$ +19.3$&$ -17.9$&$  +4.1$&$  -6.3$&$ -25.3$&$  +3.0$&$  +1.2$ \\
$ 7$&        &        &        &        &        &        &$+100.0$&$ -29.9$&$  -4.8$&$ -16.5$&$  -6.5$&$  -3.5$&$  +4.8$&$  +0.5$&$  -8.7$&$  +4.6$ \\
\end{tabular}
}
\end{table*}
 \begin{table*}
\centering
\topcaption{Sources and values of the relative systematic uncertainties in percent of the measured normalized \ttbar double-differential cross sections as a function of \yt and \ptt.
For bin indices see Table~\ref{tab:yt_ptt_xsec}.}
\label{tab:yt_ptt_syst}
\renewcommand*{\arraystretch}{1.60}
\cmsTable{
\begin{tabular}{l|rrrrrrrrrrrrrrrr}
            syst. source / bin&$                 1$&$                 2$&$                 3$&$                 4$&$                 5$&$                 6$&$                 7$&$                 8$&$                 9$&$                10$&$                11$&$                12$&$                13$&$                14$&$                15$&$                16$\\
\hline
              Jet energy scale&${}^{ +2.7}_{ -0.5}$&${}^{ -1.9}_{ -1.0}$&${}^{ -0.1}_{ -0.5}$&${}^{ +2.6}_{ -1.4}$&${}^{ +0.8}_{ -1.5}$&${}^{ -1.4}_{ +0.3}$&${}^{ +2.3}_{ +0.8}$&${}^{ +0.3}_{ -3.8}$&${}^{ -1.8}_{ -1.0}$&${}^{ -1.9}_{ +2.6}$&${}^{ -2.7}_{ -1.0}$&${}^{ +3.8}_{ -1.8}$&${}^{ +2.1}_{ +3.0}$&${}^{ +1.7}_{ +1.2}$&${}^{ +0.6}_{ -1.1}$&${}^{ +1.5}_{ -5.0}$\\
         Jet energy resolution&${}^{ +2.1}_{ +0.4}$&${}^{ -0.6}_{ -0.8}$&${}^{ -1.3}_{ -0.5}$&${}^{ +0.0}_{ -0.0}$&${}^{ -0.8}_{ -0.2}$&${}^{ -1.4}_{ -0.6}$&${}^{ +1.7}_{ +1.5}$&${}^{ -3.0}_{ +0.4}$&${}^{ -1.4}_{ -1.4}$&${}^{ +0.2}_{ +1.2}$&${}^{ +0.0}_{ -1.7}$&${}^{ -0.1}_{ +3.5}$&${}^{ +2.7}_{ +1.0}$&${}^{ +1.8}_{ +1.2}$&${}^{ -2.9}_{ -0.5}$&${}^{ +1.9}_{ -0.4}$\\
           Kin. reconstruction&${}^{ +0.0}_{ -0.0}$&${}^{ +0.0}_{ -0.0}$&${}^{ -0.0}_{ +0.0}$&${}^{ -0.0}_{ +0.0}$&${}^{ -0.0}_{ +0.0}$&${}^{ -0.0}_{ +0.0}$&${}^{ -0.0}_{ +0.0}$&${}^{ -0.0}_{ +0.0}$&${}^{ +0.0}_{ +0.0}$&${}^{ -0.0}_{ +0.0}$&${}^{ +0.0}_{ -0.0}$&${}^{ -0.0}_{ +0.0}$&${}^{ +0.0}_{ -0.0}$&${}^{ +0.0}_{ -0.0}$&${}^{ -0.0}_{ +0.0}$&${}^{ +0.0}_{ -0.0}$\\
                        Pileup&${}^{ +0.0}_{ +0.0}$&${}^{ +0.1}_{ -0.2}$&${}^{ -0.2}_{ +0.3}$&${}^{ -0.3}_{ +0.1}$&${}^{ -0.1}_{ +0.1}$&${}^{ +0.0}_{ -0.1}$&${}^{ -0.4}_{ +0.5}$&${}^{ -0.4}_{ +0.4}$&${}^{ +0.1}_{ -0.1}$&${}^{ +0.4}_{ -0.5}$&${}^{ -0.4}_{ +0.5}$&${}^{ -0.5}_{ +0.5}$&${}^{ +0.2}_{ -0.1}$&${}^{ +0.0}_{ -0.0}$&${}^{ +0.2}_{ -0.1}$&${}^{ -0.8}_{ +0.9}$\\
\end{tabular}
}
\end{table*}

\begin{table*}
\centering
\topcaption{The measured normalized \ttbar double-differential cross sections in different bins of \mtt and \yt, along with their relative statistical and systematic uncertainties.}
\label{tab:mtt_yt_xsec}
\renewcommand*{\arraystretch}{1.40}
\begin{tabular}{*{6}{c}}
$\mtt$ [\GeVns{}] & $\abs{\yt}$  & $\frac{1}{\sigma(\ttbar)} \frac{\rd ^{2}\sigma(\ttbar)}{\rd \mtt\rd \yt}$ [\GeVns{}$^{-1}$] & stat. [\%] & syst. [\%] & bin \\
\hline
\multirow{4}{*}{$    340$--$    400$} & $      0$--$   0.35$ & $3.21 \times 10^{-3}$ & $ 4.9$ & ${}_{ -9.4}^{ +9.8}$ & $ 1$\\
                                      & $   0.35$--$   0.85$ & $2.92 \times 10^{-3}$ & $ 4.0$ & ${}_{ -5.7}^{ +3.9}$ & $ 2$\\
                                      & $   0.85$--$   1.45$ & $2.06 \times 10^{-3}$ & $ 4.5$ & ${}_{ -3.8}^{ +3.5}$ & $ 3$\\
                                      & $   1.45$--$    2.5$ & $6.58 \times 10^{-4}$ & $ 9.3$ & ${}_{ -5.8}^{ +7.4}$ & $ 4$\\
\hline
\multirow{4}{*}{$    400$--$    500$} & $      0$--$   0.35$ & $2.92 \times 10^{-3}$ & $ 3.2$ & ${}_{ -8.3}^{ +7.4}$ & $ 5$\\
                                      & $   0.35$--$   0.85$ & $2.39 \times 10^{-3}$ & $ 2.8$ & ${}_{ -2.1}^{ +2.9}$ & $ 6$\\
                                      & $   0.85$--$   1.45$ & $1.67 \times 10^{-3}$ & $ 3.2$ & ${}_{ -4.2}^{ +3.4}$ & $ 7$\\
\end{tabular}

\end{table*}
 \begin{table*}
\centering
\topcaption{The correlation matrix of statistical uncertainties for the normalized \ttbar double-differential cross sections as a function of \mtt and \yt. The values are expressed as percentages. For bin indices see Table~\ref{tab:mtt_yt_xsec}.}
\label{tab:mtt_yt_corr}
\renewcommand*{\arraystretch}{1.40}
\tabcolsep0.40mm
\cmsTable{
\begin{tabular}{c|rrrrrrrrrrrrrrrr}
\quad bin\quad &$     1$&$     2$&$     3$&$     4$&$     5$&$     6$&$     7$&$     8$&$     9$&$    10$&$    11$&$    12$&$    13$&$    14$&$    15$&$    16$ \\
\hline
$ 1$&$+100.0$&$ -29.7$&$ -31.0$&$ +11.4$&$ -21.1$&$ -26.6$&$  +0.9$&$  +6.5$&$ -16.8$&$  +6.5$&$  +6.6$&$  -7.0$&$  +5.8$&$  +0.2$&$  -3.8$&$  -2.3$ \\
$ 2$&        &$+100.0$&$ +19.8$&$ -35.7$&$ -18.0$&$  -6.1$&$ -23.7$&$  -3.6$&$  +9.9$&$ -20.9$&$  -6.3$&$ +10.0$&$  -1.1$&$  +5.5$&$  +6.8$&$  -4.7$ \\
$ 3$&        &        &$+100.0$&$ -19.4$&$  +2.7$&$ -16.6$&$ -14.8$&$ -28.3$&$  +4.6$&$  -4.2$&$ -19.4$&$  +1.3$&$  -3.1$&$  +1.3$&$  +5.7$&$  +3.1$ \\
$ 4$&        &        &        &$+100.0$&$  -2.2$&$  -1.5$&$ -13.8$&$ -19.8$&$  -9.4$&$  +3.0$&$  +3.9$&$ -32.5$&$  -1.1$&$  -6.0$&$  -6.5$&$  +7.8$ \\
$ 5$&        &        &        &        &$+100.0$&$ -10.2$&$ -24.0$&$  -0.2$&$ -10.0$&$ -23.0$&$  +0.6$&$  +4.6$&$ -10.1$&$  +6.7$&$  +3.4$&$  -5.4$ \\
$ 6$&        &        &        &        &        &$+100.0$&$ +21.6$&$ -25.3$&$ -18.7$&$  +8.8$&$ -21.5$&$  -5.3$&$  +8.0$&$ -16.0$&$  -0.7$&$  +3.7$ \\
$ 7$&        &        &        &        &        &        &$+100.0$&$  -5.9$&$  +1.2$&$  -9.8$&$  +1.5$&$ -24.1$&$  +1.4$&$  +0.8$&$ -15.4$&$  +3.3$ \\
\end{tabular}
}
\end{table*}
 \begin{table*}
\centering
\topcaption{Sources and values of the relative systematic uncertainties in percent of the measured normalized \ttbar double-differential cross sections as a function of \mtt and \yt.
For bin indices see Table~\ref{tab:mtt_yt_xsec}.}
\label{tab:mtt_yt_syst}
\renewcommand*{\arraystretch}{1.80}
\tabcolsep0.87mm
\cmsTable{
\begin{tabular}{l|rrrrrrrrrrrrrrrr}
            syst. source/bin&$                 1$&$                 2$&$                 3$&$                 4$&$                 5$&$                 6$&$                 7$&$                 8$&$                 9$&$                10$&$                11$&$                12$&$                13$&$                14$&$                15$&$                16$\\
\hline
              Jet energy scale&${}^{ +2.9}_{ -2.9}$&${}^{ +0.3}_{ -1.4}$&${}^{ +1.8}_{ +0.5}$&${}^{ +2.4}_{ -0.4}$&${}^{ -0.8}_{ +1.1}$&${}^{ -1.1}_{ +1.5}$&${}^{ -3.5}_{ +1.3}$&${}^{ +0.1}_{ +2.0}$&${}^{ -0.6}_{ -0.5}$&${}^{ +1.7}_{ +0.5}$&${}^{ -0.7}_{ +0.6}$&${}^{ +1.6}_{ +1.2}$&${}^{ +4.6}_{ -2.1}$&${}^{ -0.5}_{ -2.6}$&${}^{ -0.7}_{ -1.1}$&${}^{ -1.7}_{ -5.0}$\\
         Jet energy resolution&${}^{ +0.7}_{ +0.9}$&${}^{ -0.4}_{ -0.7}$&${}^{ -0.3}_{ -0.9}$&${}^{ +3.2}_{ +1.4}$&${}^{ -1.4}_{ -0.4}$&${}^{ +0.8}_{ +0.1}$&${}^{ -1.3}_{ -0.9}$&${}^{ +0.6}_{ +1.1}$&${}^{ -0.3}_{ -0.6}$&${}^{ +1.7}_{ +0.6}$&${}^{ +1.1}_{ +1.6}$&${}^{ +1.0}_{ -0.8}$&${}^{ +1.0}_{ +0.7}$&${}^{ -4.4}_{ +0.1}$&${}^{ +1.2}_{ -0.0}$&${}^{ -3.3}_{ -0.3}$\\
           Kin. reconstruction&${}^{ +0.0}_{ -0.0}$&${}^{ -0.0}_{ +0.0}$&${}^{ -0.0}_{ +0.0}$&${}^{ +0.0}_{ -0.0}$&${}^{ +0.0}_{ -0.0}$&${}^{ -0.0}_{ +0.0}$&${}^{ +0.0}_{ -0.0}$&${}^{ +0.0}_{ -0.0}$&${}^{ -0.0}_{ +0.0}$&${}^{ -0.0}_{ +0.0}$&${}^{ -0.0}_{ +0.0}$&${}^{ -0.0}_{ +0.0}$&${}^{ +0.0}_{ -0.0}$&${}^{ +0.0}_{ -0.0}$&${}^{ +0.0}_{ -0.0}$&${}^{ +0.0}_{ -0.0}$\\
                        Pileup&${}^{ +0.4}_{ -0.6}$&${}^{ -0.3}_{ +0.2}$&${}^{ +0.1}_{ -0.1}$&${}^{ +0.6}_{ -0.7}$&${}^{ +0.0}_{ -0.1}$&${}^{ -0.1}_{ +0.1}$&${}^{ +0.2}_{ -0.1}$&${}^{ -0.3}_{ +0.6}$&${}^{ -0.3}_{ +0.3}$&${}^{ -0.4}_{ +0.6}$&${}^{ +0.2}_{ -0.2}$&${}^{ +0.5}_{ -0.7}$&${}^{ -0.0}_{ -0.0}$&${}^{ +0.2}_{ -0.3}$&${}^{ -0.4}_{ +0.3}$&${}^{ -0.4}_{ +0.5}$\\
\end{tabular}
}
\end{table*}

\begin{table*}
\centering
\topcaption{The measured normalized \ttbar double-differential cross sections in different bins of \mtt and \ytt, along with their relative statistical and systematic uncertainties.}
\label{tab:mtt_ytt_xsec}
\renewcommand*{\arraystretch}{1.40}
\begin{tabular}{*{6}{c}}
$\mtt$ [\GeVns{}] & $\abs{\ytt}$  & $\frac{1}{\sigma(\ttbar)} \frac{\rd ^{2}\sigma(\ttbar)}{\rd \mtt\rd \ytt}$ [\GeVns{}$^{-1}$] & stat. [\%] & syst. [\%] & bin \\
\hline
\multirow{4}{*}{$    340$--$    400$} & $      0$--$   0.35$ & $3.17 \times 10^{-3}$ & $ 4.5$ & ${}_{ -5.9}^{ +6.6}$ & $ 1$\\
                                      & $   0.35$--$   0.75$ & $3.07 \times 10^{-3}$ & $ 4.4$ & ${}_{ -5.0}^{ +3.6}$ & $ 2$\\
                                      & $   0.75$--$   1.15$ & $2.44 \times 10^{-3}$ & $ 5.1$ & ${}_{ -5.1}^{ +6.1}$ & $ 3$\\
                                      & $   1.15$--$    2.5$ & $9.14 \times 10^{-4}$ & $ 4.9$ & ${}_{ -3.1}^{ +3.9}$ & $ 4$\\
\hline
\multirow{4}{*}{$    400$--$    500$} & $      0$--$   0.35$ & $3.06 \times 10^{-3}$ & $ 2.8$ & ${}_{ -5.5}^{ +6.4}$ & $ 5$\\
                                      & $   0.35$--$   0.75$ & $2.76 \times 10^{-3}$ & $ 2.8$ & ${}_{ -5.4}^{ +4.7}$ & $ 6$\\
                                      & $   0.75$--$   1.15$ & $2.05 \times 10^{-3}$ & $ 3.6$ & ${}_{ -3.9}^{ +3.2}$ & $ 7$\\
\end{tabular}

\end{table*}
 \begin{table*}
\centering
\topcaption{The correlation matrix of statistical uncertainties for the normalized \ttbar double-differential cross sections as a function of \mtt and \ytt. The values are expressed as percentages. For bin indices see Table~\ref{tab:mtt_ytt_xsec}.}
\label{tab:mtt_ytt_corr}
\renewcommand*{\arraystretch}{1.40}
\tabcolsep0.40mm
\cmsTable{
\begin{tabular}{c|rrrrrrrrrrrrrrrr}
\quad bin\quad &$     1$&$     2$&$     3$&$     4$&$     5$&$     6$&$     7$&$     8$&$     9$&$    10$&$    11$&$    12$&$    13$&$    14$&$    15$&$    16$ \\
\hline
$ 1$&$+100.0$&$ -23.2$&$ -23.1$&$  +5.4$&$ -31.7$&$ -19.6$&$  +8.4$&$  +0.6$&$ -21.0$&$  +5.6$&$  +4.7$&$  -7.2$&$  +8.3$&$  -0.8$&$  -4.7$&$  +0.4$ \\
$ 2$&        &$+100.0$&$  +7.6$&$ -29.8$&$ -17.6$&$ -11.5$&$ -24.6$&$  +3.5$&$  +6.5$&$ -22.2$&$  -2.5$&$  +9.2$&$  -0.5$&$  +6.2$&$  +3.7$&$  -5.8$ \\
$ 3$&        &        &$+100.0$&$ -12.6$&$  +7.8$&$ -24.0$&$ -19.9$&$ -22.4$&$  +4.2$&$  -2.5$&$ -19.2$&$  +1.2$&$  -4.3$&$  +2.5$&$  +6.6$&$  +1.2$ \\
$ 4$&        &        &        &$+100.0$&$  -2.1$&$  +2.1$&$ -17.2$&$ -24.4$&$  -8.3$&$  +4.9$&$  +2.2$&$ -38.7$&$  -2.6$&$  -7.1$&$  -2.5$&$ +13.1$ \\
$ 5$&        &        &        &        &$+100.0$&$  -5.2$&$ -18.6$&$  -4.1$&$  -6.5$&$ -18.7$&$  +1.2$&$  +1.3$&$ -14.1$&$  +3.9$&$  +0.8$&$  -3.4$ \\
$ 6$&        &        &        &        &        &$+100.0$&$ +17.6$&$ -24.7$&$ -18.9$&$  +8.9$&$ -17.7$&$  -5.0$&$  +5.3$&$ -17.6$&$  +1.5$&$  +3.5$ \\
$ 7$&        &        &        &        &        &        &$+100.0$&$  -2.6$&$  +0.3$&$ -15.0$&$  +0.3$&$ -18.4$&$  +1.2$&$  +0.6$&$ -13.5$&$  +4.4$ \\
\end{tabular}
}
\end{table*}
 \begin{table*}
\centering
\topcaption{Sources and values of the relative systematic uncertainties in percent of the measured normalized \ttbar double-differential cross sections as a function of \mtt and \ytt.
For bin indices see Table~\ref{tab:mtt_ytt_xsec}.}
\label{tab:mtt_ytt_syst}
\renewcommand*{\arraystretch}{1.80}
\tabcolsep0.87mm
\cmsTable{
\begin{tabular}{l|rrrrrrrrrrrrrrrr}
            syst. source / bin&$                 1$&$                 2$&$                 3$&$                 4$&$                 5$&$                 6$&$                 7$&$                 8$&$                 9$&$                10$&$                11$&$                12$&$                13$&$                14$&$                15$&$                16$\\
\hline
              Jet energy scale&${}^{ +1.1}_{ -3.1}$&${}^{ +2.1}_{ -2.5}$&${}^{ +1.7}_{ -1.1}$&${}^{ +1.1}_{ +1.1}$&${}^{ +0.1}_{ +3.0}$&${}^{ -2.2}_{ +1.1}$&${}^{ -1.7}_{ +2.3}$&${}^{ -1.1}_{ +0.2}$&${}^{ -0.5}_{ -0.1}$&${}^{ -0.4}_{ +0.3}$&${}^{ -1.5}_{ +0.3}$&${}^{ +2.1}_{ -0.8}$&${}^{ -0.4}_{ -2.8}$&${}^{ +1.3}_{ -0.6}$&${}^{ +3.2}_{ -2.8}$&${}^{ +0.5}_{ -4.9}$\\
         Jet energy resolution&${}^{ -0.8}_{ +0.1}$&${}^{ +1.0}_{ +0.7}$&${}^{ -0.3}_{ -0.8}$&${}^{ +0.3}_{ +0.5}$&${}^{ +1.0}_{ -0.2}$&${}^{ -0.7}_{ -0.3}$&${}^{ -0.2}_{ -0.3}$&${}^{ +0.3}_{ +0.2}$&${}^{ -0.6}_{ -0.1}$&${}^{ +0.1}_{ -0.5}$&${}^{ -0.3}_{ -0.1}$&${}^{ +2.8}_{ +0.5}$&${}^{ -0.4}_{ +0.5}$&${}^{ -0.9}_{ +1.1}$&${}^{ -1.1}_{ -1.3}$&${}^{ -3.6}_{ +0.1}$\\
           Kin. reconstruction&${}^{ -0.0}_{ +0.0}$&${}^{ +0.0}_{ -0.0}$&${}^{ -0.0}_{ +0.0}$&${}^{ +0.0}_{ -0.0}$&${}^{ +0.0}_{ -0.0}$&${}^{ +0.0}_{ -0.0}$&${}^{ -0.0}_{ +0.0}$&${}^{ +0.0}_{ -0.0}$&${}^{ -0.0}_{ +0.0}$&${}^{ -0.0}_{ +0.0}$&${}^{ -0.0}_{ +0.0}$&${}^{ -0.0}_{ +0.0}$&${}^{ +0.0}_{ -0.0}$&${}^{ +0.0}_{ -0.0}$&${}^{ +0.0}_{ -0.0}$&${}^{ +0.0}_{ -0.0}$\\
                        Pileup&${}^{ +0.1}_{ -0.0}$&${}^{ +0.0}_{ -0.2}$&${}^{ +0.1}_{ -0.2}$&${}^{ +0.4}_{ -0.7}$&${}^{ +0.1}_{ -0.1}$&${}^{ -0.1}_{ +0.1}$&${}^{ +0.1}_{ -0.0}$&${}^{ -0.1}_{ +0.3}$&${}^{ -0.1}_{ +0.0}$&${}^{ -0.2}_{ +0.2}$&${}^{ +0.1}_{ -0.1}$&${}^{ -0.1}_{ +0.5}$&${}^{ -0.2}_{ -0.0}$&${}^{ -0.5}_{ +0.7}$&${}^{ +0.5}_{ -0.7}$&${}^{ -0.8}_{ +0.9}$\\
\end{tabular}
}
\end{table*}

\begin{table*}
\centering
\topcaption{The measured normalized \ttbar double-differential cross sections in different bins of \mtt and \detatt, along with their relative statistical and systematic uncertainties.}
\label{tab:mtt_detatt_xsec}
\renewcommand*{\arraystretch}{1.40}
\begin{tabular}{*{6}{c}}
$\mtt$ [\GeVns{}] & $\detatt$  & $\frac{1}{\sigma(\ttbar)} \frac{\rd ^{2}\sigma(\ttbar)}{\rd \mtt\rd \detatt}$ [\GeVns{}$^{-1}$] & stat. [\%] & syst. [\%] & bin \\
\hline
\multirow{3}{*}{$    340$--$    400$} & $      0$--$    0.4$ & $3.35 \times 10^{-3}$ & $ 7.3$ & ${}_{ -7.5}^{ +9.7}$ & $ 1$\\
                                      & $    0.4$--$    1.2$ & $2.53 \times 10^{-3}$ & $ 3.2$ & ${}_{ -6.1}^{ +5.7}$ & $ 2$\\
                                      & $    1.2$--$      6$ & $2.31 \times 10^{-4}$ & $ 8.4$ & ${}_{ -9.5}^{+10.4}$ & $ 3$\\
\hline
\multirow{3}{*}{$    400$--$    500$} & $      0$--$    0.4$ & $1.60 \times 10^{-3}$ & $ 6.2$ & ${}_{ -5.9}^{ +4.8}$ & $ 4$\\
                                      & $    0.4$--$    1.2$ & $1.69 \times 10^{-3}$ & $ 3.3$ & ${}_{ -3.0}^{ +3.9}$ & $ 5$\\
                                      & $    1.2$--$      6$ & $3.87 \times 10^{-4}$ & $ 2.3$ & ${}_{ -4.1}^{ +4.0}$ & $ 6$\\
\hline
\multirow{3}{*}{$    500$--$    650$} & $      0$--$    0.4$ & $3.56 \times 10^{-4}$ & 10.7 & ${}_{ -9.0}^{ +8.3}$ & $ 7$\\
                                      & $    0.4$--$    1.2$ & $3.55 \times 10^{-4}$ & $ 7.7$ & ${}_{ -9.8}^{+15.0}$ & $ 8$\\
                                      & $    1.2$--$      6$ & $2.29 \times 10^{-4}$ & $ 2.4$ & ${}_{ -6.2}^{ +2.9}$ & $ 9$\\
\hline
\multirow{3}{*}{$    650$--$   1500$} & $      0$--$    0.4$ & $2.08 \times 10^{-5}$ & 13.3 & ${}_{-11.4}^{ +8.8}$ & 10\\
                                      & $    0.4$--$    1.2$ & $2.42 \times 10^{-5}$ & $ 8.6$ & ${}_{-12.9}^{ +9.1}$ & 11\\
                                      & $    1.2$--$      6$ & $2.31 \times 10^{-5}$ & $ 3.1$ & ${}_{ -5.9}^{ +7.4}$ & 12\\
\end{tabular}
\end{table*}

 \begin{table*}
\centering
\topcaption{The correlation matrix of statistical uncertainties for the normalized \ttbar double-differential cross sections as a function of \mtt and \detatt. The values are expressed as percentages. For bin indices see Table~\ref{tab:mtt_detatt_xsec}.}
\label{tab:mtt_detatt_corr}
\renewcommand*{\arraystretch}{1.40}
\tabcolsep0.40mm
\cmsTable{
\begin{tabular}{c|rrrrrrrrrrrr}
\quad bin\quad &$     1$&$     2$&$     3$&$     4$&$     5$&$     6$&$     7$&$     8$&$     9$&$    10$&$    11$&$    12$ \\
\hline
$ 1$&$+100.0$&$  +1.8$&$ -71.5$&$  +2.0$&$ -36.9$&$ -15.9$&$ -13.0$&$  +3.4$&$ +20.0$&$  +3.5$&$  -0.3$&$ -11.8$ \\
$ 2$&        &$+100.0$&$ -13.1$&$ -34.2$&$  +6.4$&$ -33.6$&$ +15.6$&$ -19.6$&$ -13.1$&$  -5.3$&$  +3.5$&$  +1.6$ \\
$ 3$&        &        &$+100.0$&$  +4.3$&$  -2.1$&$  -7.7$&$  +5.8$&$  +6.9$&$ -33.8$&$  -2.7$&$  -1.0$&$ +13.4$ \\
$ 4$&        &        &        &$+100.0$&$ -26.3$&$ -25.5$&$ -16.1$&$ -24.2$&$  +4.1$&$  -0.3$&$ +10.9$&$  +2.0$ \\
$ 5$&        &        &        &        &$+100.0$&$  +5.4$&$ -16.5$&$  +6.3$&$ -33.9$&$  +8.2$&$ -11.2$&$  +1.4$ \\
$ 6$&        &        &        &        &        &$+100.0$&$ +10.5$&$ -11.5$&$  +3.4$&$  -3.0$&$  +5.6$&$ -28.9$ \\
$ 7$&        &        &        &        &        &        &$+100.0$&$ -37.6$&$  -3.1$&$ -42.6$&$  +6.0$&$  +4.2$ \\
$ 8$&        &        &        &        &        &        &        &$+100.0$&$ -10.3$&$  +9.6$&$ -35.8$&$  -7.7$ \\
$ 9$&        &        &        &        &        &        &        &        &$+100.0$&$  +2.2$&$  +0.6$&$ -40.7$ \\
10&        &        &        &        &        &        &        &        &        &$+100.0$&$ -31.1$&$  +0.4$ \\
11&        &        &        &        &        &        &        &        &        &        &$+100.0$&$ -12.2$ \\
12&        &        &        &        &        &        &        &        &        &        &        &$+100.0$ \\
\end{tabular}
}
\end{table*}
 \begin{table*}
\centering
\topcaption{Sources and values of the relative systematic uncertainties in percent of the measured normalized \ttbar double-differential cross sections as a function of \mtt and \detatt.
For bin indices see Table~\ref{tab:mtt_detatt_xsec}.}
\label{tab:mtt_detatt_syst}
\renewcommand*{\arraystretch}{1.80}
\cmsTable{
\begin{tabular}{l|rrrrrrrrrrrr}
            syst. source / bin&$                 1$&$                 2$&$                 3$&$                 4$&$                 5$&$                 6$&$                 7$&$                 8$&$                 9$&$                10$&$                11$&$                12$\\
\hline
              Jet energy scale&${}^{ +3.6}_{ -0.1}$&${}^{ +2.4}_{ -2.4}$&${}^{ -0.1}_{ -1.5}$&${}^{ -2.9}_{ +0.6}$&${}^{ -1.6}_{ +1.0}$&${}^{ -1.3}_{ +2.0}$&${}^{ +3.0}_{ +0.3}$&${}^{ +3.2}_{ -0.2}$&${}^{ -1.1}_{ +0.1}$&${}^{ +1.8}_{ -2.1}$&${}^{ +2.5}_{ -4.9}$&${}^{ -0.0}_{ -0.6}$\\
         Jet energy resolution&${}^{ -1.3}_{ +0.8}$&${}^{ +0.2}_{ -0.7}$&${}^{ +1.1}_{ -0.0}$&${}^{ -1.4}_{ -1.2}$&${}^{ +1.6}_{ +0.2}$&${}^{ -0.4}_{ -0.0}$&${}^{ +1.2}_{ +2.2}$&${}^{ +0.5}_{ -1.0}$&${}^{ -0.4}_{ -0.6}$&${}^{ -1.1}_{ -1.4}$&${}^{ -2.4}_{ +1.3}$&${}^{ +0.4}_{ +1.7}$\\
           Kin. reconstruction&${}^{ +0.0}_{ -0.0}$&${}^{ -0.0}_{ +0.0}$&${}^{ -0.0}_{ +0.0}$&${}^{ +0.0}_{ -0.0}$&${}^{ -0.0}_{ +0.0}$&${}^{ +0.0}_{ -0.0}$&${}^{ +0.0}_{ -0.0}$&${}^{ -0.1}_{ +0.1}$&${}^{ +0.0}_{ -0.0}$&${}^{ -0.0}_{ +0.0}$&${}^{ +0.0}_{ -0.0}$&${}^{ +0.0}_{ -0.0}$\\
                        Pileup&${}^{ +0.4}_{ -0.3}$&${}^{ -0.0}_{ -0.1}$&${}^{ +0.1}_{ -0.2}$&${}^{ +0.1}_{ -0.2}$&${}^{ -0.2}_{ +0.4}$&${}^{ +0.2}_{ -0.2}$&${}^{ -0.3}_{ +0.1}$&${}^{ -0.5}_{ +0.8}$&${}^{ +0.2}_{ -0.3}$&${}^{ -0.2}_{ +0.1}$&${}^{ +0.2}_{ -0.2}$&${}^{ -0.6}_{ +0.6}$\\
                       Trigger&${}^{ -0.0}_{ +0.0}$&${}^{ -0.1}_{ +0.1}$&${}^{ +0.1}_{ -0.1}$&${}^{ +0.2}_{ -0.2}$&${}^{ -0.1}_{ +0.1}$&${}^{ +0.1}_{ -0.1}$&${}^{ +0.3}_{ -0.3}$&${}^{ -0.2}_{ +0.2}$&${}^{ -0.0}_{ +0.0}$&${}^{ +0.0}_{ -0.0}$&${}^{ +0.3}_{ -0.3}$&${}^{ -0.1}_{ +0.1}$\\
\end{tabular}
}
\end{table*}

\begin{table*}
\centering
\topcaption{The measured normalized \ttbar double-differential cross sections in different bins of \mtt and \pttt, along with their relative statistical and systematic uncertainties.}
\label{tab:mtt_pttt_xsec}
\renewcommand*{\arraystretch}{1.40}
\tabcolsep2.0mm
\begin{tabular}{*{6}{c}}
$\mtt$ [\GeVns{}] & $\pttt$ [\GeVns{}] & $\frac{1}{\sigma(\ttbar)} \frac{\rd ^{2}\sigma(\ttbar)}{\rd \mtt\rd \pttt}$ [\GeVns{}$^{-2}$] & stat. [\%] & syst. [\%] & bin \\
\hline
\multirow{4}{*}{$    340$--$    400$} & $      0$--$     30$ & $6.63 \times 10^{-5}$ & $ 3.0$ & ${}_{ -4.0}^{ +6.7}$ & $ 1$\\
                                      & $     30$--$     75$ & $3.51 \times 10^{-5}$ & $ 3.5$ & ${}_{ -7.5}^{ +5.9}$ & $ 2$\\
                                      & $     75$--$    150$ & $9.61 \times 10^{-6}$ & $ 6.6$ & ${}_{-10.8}^{+13.1}$ & $ 3$\\
                                      & $    150$--$    500$ & $6.19 \times 10^{-7}$ & 14.8 & ${}_{-17.2}^{ +8.6}$ & $ 4$\\
\hline
\multirow{4}{*}{$    400$--$    500$} & $      0$--$     30$ & $5.57 \times 10^{-5}$ & $ 2.0$ & ${}_{ -6.0}^{ +6.7}$ & $ 5$\\
                                      & $     30$--$     75$ & $2.82 \times 10^{-5}$ & $ 2.5$ & ${}_{ -5.5}^{ +5.4}$ & $ 6$\\
                                      & $     75$--$    150$ & $8.34 \times 10^{-6}$ & $ 4.7$ & ${}_{-10.9}^{+14.3}$ & $ 7$\\
\end{tabular}

\end{table*}
 \begin{table*}
\centering
\topcaption{The correlation matrix of statistical uncertainties for the normalized \ttbar double-differential cross sections as a function of \mtt and \pttt. The values are expressed as percentages. For bin indices see Table~\ref{tab:mtt_pttt_xsec}.}
\label{tab:mtt_pttt_corr}

\renewcommand*{\arraystretch}{1.40}
\tabcolsep0.40mm
\cmsTable{
\begin{tabular}{c|rrrrrrrrrrrrrrrr}
\quad bin\quad &$     1$&$     2$&$     3$&$     4$&$     5$&$     6$&$     7$&$     8$&$     9$&$    10$&$    11$&$    12$&$    13$&$    14$&$    15$&$    16$ \\
\hline
$ 1$&$+100.0$&$ -20.4$&$ -22.7$&$  +1.5$&$ -25.9$&$ -21.0$&$  +8.9$&$  -1.5$&$ -37.8$&$  +5.1$&$  +6.6$&$  -4.3$&$  +6.9$&$  -1.4$&$  -7.9$&$  -0.2$ \\
$ 2$&        &$+100.0$&$  +5.3$&$ -10.6$&$ -26.3$&$ -17.3$&$ -21.5$&$  +7.5$&$  +4.0$&$ -30.1$&$  +0.9$&$  -0.1$&$  +0.6$&$  +4.5$&$  +2.2$&$  -3.1$ \\
$ 3$&        &        &$+100.0$&$  -8.5$&$  +6.4$&$ -32.2$&$ -36.0$&$  -3.0$&$  +9.5$&$  -6.0$&$ -17.9$&$  +5.8$&$  -5.7$&$  +5.4$&$  +8.6$&$  -3.4$ \\
$ 4$&        &        &        &$+100.0$&$  -0.2$&$  +8.2$&$  -9.0$&$ -59.8$&$  -4.0$&$  +1.2$&$  +3.5$&$  +7.5$&$  -0.1$&$  -2.2$&$  -1.1$&$  +1.6$ \\
$ 5$&        &        &        &        &$+100.0$&$  -7.6$&$ -22.5$&$  -1.6$&$ +13.7$&$ -24.7$&$  -0.8$&$  +1.3$&$ -26.9$&$  +7.1$&$  +2.0$&$  -4.1$ \\
$ 6$&        &        &        &        &        &$+100.0$&$ +18.0$&$ -13.9$&$ -28.2$&$ +19.3$&$ -16.3$&$  +5.1$&$  +8.8$&$ -24.1$&$  +3.0$&$  -0.2$ \\
$ 7$&        &        &        &        &        &        &$+100.0$&$  +1.2$&$  -5.8$&$ -17.1$&$ -13.4$&$  -6.5$&$  +4.9$&$  +0.4$&$ -10.5$&$  +4.7$ \\
\end{tabular}
}
\end{table*}
 \begin{table*}
\centering
\topcaption{Sources and values of the relative systematic uncertainties in percent of the measured normalized \ttbar double-differential cross sections as a function of \mtt and \pttt.
For bin indices see Table~\ref{tab:mtt_pttt_xsec}.}
\label{tab:mtt_pttt_syst}
\renewcommand*{\arraystretch}{1.60}
\cmsTable{
\begin{tabular}{l|rrrrrrrrrrrrrrrr}
            syst. source / bin&$                 1$&$                 2$&$                 3$&$                 4$&$                 5$&$                 6$&$                 7$&$                 8$&$                 9$&$                10$&$                11$&$                12$&$                13$&$                14$&$                15$&$                16$\\
\hline
              Jet energy scale&${}^{ +4.3}_{ -1.7}$&${}^{ -0.3}_{ -1.3}$&${}^{ -0.1}_{ +0.3}$&${}^{ +6.4}_{ +0.9}$&${}^{ -0.6}_{ -0.4}$&${}^{ -2.3}_{ +2.2}$&${}^{ -2.5}_{ +3.1}$&${}^{ -4.5}_{ +0.4}$&${}^{ -0.3}_{ -0.8}$&${}^{ +1.5}_{ +0.9}$&${}^{ -1.5}_{ +3.4}$&${}^{ -0.5}_{ +3.5}$&${}^{ +2.4}_{ -1.8}$&${}^{ +0.4}_{ -4.5}$&${}^{ -1.2}_{ -0.6}$&${}^{ +0.5}_{ -0.7}$\\
         Jet energy resolution&${}^{ +1.1}_{ +0.1}$&${}^{ -0.7}_{ -0.7}$&${}^{ -1.0}_{ +0.5}$&${}^{ +2.0}_{ -0.2}$&${}^{ -0.3}_{ -1.7}$&${}^{ -0.2}_{ +1.0}$&${}^{ +1.6}_{ +1.9}$&${}^{ -1.7}_{ -2.2}$&${}^{ -0.2}_{ -1.6}$&${}^{ +0.4}_{ +4.0}$&${}^{ +0.1}_{ +0.4}$&${}^{ +2.5}_{ +2.9}$&${}^{ +0.1}_{ +0.4}$&${}^{ -1.3}_{ -1.5}$&${}^{ -0.2}_{ -1.4}$&${}^{ -1.4}_{ +1.9}$\\
           Kin. reconstruction&${}^{ -0.0}_{ +0.0}$&${}^{ +0.0}_{ -0.0}$&${}^{ -0.0}_{ +0.0}$&${}^{ -0.0}_{ +0.0}$&${}^{ +0.0}_{ -0.0}$&${}^{ -0.0}_{ +0.0}$&${}^{ +0.0}_{ -0.0}$&${}^{ +0.0}_{ -0.0}$&${}^{ -0.0}_{ +0.0}$&${}^{ -0.0}_{ +0.0}$&${}^{ -0.0}_{ +0.0}$&${}^{ -0.0}_{ +0.0}$&${}^{ +0.0}_{ -0.0}$&${}^{ +0.0}_{ -0.0}$&${}^{ +0.0}_{ -0.0}$&${}^{ -0.0}_{ +0.0}$\\
                        Pileup&${}^{ +0.5}_{ -0.6}$&${}^{ -0.5}_{ +0.5}$&${}^{ +0.5}_{ -0.4}$&${}^{ -0.8}_{ -0.4}$&${}^{ +0.4}_{ -0.3}$&${}^{ -0.3}_{ +0.4}$&${}^{ +0.0}_{ +0.1}$&${}^{ -0.5}_{ +0.7}$&${}^{ +0.5}_{ -0.4}$&${}^{ -0.6}_{ +0.5}$&${}^{ -0.6}_{ +0.6}$&${}^{ -0.7}_{ +0.9}$&${}^{ +0.8}_{ -0.7}$&${}^{ -0.9}_{ +0.6}$&${}^{ +0.1}_{ -0.4}$&${}^{ -0.7}_{ +0.7}$\\
\end{tabular}
}
\end{table*}

\begin{table*}
\centering
\topcaption{The measured normalized \ttbar double-differential cross sections in different bins of \mtt and \dphitt, along with their relative statistical and systematic uncertainties.}
\label{tab:mtt_dphitt_xsec}

\renewcommand*{\arraystretch}{1.40}
\begin{tabular}{*{6}{c}}
$\mtt$ [\GeVns{}] & $\dphitt$ [rad] & $\frac{1}{\sigma(\ttbar)} \frac{\rd ^{2}\sigma(\ttbar)}{\rd \mtt\rd \dphitt}$ [\GeVns{}$^{-1}$rad$^{-1}$] & stat. [\%] & syst. [\%] & bin \\
\hline
\multirow{3}{*}{$    340$--$    400$} & $      0$--$    2.2$ & $5.68 \times 10^{-4}$ & $ 5.5$ & ${}_{ -8.1}^{+10.1}$ & $ 1$\\
                                      & $    2.2$--$   2.95$ & $2.68 \times 10^{-3}$ & $ 3.3$ & ${}_{ -6.4}^{ +5.3}$ & $ 2$\\
                                      & $   2.95$--$    \pi$ & $6.67 \times 10^{-3}$ & $ 5.6$ & ${}_{-13.7}^{+22.1}$ & $ 3$\\
\hline
\multirow{3}{*}{$    400$--$    500$} & $      0$--$    2.2$ & $2.59 \times 10^{-4}$ & $ 5.9$ & ${}_{-10.2}^{+12.1}$ & $ 4$\\
                                      & $    2.2$--$   2.95$ & $2.09 \times 10^{-3}$ & $ 2.5$ & ${}_{ -6.8}^{ +5.2}$ & $ 5$\\
                                      & $   2.95$--$    \pi$ & $8.90 \times 10^{-3}$ & $ 2.3$ & ${}_{ -3.3}^{ +4.9}$ & $ 6$\\
\hline
\multirow{3}{*}{$    500$--$    650$} & $      0$--$    2.2$ & $7.04 \times 10^{-5}$ & 11.0 & ${}_{-16.0}^{+19.6}$ & $ 7$\\
                                      & $    2.2$--$   2.95$ & $7.23 \times 10^{-4}$ & $ 4.4$ & ${}_{-13.8}^{+10.5}$ & $ 8$\\
                                      & $   2.95$--$    \pi$ & $4.11 \times 10^{-3}$ & $ 2.9$ & ${}_{ -7.4}^{ +7.0}$ & $ 9$\\
\hline
\multirow{3}{*}{$    650$--$   1500$} & $      0$--$    2.2$ & $6.06 \times 10^{-6}$ & 13.3 & ${}_{-22.1}^{+16.8}$ & 10\\
                                      & $    2.2$--$   2.95$ & $6.40 \times 10^{-5}$ & $ 6.0$ & ${}_{ -8.3}^{+10.5}$ & 11\\
                                      & $   2.95$--$    \pi$ & $4.21 \times 10^{-4}$ & $ 3.6$ & ${}_{ -5.2}^{ +4.1}$ & 12\\
\end{tabular}

\end{table*}
 \begin{table*}
\centering
\topcaption{The correlation matrix of statistical uncertainties for the normalized \ttbar double-differential cross sections as a function of \mtt and \dphitt. The values are expressed as percentages. For bin indices see Table~\ref{tab:mtt_dphitt_xsec}.}
\label{tab:mtt_dphitt_corr}
\cmsTableSwitch{
\renewcommand*{\arraystretch}{1.40}
\tabcolsep0.40mm
\begin{tabular}{c|rrrrrrrrrrrr}
\quad bin\quad &$     1$&$     2$&$     3$&$     4$&$     5$&$     6$&$     7$&$     8$&$     9$&$    10$&$    11$&$    12$ \\
\hline
$ 1$&$+100.0$&$  -2.3$&$ -44.6$&$ -29.0$&$ -31.3$&$  +4.7$&$ -15.6$&$  -5.0$&$ +11.8$&$  +7.6$&$  +5.1$&$  -6.0$ \\
$ 2$&        &$+100.0$&$ -24.4$&$ -18.1$&$  +2.6$&$ -43.1$&$  +6.8$&$ -30.5$&$  +2.1$&$  -2.8$&$  +5.2$&$  +2.6$ \\
$ 3$&        &        &$+100.0$&$ +16.1$&$ -24.6$&$ -11.2$&$  +3.4$&$ +11.4$&$ -30.7$&$  -5.4$&$  -3.5$&$  +5.1$ \\
$ 4$&        &        &        &$+100.0$&$  -2.5$&$ -18.5$&$ -25.5$&$ -19.7$&$  -1.2$&$  -4.4$&$  +2.3$&$  +2.9$ \\
$ 5$&        &        &        &        &$+100.0$&$ -10.5$&$  -8.8$&$ +14.5$&$ -33.5$&$  +4.1$&$ -23.5$&$  +6.2$ \\
$ 6$&        &        &        &        &        &$+100.0$&$  +4.9$&$ -20.4$&$  +6.9$&$  -2.3$&$  +7.1$&$ -22.7$ \\
$ 7$&        &        &        &        &        &        &$+100.0$&$  -7.5$&$  -9.2$&$ -39.9$&$  -9.2$&$  +4.5$ \\
$ 8$&        &        &        &        &        &        &        &$+100.0$&$ -18.5$&$  +0.6$&$ -29.5$&$ -10.3$ \\
$ 9$&        &        &        &        &        &        &        &        &$+100.0$&$  +3.7$&$  -0.0$&$ -31.9$ \\
10&        &        &        &        &        &        &        &        &        &$+100.0$&$ -18.3$&$  -3.3$ \\
11&        &        &        &        &        &        &        &        &        &        &$+100.0$&$ -23.0$ \\
12&        &        &        &        &        &        &        &        &        &        &        &$+100.0$ \\
\end{tabular}
}
\end{table*}
 \begin{table*}
\centering
\topcaption{Sources and values of the relative systematic uncertainties in percent of the measured normalized \ttbar double-differential cross sections as a function of \mtt and \dphitt.
For bin indices see Table~\ref{tab:mtt_dphitt_xsec}.}
\label{tab:mtt_dphitt_syst}
\renewcommand*{\arraystretch}{1.80}
\tabcolsep0.87mm
\cmsTableSwitch{
\begin{tabular}{l|rrrrrrrrrrrr}
            syst. source / bin&$                 1$&$                 2$&$                 3$&$                 4$&$                 5$&$                 6$&$                 7$&$                 8$&$                 9$&$                10$&$                11$&$                12$\\
\hline
              Jet energy scale&${}^{ +1.9}_{ +0.9}$&${}^{ +0.5}_{ -2.4}$&${}^{ +2.8}_{ -2.9}$&${}^{ -1.5}_{ +3.3}$&${}^{ -2.7}_{ +2.4}$&${}^{ +0.3}_{ +0.2}$&${}^{ -1.8}_{ +0.8}$&${}^{ -2.5}_{ +2.3}$&${}^{ +0.9}_{ -0.5}$&${}^{ -6.9}_{ -3.1}$&${}^{ +3.9}_{ -1.2}$&${}^{ +1.2}_{ -3.0}$\\
         Jet energy resolution&${}^{ +1.1}_{ +2.1}$&${}^{ -0.8}_{ +0.4}$&${}^{ -0.9}_{ -2.6}$&${}^{ -2.1}_{ -1.7}$&${}^{ +0.3}_{ -0.1}$&${}^{ +0.7}_{ -0.0}$&${}^{ +5.5}_{ +4.0}$&${}^{ -0.4}_{ +0.3}$&${}^{ +0.1}_{ -0.3}$&${}^{ -3.6}_{ -4.8}$&${}^{ +0.5}_{ +2.4}$&${}^{ -1.0}_{ -0.1}$\\
           Kin. reconstruction&${}^{ +0.0}_{ -0.0}$&${}^{ -0.0}_{ +0.0}$&${}^{ -0.0}_{ +0.0}$&${}^{ +0.0}_{ -0.0}$&${}^{ -0.0}_{ +0.0}$&${}^{ +0.0}_{ -0.0}$&${}^{ +0.0}_{ -0.0}$&${}^{ -0.0}_{ +0.0}$&${}^{ -0.0}_{ +0.0}$&${}^{ -0.0}_{ +0.0}$&${}^{ +0.0}_{ -0.0}$&${}^{ +0.0}_{ -0.0}$\\
                        Pileup&${}^{ -0.2}_{ +0.1}$&${}^{ -0.2}_{ +0.1}$&${}^{ +1.1}_{ -1.3}$&${}^{ +0.1}_{ -0.0}$&${}^{ -0.4}_{ +0.4}$&${}^{ +0.4}_{ -0.3}$&${}^{ -1.0}_{ +1.3}$&${}^{ -0.6}_{ +0.5}$&${}^{ +0.2}_{ -0.1}$&${}^{ -0.9}_{ +0.8}$&${}^{ +0.1}_{ -0.2}$&${}^{ -0.0}_{ -0.0}$\\
                       Trigger&${}^{ +0.0}_{ -0.0}$&${}^{ -0.0}_{ +0.0}$&${}^{ -0.1}_{ +0.1}$&${}^{ +0.0}_{ -0.0}$&${}^{ -0.0}_{ +0.0}$&${}^{ +0.1}_{ -0.1}$&${}^{ +0.1}_{ -0.1}$&${}^{ -0.1}_{ +0.2}$&${}^{ -0.0}_{ +0.0}$&${}^{ -0.3}_{ +0.3}$&${}^{ +0.0}_{ +0.0}$&${}^{ +0.1}_{ -0.1}$\\
\end{tabular}
}
\end{table*}

\section{PDF fit of single-differential \texorpdfstring{\ttbar}{ttbar} measurement at NNLO}
\label{sec:appb}
\begin{figure*}
  \centering
  \includegraphics[width=0.495\textwidth]{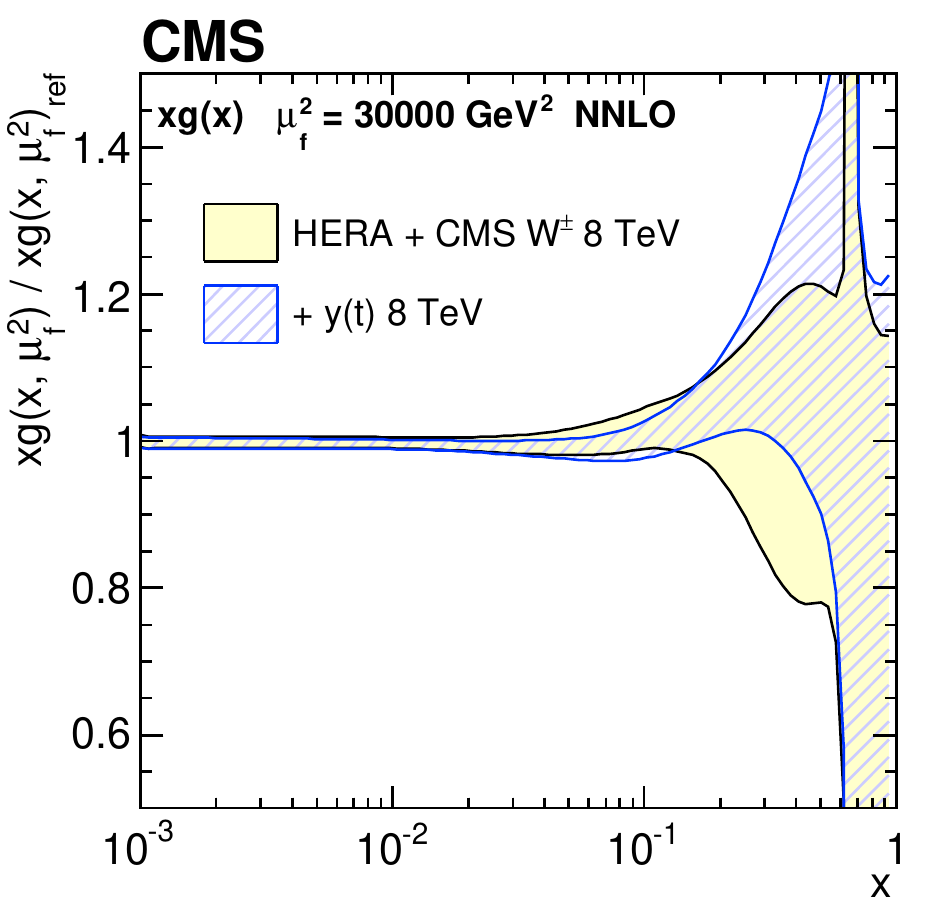}
  \includegraphics[width=0.495\textwidth]{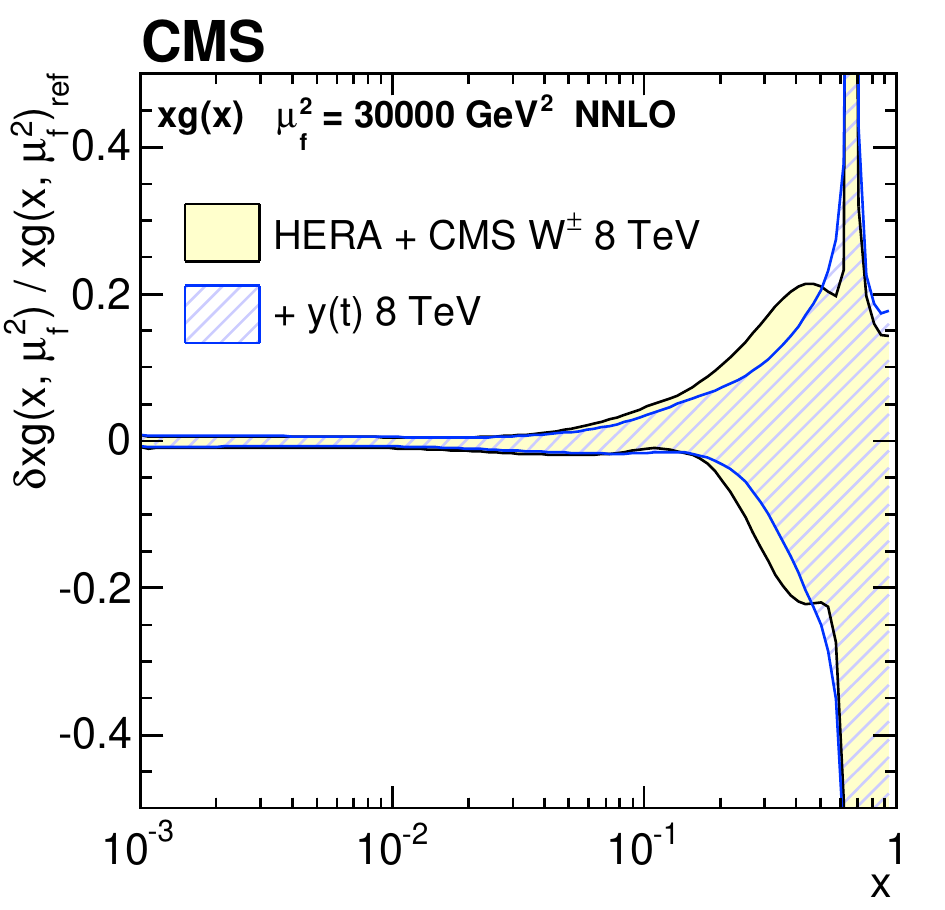}
  \caption{The gluon distribution (left) and its fractional total uncertainty (right) at $\mu_\mathrm{f}^2=30\,000\GeV^2$,
    as obtained in the PDF fit at NNLO using the DIS and \Wasymm data only,
    as well as \yt cross sections. The distributions
    shown in the left panel are normalized to the results from the fit using the DIS and \Wasymm data only.
    The total uncertainty of each distribution is shown by a shaded (or hatched) band.}
  \label{fig:pdf_nnlo}
\end{figure*}

Approximate NNLO predictions~\cite{Kidonakis:2001nj} for the \yt single-differential cross section are obtained using the \difftop program,
which is interfaced to \fastnlo~\cite{ORIG:Britzger:2012bs} (version 2.1).
The results are used in a PDF fit at NNLO.
The procedure follows the determination of the PDFs at NLO described in Section~\ref{sec:qcdanalysis:details}.
In the NNLO fit, the scales for \ttbar production are set to $\mu_\mathrm{r} = \mu_\mathrm{f} = m_{\PQt}$,
with $m_{\PQt} = 173$\GeV being the top quark pole mass.
The scale evolution of partons is calculated through the DGLAP equations at NNLO.
The DIS and \Wasymm theoretical predictions are calculated at NNLO accuracy.
For the \Wasymm predictions, the NNLO corrections are obtained by using K-factors, defined as the ratios of the predictions at NNLO to the
ones at NLO, both calculated with the \FEWZ~\cite{Li:2012wna} program (version 3.1), using the NNLO CT10~\cite{Lai:2010vv} PDFs.
As in Ref.~\cite{Abramowicz:2015mha}, the charm quark mass parameter is set to $M_{\PQc}= 1.43$\GeV for a fit at NNLO.
To stabilise for the comparison, the fit of the gluon distribution at NNLO, which suffers from
insufficient constraints when using the inclusive HERA DIS and \Wasymm data alone,
the $Q^2$ range of the HERA data is further restricted to $Q^2 > Q^2_\text{min} = 7.5\GeV^2$.
In addition, a reduced set of 15 parameters is used for the PDFs, which are
parametrized at the initial scale of the QCD evolution as:
\begin{eqnarray}
x\Pg(x) &=& A_{\Pg} x^{B_{\Pg}}\,(1-x)^{C_{\Pg}}\, - A'_{\Pg} x^{B'_{\Pg}}\,(1-x)^{C'_{\Pg}},
\label{eqNNLO:g} \nonumber\\
x\cPqu_v(x) &=& A_{\cPqu_v}x^{B_{\cPqu_v}}\,(1-x)^{C_{\cPqu_v}}\,(1+E_{\cPqu_v}x^2) ,
\label{eqNNLO:uv} \nonumber\\
x\cPqd_v(x) &=& A_{\cPqd_v}x^{B_{\cPqd_v}}\,(1-x)^{C_{\cPqd_v}},
\label{eqNNLO:dv:nnlo}\\
x\overline{U}(x)&=& A_{\overline{U}}x^{B_{\overline{U}}}\, (1-x)^{C_{\overline{U}}}\, (1+E_{\overline{U}}x^{2}),
\label{eqNNLO:Ubar} \nonumber\\
x\overline{D}(x)&=& A_{\overline{D}}x^{B_{\overline{D}}}\, (1-x)^{C_{\overline{D}}}\, (1+E_{\overline{D}}x^{2}).
\label{eqNNLO:Dbar} \nonumber
\end{eqnarray}
The PDF uncertainty estimation follows the NLO fit procedure described in Section~\ref{sec:qcdanalysis:details},
except for the model parameter variations of $5 \leq Q^2_{\text{min}}\leq 10\GeV^2$ and $1.37\leq M_{\PQc}\leq 1.49\GeV$.
The resulting gluon distribution at a scale of $\mu_\mathrm{f}^2=30\,000\GeV^2 \simeq m_{\PQt}^2$ is shown in Fig.~\ref{fig:pdf_nnlo}, together with its uncertainty band.
The reduction of the total gluon PDF uncertainty is noticeable at large $x$, once the \ttbar cross sections are included in the fit.
This impact is smaller compared to the one observed in the 18-parameter fit at NLO (Fig.~\ref{fig:pdf_relunc_1d}).

\cleardoublepage \section{The CMS Collaboration \label{app:collab}}\begin{sloppypar}\hyphenpenalty=5000\widowpenalty=500\clubpenalty=5000\textbf{Yerevan Physics Institute,  Yerevan,  Armenia}\\*[0pt]
A.M.~Sirunyan, A.~Tumasyan
\vskip\cmsinstskip
\textbf{Institut f\"{u}r Hochenergiephysik,  Wien,  Austria}\\*[0pt]
W.~Adam, E.~Asilar, T.~Bergauer, J.~Brandstetter, E.~Brondolin, M.~Dragicevic, J.~Er\"{o}, M.~Flechl, M.~Friedl, R.~Fr\"{u}hwirth\cmsAuthorMark{1}, V.M.~Ghete, C.~Hartl, N.~H\"{o}rmann, J.~Hrubec, M.~Jeitler\cmsAuthorMark{1}, A.~K\"{o}nig, I.~Kr\"{a}tschmer, D.~Liko, T.~Matsushita, I.~Mikulec, D.~Rabady, N.~Rad, B.~Rahbaran, H.~Rohringer, J.~Schieck\cmsAuthorMark{1}, J.~Strauss, W.~Waltenberger, C.-E.~Wulz\cmsAuthorMark{1}
% ...
\vskip\cmsinstskip
% ...
1:~~Also at Vienna University of Technology, Vienna, Austria\\
% ...

\end{sloppypar}
\end{document}